\documentclass[prd,preprintnumbers,aps,showpacs,tightenlines,notitlepage,nofootinbib,superscriptaddress]{revtex4-1}
\usepackage{mathrsfs}
\usepackage{amsfonts}
\usepackage{amsmath}
\usepackage{array}
\usepackage{verbatim}
\usepackage{epsfig}
\usepackage{graphicx} 
\usepackage{color} 
\usepackage{cancel}
\usepackage{here}

\newcommand*{\FigPath}{./figs}%
\newcommand*{\BibPath}{.}%

\def\bea#1\eea{\begin{align}#1\end{align}}

\begin{document}

\preprint{JLAB-THY-15-2044}

\title{Extraction of Quark Transversity Distribution and Collins Fragmentation Functions
with QCD Evolution}

\author{Zhong-Bo Kang}
\email{zkang@lanl.gov}
\affiliation{Theoretical Division, 
                   Los Alamos National Laboratory, 
                   Los Alamos, NM 87545, USA}

\author{Alexei Prokudin}
\email{prokudin@jlab.org}
\affiliation{Jefferson Lab, 
                   12000 Jefferson Avenue, 
                   Newport News, VA 23606, USA}

\author{Peng Sun}
\email{psun@lbl.gov}
\affiliation{Nuclear Science Division, 
                   Lawrence Berkeley National Laboratory, 
                   Berkeley, CA 94720, USA}

\author{Feng Yuan}
\email{fyuan@lbl.gov}
\affiliation{Nuclear Science Division, 
                   Lawrence Berkeley National Laboratory, 
                   Berkeley, CA 94720, USA}

\begin{abstract}
We study the transverse momentum dependent (TMD) evolution of the 
Collins azimuthal asymmetries in $e^+e^-$ annihilations and semi-inclusive 
hadron production in deep inelastic scattering (SIDIS) processes. 
All the relevant coefficients are calculated
up to the next-to-leading logarithmic (NLL) order accuracy. 
By applying the TMD evolution at the approximate NLL order
in the Collins-Soper-Sterman (CSS) formalism, 
we extract transversity distributions for $u$ and $d$ quarks and 
Collins fragmentation functions from current experimental data 
by a global analysis of the Collins asymmetries in back-to-back di-hadron
productions in $e^+e^-$ annihilations measured by BELLE and 
{\em BABAR} Collaborations and SIDIS data from HERMES, COMPASS, 
and JLab HALL A experiments. The impact of the evolution 
effects and the relevant theoretical uncertainties are discussed.
We further discuss the TMD interpretation for our results, and illustrate 
the unpolarized quark distribution, transversity distribution, unpolarized quark
fragmentation and Collins fragmentation functions depending on
the transverse momentum and the hard momentum scale.
We make detailed predictions for future experiments and discuss their impact.
\end{abstract}

\maketitle

\section{Introduction \label{introduction}}
The transversity distribution function is one of the three leading-twist quark distributions of 
nucleon  that describe completely
spin-1/2 nucleon~\cite{Ralston:1979ys,Jaffe:1991kp,Barone:2001sp,Boer:2011fh}. 
Different from the other two, unpolarized and helicity distributions, the quark transversity is
difficult to measure in experiment because of its chiral-odd nature~\cite{Jaffe:1991kp}. 
In order to study it in a physical process, one has to couple it to another chiral-odd function. 
The first moments (integral over the longitudinal momentum fraction) of the
quark transversity distributions lead to the quark contributions to the nucleon tensor
charge, which is a fundamental property of the nucleon. 

An important channel to investigate the quark transversity distribution is to measure the
Collins azimuthal spin asymmetries in semi-inclusive hadron
production in deep inelastic scattering (SIDIS)~\cite{Collins:1992kk}. 
Measurements have been made by the HERMES Collaboration~\cite{Airapetian:2004tw,Airapetian:2010ds}, the COMPASS Colaboration~\cite{Adolph:2012sn}, 
and JLab HALL A~\cite{Qian:2011py} experiments.
However, the extraction of the quark transversity distributions requires
the knowledge of the Collins fragmentation functions, which
are different from the usual unpolarized fragmentation functions. It was further
suggested to measure the Collins fragmentation functions from the 
azimuthal angular asymmetries of two back-to-back hadron productions in $e^+e^-$ annihilations~\cite{Boer:1997mf}. 
Recently both BELLE and {\em BABAR} Collaborations have studied these
asymmetries at the B-factories at center of mass energy around
$\sqrt{s}\simeq 10.6$ GeV~\cite{Abe:2005zx,Seidl:2008xc,Garzia:2012za}. 
Thanks to the universality of the Collins fragmentation 
functions~\cite{Metz:2002iz}, we will be able to combine the
analysis of these two processes to constrain the quark 
transversity distributions. 
The effort to extract the transversity distributions and Collins fragmentation functions 
has been carried out by the Torino-Cagliari-JLab group extensively in the last 
few years~\cite{Anselmino:2007fs,Anselmino:2008jk,Anselmino:2013vqa}.  
Transversity coupled to the so-called dihadron interference fragmentation 
functions is employed to study transversity in its collinear version in Ref.~\cite{Radici:2015mwa}.
These results have demonstrated the powerful capability of the Collins
asymmetry measurements in constraining the quark transversity distributions 
and hence the nucleon tensor charge in high energy scattering experiments. 
In this study we will implement, for the first time, the appropriate QCD evolution
for the phenomenological studies of Refs.~\cite{Anselmino:2007fs,Anselmino:2008jk,Anselmino:2013vqa}
and thus improve significantly our understanding of transversity distribution and Collins 
fragmentation functions. We will also show the consistency with previous phenomenological results. 
A brief summary of our results has been published in Ref.~\cite{Kang:2014zza}. 

The appropriate QCD evolution for these low transverse momentum
hard processes is the so-called transverse momentum dependent (TMD) evolution, 
which follows from factorization theorems and has been well developed in 
recent years, following the pioneering works by Collins-Soper-Sterman 
(CSS)~\cite{Collins:1981uk,Collins:1984kg}.  
 In particular, the Collins 2011 formalism of Ref.~\cite{Collins:2011zzd} constructs 
a well defined universal TMDs that ``absorb" effects of soft gluon radiation
which was traditionally signed out in a separate factor in 
Refs.~\cite{Collins:1981uk,Collins:1984kg}, and defines a hard part function
that contains the process dependence. 
This allows for an explicit universality of the TMDs in the phenomenological
applications of the hard scattering processes mentioned above. 
 
The TMD evolution effects in the Collins asymmetries in the above processes 
have been estimated in Refs.~\cite{Boer:2001he}.
The TMD factorization is an important step to derive the results for the physical observables
we are interested in, and has been shown to be valid for 
processes with two separate measured momentum scales $Q_T \ll Q$, such 
as SIDIS, Drell-Yan and $e^+e^-$ annihilation into back-to-back 
hadrons. Here the small scale $Q_T$ corresponds to the measured transverse 
momentum of, for instance, produced hadron in SIDIS, lepton pair in 
Drell-Yan. 
TMD factorization is formulated in such a way that one can calculate cross sections 
up to the values $Q_T\sim \Lambda_{\rm QCD}$ and thus 
incorporates non-perturbative information on the hadron structure. 
Schematically the TMD factorization expresses the transverse-momentum-dependent
differential cross section as a convolution of a so-called hard part $H$, which is specific for the process and 
thus process dependent, and universal TMD parton 
distributions and/or TMD parton fragmentation functions, collectively called TMDs.
These TMDs are universal (for the ``naively time reversal odd" functions such as 
Sivers function \cite{Sivers:1990cc,Sivers:1991fh} and Boer-Mulders 
function~\cite{Boer:1997nt} the universality is generalized~\cite{Brodsky:2002cx,Collins:2002kn}) 
and can be associated with properties of specific hadrons. In this sense 
TMDs represent the three-dimensional partonic structure of the incoming nucleons 
as well as outgoing hadrons. 
Evolution equations are used to calculate the dependence of TMDs on 
the hard scale $Q$. Since the definition of TMDs contains the so-called 
light-cone singularity~\cite{Collins:1981uk}, the detailed calculations depend on 
the scheme to regulate this singularity~\cite{Collins:1981uk,
Collins:1984kg,Ji:2004wu,Collins:2004nx,Collins:2011zzd,Mantry:2009qz,Becher:2010tm,
GarciaEchevarria:2011rb,Chiu:2012ir,Ji:2014hxa}, which leads to the scheme dependence in the TMD factorization.
Although there are different ways to formulate the TMD factorization
and to define the TMD distribution and fragmentation functions, 
the energy evolution (historically called ``resummation")  for the {\em physical observables} (including the 
transverse momentum dependent differential cross
sections and spin asymmetries) will take the same form in all schemes. 
In particular, after solving the evolution equations, the final results are identical to each other
in all TMD factorization schemes,
where the TMDs are expressed in terms of their collinear counterparts
with perturbatively calculable coefficients, and the evolution effects are 
included in the exponential factor -- the so-called Sudakov-like form factors. 
Therefore, in terms of a phenomenological study,
all TMD factorization and evolution calculations will be identical 
to that originally proven in the form of CSS~\cite{Collins:1984kg}. 
Interpretation of results and individual functions depends of course 
on the scheme and one should be very careful when giving interpretations.

TMD evolution is performed in coordinate $b$-space, where $b$ 
is conjugate to the $k_\perp$ in momentum space through the Fourier transformation
and corresponds to the transverse distance separating the quark/gluon fields. The usage of 
$b$-space highly simplifies the expressions for the cross sections which 
become simple products of $b$ dependent TMDs in contrast to 
convolutions in $k_\perp$ space. In order to calculate the measured 
cross-sections (and individual TMDs) one performs a two dimensional 
Fourier transform to the physical $Q_T$ (or $k_\perp$) space. 
A very unique feature of TMD/CSS formalism is the fact that the evolution 
kernel becomes non-perturbative at large separation distances 
$b$; while at small $b \ll 1/\Lambda_{\rm QCD}$ it is perturbative 
and can be calculated order by order in strong coupling constant 
$\alpha_s (1/b)$. Over short transverse distance scales, $1/b$ 
becomes a legitimate hard scale, and the $b$ 
dependence of TMDs can be calculated in perturbation theory 
and related to their {\em collinear} counterparts, such as
collinear parton distribution (PDFs), 
fragmentation functions (FFs) or multiparton correlation 
functions. The important non-perturabrive part of the so-called 
Soft factor that corresponds to vacuum expectation value of 
Wilson loops is predicted \cite{Collins:2011zzd} to be process 
independent, it is also independent of the fact that the individual 
TMDs is distribution or fragmentation function and independent 
of the particular value of momentum fractions $x_B$ or $z_h$ 
measured. It may depend on the parton type, quark or a gluon, 
in this paper we are going to consider only quark distribution and 
fragmentation TMDs. The information on intrinsic non-perturbative 
motion of patrons associated with the hadron wave function is 
encoded in non-perturbative inputs for TMD PDFs and FFs and 
in turn universal in different processes but in principle dependent 
on the parton/hadron type and on value of $x_B$ or $z_h$. 

The implementation of the TMD formalism requires parametrization of the 
non-perturbative inputs~\cite{Landry:2002ix,Qiu:2000ga,Aidala:2014hva,
Kang:2011mr,Sun:2013hua,Echevarria:2014xaa,Collins:2014jpa} 
for the TMDs. The growth of $\alpha_s (1/b)$ at large values of $b$ 
can be tamed by the so-called $b_*$-prescription (which we will follow in this paper) 
originally  introduced in the CSS formalism~\cite{Collins:1984kg} that allows 
one to avoid Landau pole in strong coupling constant and 
provides a smooth transition from perturbative to non-perturbative 
regimes. Fits of experimental data utilizing $b_*$ prescription have been well 
developed in the literature, in particular, in the publications
of the BLNY-type of parameterizations~\cite{Landry:2002ix,Su:2014wpa}. Other choices have been
made in the literature, see, for example, 
Refs.~\cite{Qiu:2000ga,Kulesza:2002rh,Catani:2000vq, Catani:2003zt,Bozzi:2003jy}.
However, in all these implementations of the TMDs in the CSS formalism, 
an important step is to verify that they provide a robust method of treating  
non-perturbative physics and  can well describe the existing experimental
data~\cite{Sun:2013hua,Aidala:2014hva}. 

For the Collins asymmetries studied in this paper, we  extend the CSS
formalism to the azimuthal angular asymmetries and in the relevant hard 
processes. This involves the Collins-Soper (CS) evolution equation for the 
$k_\perp$-odd distribution and fragmentation functions, which were
derived in Refs.~\cite{Idilbi:2004vb,Ji:2006ub,Ji:2006vf,Yuan:2009dw,Kang:2011mr,Sun:2013dya,Echevarria:2014rua}.
In our calculations, we apply the TMD evolution at the approximate 
next-to-leading-logarithmic order (NLL') as specified below. The formalism follows the CSS procedure
for the unpolarized cross section, and is similar to that of
Sivers asymmetries in SIDIS and Drell-Yan processes~\cite{Kang:2011mr,Sun:2013hua,Echevarria:2014xaa,Aybat:2011ge}. 
We will derive the perturbative coefficients at one-loop order as well.

There exists a freedom (scheme-dependence) to separate out the so-called hard factor from the
splitting function contribution in the CSS formalism~\cite{Catani:2000vq}. This provides a
useful way of interpretation for the final results in terms of the TMDs~\cite{Collins:2011zzd,Aybat:2011zv,scheme}. 
It allows to interpret a part of the splitting functions in CSS as a universal TMDs splitting functions, and
the difference in the coefficients can be regarded as part of hard 
factors. Once rigorously defined, we shall have a unique interpretation
of the CSS formalism in terms of TMDs. We will elaborate this 
interpretation in details in our paper.

In applying the CSS evolution at the NLL order, we relate transversity TMD and 
Collins FF to the collinear quark transversity distribution and the collinear twist-3 
fragmentation function and include
the DGLAP-type scale evolution of the latter two collinear distributions. 
The evolution of transversity distribution is very well 
known~\cite{Barone:1997fh,Hayashigaki:1997dn,Vogelsang:1997ak} while 
the evolution of the twist-3 fragmentation function  involves 
multiparton correlation functions~\cite{Yuan:2009dw,Kang:2010xv}, 
as a common feature of higher-twist correlation functions. 
In the following calculations, we will only keep the homogenous terms
in the splitting kernel, which is an approximation to the complete
evolution equation. To differentiate from the complete NLL computation,
we denote it as NLL$'$ (an approximate NLL). To achieve this precision we 
include the most recent developments from both theory and phenomenology
sides~\cite{Collins:2011zzd,Yuan:2009dw,Kang:2010xv, Kang:2011mr,Echevarria:2012js,
Bacchetta:2013pqa,Sun:2013hua,Echevarria:2014xaa,Echevarria:2014rua,Su:2014wpa}.

The quark transversity distributions are important ingredients for several
other spin related asymmetries. For example, they contribute to the
azimuthal asymmetries of two-hadron fragmentation processes in SIDIS
and $e^+e^-$ annihilations~\cite{Bacchetta:2011ip}, and single inclusive hadron production
at large transverse momentum in $pp$ collisions~\cite{Kang:2010zzb,Anselmino:2012rq,Kanazawa:2014dca}. 
Future RHIC measurements \cite{Aschenauer:2015eha} are going to explore more phenomena related to transverse spin and ultimately to the partonic three-dimensional structure of the nucleon.
Our results will provide important cross checks and a step further
toward a global analysis to all these spin asymmetries associated
with the quark transversity distributions.

The rest of the paper is organized as follows. In Section~\ref{sec:framework}, we review 
the theoretical framework for the Collins azimuthal asymmetries
in SIDIS and $e^+e^-$ annihilations and derive the associated TMD
evolution results and the relevant perturbative coefficients. We also reformulate
the resummation formalism in an appropriate way to better connect to the recently developed TMD
formalism in Section~\ref{sec:interpretation}. In Section~\ref{sec:phenomenology}, we perform the phenomenological studies and focus
on the global fit of the quark transversity distribution and Collins fragmentation
functions from the existing experimental data.
We make predictions for future experiments and compare our results with previous analyses. 
Finally, we conclude our paper in Section~\ref{sec:summary}.

%
%
\section{The Collins Azimuthal Asymmetries in SIDIS and $e^+e^-$ Annihilation}
\label{sec:framework}

In this section, we discuss the asymmetries generated by transversity and Collins fragmentation functions in SIDIS and $e^+e^-$ annihilation. We apply TMD evolution and represent the differential cross sections,
spin-dependent and spin-independent ones, in a compact form. 

%
%
\subsection{Collins Azimuthal Asymmetries in   SIDIS \label{subsectionI}}

\begin{figure}[tbh]
\includegraphics[width=8cm]{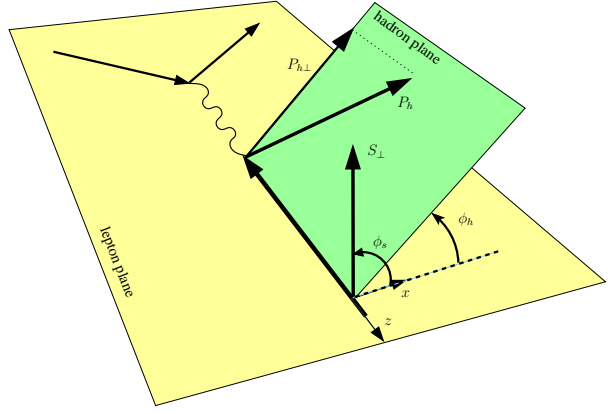}
\caption{Semi Inclusive Deep Inelastic Scattering process (SIDIS) in $\gamma^* P$ center of mass frame.}
\label{fig:sidis}
\end{figure}

In the SIDIS, see Fig.~\ref{fig:sidis}, a lepton scatters on the nucleon target, and produces
a hadron in the final state, 
\bea
e (\ell)+p(P)\to e(\ell') + h(P_h) + X\; ,
\eea
by exchanging a virtual photon $q_\mu=\ell_\mu-\ell'_\mu$ with invariant mass
$Q^2=-q^2$.  We adopt the usual SIDIS variables \cite{Meng:1991da}:
\bea
S_{ep} = (P+\ell)^2,
\quad
x_B = \frac{Q^2}{2P\cdot q},
\quad
y=\frac{P\cdot q}{P\cdot \ell} = \frac{Q^2}{x_B S_{ep}},
\quad
z_h = \frac{P\cdot P_h}{P\cdot q},
\eea
with $S_{ep} = (\ell + P)^2$ the center of mass energy square. 
The differential SIDIS cross section that includes the Collins effect, 
the $\sin\left(\phi_h +\phi_s\right)$ modulation, can be written as~\cite{Bacchetta:2006tn,Anselmino:2008sga}, 
\bea
\frac{d^5\sigma(S_\perp)}{dx_B dy dz_h d^2P_{h\perp}}
&= \sigma_0(x_B, y, Q^2)
\left[F_{UU} +    \sin(\phi_h+\phi_s)\,
\frac{2 (1-y)}{1+(1-y)^2} \, F_{UT}^{\sin\left(\phi_h +\phi_s\right)} + ... \right],
  \label{eq:aut}
\eea
where $\sigma_0 = \frac{2\pi \alpha_{\rm em}^2}{Q^2}\frac{1+(1-y)^2}{y}$, 
and $\phi_s$ and $\phi_h$ are the azimuthal angles for the nucleon spin and the transverse momentum 
of the outgoing hadron, respectively. $F_{UU}$ and $F_{UT}^{\sin(\phi_h+\phi_s)}$ are the spin-averaged and transverse spin-dependent 
structure functions. The latter is related to convolution of transversity distribution and Collins fragmentation function.
Ellipsis in Eq.\eqref{eq:aut} denote other structure functions that we do not consider in this paper.


The Collins asymmetry measured experimetnally are related to the structure functions as follows:
\bea
A_{UT}^{\sin(\phi_h+\phi_s)} \equiv 2 \langle \sin\left(\phi_h +\phi_s\right) \rangle = \frac{\sigma_0(x_B, y, Q^2)}{\sigma_0(x_B, y, Q^2)}  \frac{2 (1-y)}{1+(1-y)^2}
\frac{F_{UT}^{\sin\left(\phi_h +\phi_s\right)}}{F_{UU}}.
\eea
Note that sometimes experimental results (for instance for COMPASS collaboration) are presented by factoring out the so-called   depolarization factor $D_{NN}$:
\bea
D_{NN} = \frac{2 (1-y)}{1+(1-y)^2}\; .
\eea
Both structure functions, i.e. $F_{UT}$ and $F_{UT}^{\sin\left(\phi_h +\phi_s\right)}$, depend on kinematical variables and 
on the hard scale $Q^2$ in the reaction. It is important to realize that in order to have reliable calculations of
corresponding structure functions one needs to take into account appropriate scale dependence which is generated by
QCD evolution of TMD distribution and fragmentation functions.

Historically the solution of TMD evolution equations \cite{Collins:1984kg} is presented in the  
$b$-space, where in SIDIS $\vec{b}$ is Fourier conjugate variable to $\vec{P}_{h\perp}/z_h$. 
The $P_{h\perp}$-dependent structure functions can be formulated
in terms of the TMD factorization, and they can be (omitting $x_B$, $z$ dependencies)
written as,
\begin{eqnarray}
F_{UU}(Q;P_{h\perp})&=&\frac{1}{z_h^2}\int \frac{d^2b}{(2\pi)^{2}} e^{i \vec{P}_{h\perp}\cdot \vec{b}/z_h} \widetilde
{F}_{UU}(Q;b) +Y_{UU}(Q;P_{h\perp}) \ , \label{eq:fuu-Y} \\
F_{\rm collins}^\alpha(Q;P_{h\perp})&=&\frac{1}{z_h^2}\int \frac{d^2b}{(2\pi)^{2}} e^{i \vec{P}_{h\perp}\cdot \vec{b}/z_h} \widetilde
{F}_{\rm collins}^\alpha(Q;b) +Y_{\rm collins}^\alpha(Q;P_{h\perp}) \ ,
\label{eq:fut-Y}
\end{eqnarray}
where $F_{\rm collins}^{\alpha}$ is related to the spin-dependent structure function $F_{UT}^{\sin\left(\phi_h +\phi_s\right)}$ as follows:
\begin{eqnarray}
   \sin(\phi_h+\phi_s)\,
 F_{UT}^{\sin\left(\phi_h +\phi_s\right)}  = \
      \epsilon^{\alpha\beta}S_\perp^\alpha\left[g_\perp^{\beta\rho}-2\hat{e}_x^\beta\hat{e}_x^\rho\right]
 F_{\rm collins}^{\rho}\, ,
 \label{eq:fut_collins}
\end{eqnarray}
with the unit vector $\hat{e}_x$ defined in Fig.~\ref{fig:sidis}. In Eqs.~(\ref{eq:fuu-Y}) and (\ref{eq:fut-Y}), the first TMD term dominates in $P_{h\perp}\ll Q$ region, and the
second so-called Y-factor term dominates in the region of $P_{h\perp}\gtrsim Q$ and assures accuracy of the 
formula in the wide region of $P_{h\perp}$.
 We will neglect the corresponding $Y$ factors as we will consider only the region of low $\vec{P}_{h\perp}/z$, and thus
 for spin-averaged and transverse spin-dependent structure functions one has
 \bea
F_{UU}(Q;P_{h\perp}) &= \frac{1}{z_h^2}\int \frac{d^2b}{(2\pi)^{2}} e^{i \vec{P}_{h\perp}\cdot \vec{b}/z_h} \widetilde
{F}_{UU}(Q;b) \; \label{fuu},\\
F_{UT}^{\sin\left(\phi_h +\phi_s\right)}(Q;P_{h\perp}) &= \frac{1}{z_h^2}\int \frac{d^2b}{(2\pi)^{2}} e^{i \vec{P}_{h\perp}\cdot \vec{b}/z_h} \hat P^\alpha_{h\perp} \widetilde
{F}_{\rm collins}^\alpha(Q;b)\; , \label{fut}
\eea 
one notices that while spin independent structure function is a scalar quantity, the spin dependent structure function depends on the transverse direction $\alpha=1,2$; see Eqs.~(\ref{eq:fut_collins},\ref{fut}).

%
%
\subsubsection{Unpolarized structure function with evolution}

The factorization formula for unpolarized structure function $F_{UU}(Q;P_{h\perp})$ is well known and has the following interpretation (we choose  Ji-Ma-Yuan \cite{Ji:2004wu,Ji:2004xq} scheme for the moment) in terms of 
unpolarized distribution and fragmentation functions in the $b$ space \cite{Idilbi:2004vb}:

\begin{equation}
\widetilde{F}_{UU}(Q;b) =\sum_q e_q^2 \, \tilde f_1^q(x_B,b;\rho,\zeta,\mu)\widetilde D_{h/q}(z_h,b;\rho,\hat\zeta,\mu)H(Q/\mu,\rho)
S(b,\rho;\mu) \ ,
\label{eq:fuu_bspace}
\end{equation}
where $\tilde f_1^q$ is the unpolarized TMD   distribution, $\widetilde D_{h/q}$ is the unpolarized TMD  
fragmentation function, $\zeta^2=2(v\cdot P_A)^2/v^2$, $\hat \zeta^2=(2\tilde v\cdot P_h)^2/\tilde v^2$,
and $\rho^2=(2v\cdot \tilde v)^2/v^2\tilde v^2$ represent the light-cone singularity regulation parameters.
$H$ is the  hard factor associated with hard scattering and $S$ is the so-called soft function associated with emission of 
soft gluons. 
Renormalisation group scale $\mu$ is arbitrary in full QCD, however in truncated perturbative series it is chosen to optimize the convergence in such a way that
$H$ does not have large logarithmic contributions, $\log(Q/\mu)$, and generically 
$\mu = C_1 Q$ with $C_1$ a parameter of order of $1$. We will utilize $C_1 =1$ and thus $\mu = Q$ in our calculations.
Depending on different schemes; such as Ji-Ma-Yuan \cite{Ji:2004wu,Ji:2004xq}, CSS \cite{Collins:1981uk,Collins:1984kg}, or Collins-11 \cite{Collins:2011zzd}, the dependence on these
parameters will be different. However, the final results for the structure
functions are independent of the schemes, as the actual cross-sections do not depend on the auxiliary parameters. Note that historically $H$ factor 
is absorbed in CSS formulation into the definition of Wilson coefficient functions that relate TMDs 
to the corresponding collinear distributions. The final results for the cross-sections are the same in all schemes.
However, a slight difference stems from the fact that $H$ functions contain 
$\alpha_s(\mu)$ with renormalization group scale $\mu$ while coefficient 
functions, as will be explained below, contain $\alpha_s(\mu_b)$ with a 
dynamical scale $\mu_b$. At each order of perturbation series, these 
differences are of a higher order in $\alpha_s$.
We will dedicate a separate Section~\ref{sec:interpretation} where we 
will discuss TMD interpretation of our results and give explicit TMD 
formulas in TMD Collins-11 \cite{Collins:2011zzd} formulation for all functions and structure functions considered in this paper.

Let us review the definition and the need of different factors. 
The TMD quark distributions in SIDIS is defined
through the following matrix:
\begin{eqnarray}
    {\cal M}^{\alpha\beta} &=&   P^+\int
        \frac{d\xi^-}{2\pi}e^{-ix\xi^-P^+}\int
        \frac{d^2b}{(2\pi)^2} e^{i\vec{b}\cdot
        \vec{k}_\perp}  \left\langle
PS\left|\overline\psi^\beta(\xi^-,0,\vec{b}){\cal L}_{v}^\dagger(\infty;\xi){\cal L}_{v}(\infty;0)
        \psi^\alpha(0)\right|PS\right\rangle\ ,
\end{eqnarray}
with the gauge link
\begin{equation}
 {\cal L}_{v}(\infty;\xi) \equiv \exp\left(-ig\int^{\infty}_0 d\lambda
\, v\cdot A(\lambda v +\xi)\right) \ .\label{glink}
\end{equation}
This gauge link goes to $+\infty$, indicating that we adopt the
definition for the TMD quark distributions for the SIDIS
process. The unpolarized quark distribution is projected out 
from the above matrix as,
\begin{align}
{\cal M} &= \frac{1}{2}
\left[f_1^q(x,k_\perp)\gamma_\mu P^\mu
 +
\ldots\right] \label{matrixexp} \  , \\
f_1^q(x,k_\perp) &= \frac{1}{4P^+}  {\rm Tr} [\gamma^+ {\cal M}]\ .
\end{align}
However, the above definition of the quark distribution contains soft
gluon contribution, which has to be subtracted from the naive definition.
In addition, there is light-cone singularity if we take the gauge link along 
the light-front direction $v$ with $v^2=0$. The way to regularize this singularity and
subtract soft gluon contribution defines the scheme for the TMD 
factorization.

In the Ji-Ma-Yuan scheme, the gauge link in the TMD definition
is chosen to be slightly off-light-cone, $n=(1^-,0^+,0_\perp)\to v=(v^-,v^+,0_\perp)$
with $v^-\gg v^+$. Similarly, for the TMD fragmentation function, $\tilde v$
was introduced, $\tilde v=(\tilde v^-,\tilde v^+,0_\perp)$ with $\tilde v^+\gg \tilde v^-$.
Because of the additional directions $v$ and $\tilde v$, there are additional invariants:
$\zeta^2=(2v\cdot P)^2/v^2$, $\hat \zeta^2=(2\tilde{v}\cdot P_h)^2/\tilde{v}^2$,
and $\rho^2=(2v\cdot \tilde v)^2/v^2\tilde{v}^2$. Accordingly, the soft factor
is defined as,
\begin{equation}
S^{v,\bar v}(b )={\langle 0|{\cal L}_{\tilde
vcb'}^\dagger(b ) {\cal
L}_{vb'a}^\dagger(b ){\cal L}_{vab}(0){\cal
L}_{\tilde vbc}(0)  |0\rangle   }\, . \label{softg}
\end{equation}
With soft factor subtraction, the TMD factorization for the unpolarized structure function can be rewritten as
\begin{equation}
{F}_{UU}(Q;b) =\sum_q e_q^2\, \tilde f_{1\rm JMY}^{q\; (sub)}(x_B,b;\rho,\zeta,\mu) \widetilde D_{q\rm JMY}^{(sub)}(z_h,b ;\rho,\hat\zeta,\mu)H_{UU}^{JMY}(Q/\mu,\rho)\ ,
\label{eq:fuu_b}
\end{equation}
 where the subtracted quark distribution and fragmentation
functions are defined as
\begin{eqnarray}
\tilde f_{1\rm JMY}^{q\; (sub)}(x_B,b ;\rho,\zeta,\mu)&=& \frac{\tilde f_1^q(z_h,b ;\rho,\zeta,\mu)}{\sqrt{S(b ,\rho;\mu)}}\ ,
\label{eq:sqrt0}
\\
\widetilde D_{q\rm JMY}^{(sub)}(z_h,b ;\rho,\hat\zeta,\mu)&=&\frac{\widetilde D_q(z_h,b ;\rho,\hat\zeta,\mu)} {\sqrt{S(b ,\rho;\mu)}} \ ,
\label{eq:sqrt}
\end{eqnarray}
with the soft factor $S$ subtracted from the original TMDs. After solving the evolution
equations and expressing the TMDs in terms of the integrated parton
distributions, the final expressions for TMDs are obtained by setting $\zeta^2=\hat \zeta^2=\rho Q^2$. 
Note that in Eqs.~(\ref{eq:sqrt0},\ref{eq:sqrt}) we understand the square root in the perturbative sense, 
i.e. for any quantity $A = 1+ a_1 \, \alpha_s +    ...$ one has $1/\sqrt{A} = 1- 1/2 \, a_1 \alpha_s -  ...\;\;$.

On the other hand, as explained in Introduction the new Collins-11 approach~\cite{Collins:2011zzd} is an important improvement of the original CSS formalism and includes now operator definition of TMDs, and the soft factor subtraction is taken to ensure 
the absence of light-cone singularities in the TMDs and the self-energy divergencies of the
soft factors. According to this new scheme, the TMD distribution is defined as
\begin{equation}
\tilde f_1^{q \; \; \rm
JCC}(x,b; \zeta_F, \mu)=\tilde f_1^{q}(x,b; \zeta_F, \mu)\sqrt{\frac{S^{\bar
n,v}(b)}{S^{n,\bar n}(b)S^{n,v}(b)}} \ ,
\end{equation}
where $\zeta_F^2=x^2(2v\cdot P_A)^2/v^2=2(xP_A^+)^2e^{-2y_n}$ with
$y_n$ the rapidity cut-off in Collins-11 scheme.  Fragmentation functions are defined analogously. 
The unpolarized structure function takes the form
\begin{equation}
{F}_{UU}(Q;b) =\sum_q e_q^2\, \tilde f_1^{q\; JCC}(x_B,b ;\zeta_F,\mu) \widetilde D_q^{JCC}(z_h,b ,\zeta_D;\mu)H^{JCC}_{UU}(Q/\mu)\ .
\label{eq:fuu_b_jcc}
\end{equation}

One can see from Eqs.~(\ref{eq:fuu_b},\ref{eq:fuu_b_jcc}) the formal expression for structure functions take a very simple ``parton model"-like form. The underlying TMDs, however, unlike the parton model expression are computed with an appropriate QCD evolution procedure.
In spite of differences in the schemes to define TMDs in Eqs.~(\ref{eq:fuu_b},\ref{eq:fuu_b_jcc}) 
the final result for the structure functions and the cross-sections is scheme independent and 
reduces to that of the original CSS. One of the advantages of the TMD schemes, such as 
Collins-11 approach~\cite{Collins:2011zzd} or Ji-Ma-Yuan \cite{Ji:2004wu,Ji:2004xq}, is a 
possibility to define process-independent unpolarized TMDs or account for process 
dependence in $k_\perp$-odd TMDs directly and study individual TMD functions. 
An important process-independent universal non-perturbative contributions~\cite{Collins:2014jpa} 
can be also studied and global TMD fits that include different processes are possible.

The corresponding TMDs depend on two scales $\zeta_F$ (or $\zeta_D$) and $\mu$, with their dependence encoded in 
the TMD evolution equations. The rapidity evolution with respect to $\zeta$ is given by the Collins-Soper (CS) equation \cite{Collins:1981uk}:
\begin{align}
\frac{\partial \ln \tilde f_q(x_B,b ;\zeta,\mu)}{\partial \ln \sqrt{\zeta_F}}= \frac{\partial \ln \widetilde D_q (z_h,b ,\zeta_D;\mu)}{\partial \ln \sqrt{\zeta_D}} = \tilde K(b, \mu) \; ,
\end{align}
where $\tilde K(b, \mu)$ is the so-called CS kernel \cite{Collins:1981uk}. It can be computed perturbatively for small values of $b$.
The dependence on the scale $\mu$ arises from renormalization group equations for   $\tilde f_q$,  $\widetilde D_q$, and $\tilde K$:
\begin{align}
\frac{d \tilde K(b, \mu)}{d \ln \mu} &= -\gamma_K(\alpha_s(\mu)) \; , \\
\frac{d \ln \tilde f(x_B,b ;\zeta,\mu)}{d \ln \mu}&= \gamma_F(\alpha_s(\mu),\zeta_F/\mu^2) \; , \\
\frac{d \ln \widetilde D_q (z_h,b ,\zeta_D;\mu)}{d \ln \mu}&= \gamma_D(\alpha_s(\mu),\zeta_D/\mu^2) \; ,
\end{align}
where functions $\gamma_K$, $\gamma_F$,  and $\gamma_D$ are anomalous 
dimensions of $\tilde K$, $\tilde f_q$, and $\widetilde D_q$, respectively. 
Note that the solution of these evolution equations does not depend on the scheme to define TMDs.

The equations and solutions are discussed at length in Refs.~\cite{Ji:2004wu,
Ji:2004xq,Collins:1981uk,Collins:1984kg,Collins:2011zzd,Collins:2014jpa}. 
Here we will present and discuss the final solution.
 At low values of $b\ll 1/\Lambda_{QCD}$, $1/b$ becomes a 
legitimate hard scale. One introduces \cite{Collins:1981uk} an 
auxiliary scale $\mu_b = c_0/b$, with $c_0 = 2 e^{-\gamma_E}$ and 
$\gamma_E \approx 0.577$ the Euler's constant. The $b$ 
dependence of TMDs can be computed  in the perturbative   
$1/Q \ll b\ll 1/\Lambda_{QCD}$ region in terms of the collinear parton distribution and fragmentation functions. 
This region corresponds to the transverse momentum 
which is large compared to hadronic scale, but still small 
compared to the hard scale (i.e  $\Lambda_{QCD} \ll k_\perp\ll Q$). 
That is TMDs in this region are expressed in terms of collinear distributions.
This sort of relation will be explained later in the paper. Let us mention that 
usage of such a relation helps to obtain a reliable description of the experimental data.

 The  energy evolution of TMDs from the scale $\mu_b$ to the scale $Q$ 
is encoded in the exponential factor, $\exp[-S]$, with the so-called Sudakov-like form factor, the perturbative 
 part of which can be written as
 \bea
S_{\rm pert}(Q,b)=\int_{\mu_b^2}^{Q^2}\frac{d\bar\mu^2}{\bar\mu^2}\left[A(\alpha_s(\bar\mu))\ln\frac{Q^2}{\bar\mu^2}+B(\alpha_s(\bar\mu))\right] \ ,
\label{spert}
\eea
where $A$ and $B$ coefficients can be expanded as
perturbative series $A=\sum_{n=1}^\infty A^{(n)} \left(\alpha_s/\pi\right)^n$, $B=\sum_{n=1}^\infty B^{(n)} \left(\alpha_s/\pi\right)^n$.
In our calculations,
we will take $A^{(1)}$, $A^{(2)}$ and $B^{(1)}$ for the NLL accuracy. 
Because this part is spin-independent as explained in Introduction, these coefficients are the same as
those in unpolarized cross sections~\cite{Collins:1984kg} and are given by~\cite{Kang:2011mr,Aybat:2011zv,Echevarria:2012pw,Collins:1984kg,Qiu:2000ga,Landry:2002ix}:
\bea
A^{(1)} =C_F,
\;
A^{(2)} =\frac{C_F}{2}\left[C_A\left(\frac{67}{18}-\frac{\pi^2}{6}\right)-\frac{10}{9}T_R \, n_f\right],
\;
B^{(1)} = -\frac{3}{2} C_F
\,. 
\eea

One can see from Eqs.~(\ref{fuu},\ref{fut}) that in order to reconstruct the measured cross-section one needs to perform the Fourier transform over all values of $b$. The accuracy of the perturbative solution will deteriorate for large values of $b$. In fact $\alpha_s(\mu_b)$ will hit the so-called Landau pole which is a good indication of presence of the non-perturbative physics. Thus one needs to take into account the non-perturbative behavior of  TMDs.
The original CSS approach~\cite{Collins:1984kg} proposed the so-called $b_*$-prescription  that introduces a cut-off value $b_{max}$ and allows for a smooth transition from perturbative to non-perturbative region and  avoids the Landau pole
singularity in $\alpha_s(\mu_b)$,  
\begin{equation}
b \Rightarrow b_*=b/\sqrt{1+b^2/b_{max}^2}  \ ,~~b_{max}<1/\Lambda_{QCD}\ ,
\end{equation}
where $b_{max}$ is a parameter in the prescription. From the above definition, $b_*$ is always in
the perturbative region where $b_{max}$ is normally chosen to be around  1 GeV$^{-1}$.
With the introduction of $b_*$ in the Sudakov form factor, the total 
Sudakov-like form factor can be written as the sum of perturbatively calculable part and non-perturbative contribution
\begin{equation}
{S}_{sud}(Q;b)\Rightarrow { S}_{\rm pert}(Q;b_*)+S_{\rm NP}(Q;b) \ ,
\end{equation}
where $S_{\rm NP}(Q;b)$ is defined as the
difference
from the original form factor and the perturbative one. This difference should vanish as $b\to 0$, i.e. in the perturbative region,
and thus $S_{\rm NP}(Q;b)$ has the following generic form
\begin{equation}
S_{\rm NP}(Q;b)=g_2(b)\ln Q/Q_0 +g_1(b) \ .
\end{equation}
The non-perturbative generic functions $g_2$ and $g_1$ have very unique interpretations.  In particular $g_2$ includes the information on large $b$ behavior of the evolution kernel $\tilde K$. This function does not depend on the particular process, it does not depend on the scale and has no dependence on momentum fractions $x_B$, $z$.  
This contribution should be parametrized phenomenologically and an often-used parametrization is
\begin{equation}
g_2(b) = g_2 b^2 \, ,
\end{equation}
which proved to be very reliable to describe Drell-Yan data and $W^\pm,Z$ boson production in the BLNY-type of parameterizations~\cite{Landry:2002ix}.  This gaussian-type parametrization suggests that large $b$ region is strongly suppressed \cite{Aidala:2014hva} and in principle can be unreliable to describe data from lower energies which are more sensitive to moderate-to-high values of $b$. Other parametrizations were proposed in Refs.~\cite{Aidala:2014hva} and \cite{Su:2014wpa}. For instance that of Ref.~\cite{Su:2014wpa} has the form:
\begin{equation}
g_2(b) =  g_2 \ln\left(\frac{b}{b_*}\right) \, ,
\end{equation}
and allows to describe simultaneously 
unpolarized multiplicities from SIDIS measurements by HERMES, low energy Drell-Yan as well as $Z$ boson production
up to LHC energies. In this paper we will follow the parametrization of Ref.~\cite{Su:2014wpa} for $g_2(b)$.

The function $g_1(b)$ contains information on intrinsic non-perturbative transverse motion of bound partons, in case of distribution TMD it depends on the type of hadron and quark flavor as well as potentially on $x_B$. In case of fragmentation TMD it can depend on $z_h$ and type of the hadron produced, and quark flavor. In other words, $g_1(b)$ is tied to the particular TMD. 
 Parameters in functions $g_2(b)$ and $g_1(b)$ depend on the cut-off value $b_{max}$ in case $b_*$-prescription is used.
 The non-perturbative factors could be also defined 
using different prescriptions; such as, for example, matching 
to perturbative form factors of Ref.~\cite{Qiu:2000hf} or using 
complex $b$ plane integration method
of Ref.~\cite{Kulesza:2003wn}. In this paper we use the 
standard CSS $b_*$-prescription method that allows us to 
compare easily with existing phenomenology.

Therefore with the TMD evolution, TMDs can be expressed as~\cite{Collins:2011zzd,Aybat:2011zv,scheme},
\bea
\tilde f_1^{q\;(sub)}(x_B,b;Q^2,Q)&=   e^{-\frac{1}{2}{S}_{\rm pert}(Q,b_*)-S_{\rm NP}^{f_1}(Q,b)} \widetilde{{\cal F}}_q(\alpha_s(Q))C_{q\gets i}\otimes f_{1}^{i}(x_B,\mu_b)   ,\label{tmdf}\\
\widetilde D_q^{(sub)}(z_h,b;Q^2,Q)&= e^{-\frac{1}{2}{S}_{\rm pert}(Q,b_*)-S_{\rm NP}^{D_1}(Q,b)}  \widetilde{{\cal D}}_q(\alpha_s(Q))\hat{C}_{j\gets q}\otimes D_{h/j}(z_h,\mu_b)  \label{tmdd}\ ,
\eea
where we explicitly embed the scheme dependence of TMDs from Eqs.~(\ref{eq:sqrt0},\ref{eq:sqrt})  in the coefficients
$\widetilde{\cal F}_q$ and $\widetilde{\cal D}_q$. Details on these functions will be given in Ref.~\cite{scheme}. In  Ji-Ma-Yuan scheme, 
\bea
\widetilde{\cal F}_q &= 1+\frac{\alpha_s}{2\pi}C_F\left[\ln\rho-\frac{1}{2}\ln^2\rho-\frac{\pi^2}{2}-2\right]  \ , \\
\widetilde{\cal D}_q &= 1+\frac{\alpha_s}{2\pi}C_F\left[\ln\rho-\frac{1}{2}\ln^2\rho-\frac{\pi^2}{2}-2\right]  \ ,
\eea
while in Collins-11 scheme $\widetilde{\cal F}_q = 1+ \mathcal{O}(\alpha_s^2)$ and $\widetilde{\cal D}_q = 1+ \mathcal{O}(\alpha_s^2)$. 
The final result for structure function is $\rho$ independent for Ji-Ma-Yuan scheme, so we set $\rho =1$. 
In Eqs.~(\ref{tmdf},\ref{tmdd}), $\otimes$ represents the convolution in the momentum fraction of $x$ or $z$, 
\bea
C_{q\gets i}\otimes f_{1}^{i}(x_B,\mu_b)   &\equiv 
\sum_i 
\int_{x_B}^1 \frac{dx}{x} \,
C_{q\gets i} \left(\frac{x_B}{x}, \mu_b \right) f_{1}^{i}(x,\mu_b),\label{convx}
\\
\hat {C}_{j\gets q}\otimes D_{h/j}(z_h,\mu_b)  &\equiv 
\sum_j
 \int_{z_h}^1 \frac{d z}{z} \,
\hat {C}_{j\gets q}\left(\frac{z_h}{z}, \mu_b \right) D_{h/j}(z,\mu_b).\label{convz}
\eea
The same convolutions will be used for 
transversity and Collins fragmentation functions with appropriate coefficient functions later in the paper.
 The above coefficient functions are  
\bea
C_{q\gets q'}(x,\mu_b) &=\delta_{q'q} \left[\delta(1-x)+\frac{\alpha_s}{\pi}\left(\frac{C_F}{2}(1-x) \right) \right]\; , \label{eq:cf}
\\
C_{q\gets g}(x,\mu_b) &= \frac{\alpha_s}{\pi} {T_R} \, x (1-x)\; ,  \label{eq:cf1}
\\
\hat C_{q'\gets q}(z,\mu_b) &=\delta_{q'q} \left[\delta(1-z)+\frac{\alpha_s}{\pi}\left(\frac{C_F}{2}(1-z)  + P_{q\gets q}(z)\, \ln z\right) \right]\; ,  \label{eq:cd}
\\
\hat C_{g\gets q}(z,\mu_b) &= \frac{\alpha_s}{\pi} \left( \frac{C_F}{2} z\; +  P_{g\gets q}(z)\, \ln z \right) ,  \label{eq:cd1}
\eea
with the usual splitting functions $P_{q\gets q}$ and $P_{g\gets q}$ given by
\bea
P_{q\gets q}(z) &= C_F \left[ \frac{1+z^2}{(1-z)_+} + \frac{3}{2} \delta(1-z) \right] \, , 
\label{P_qq}
\\
P_{g\gets q}(z) &= C_F \frac{1+(1-z)^2}{z} \; .
\label{P_gq}
\eea
The   $C$-functions are chosen to be universal among different TMD schemes, whereas the functions
$\widetilde{\cal F}_q$ and $\widetilde{\cal D}_q$ depend on the schemes. In Collins-11 schemes, both
factors are equal to 1 up to one-loop order. In the Ji-Ma-Yuan scheme, they will depend on $\rho$.
Again, this $\rho$ dependence in individual TMDs will be cancelled out by the associated $\rho$ dependence in the hard factor $H$  in Eq.~(\ref{eq:fuu_b}) 
when we calculate the structure function $F_{UU}(b,Q)$.

Substituting the results of Eqs.~(\ref{tmdf},\ref{tmdd}) into the 
factorization formula Eq.~(\ref{eq:fuu_b}), we can write down 
the structure function $\widetilde{F}_{UU}$ in the $b$-space as
\begin{eqnarray}
\widetilde {F}_{UU}(Q;b)&=&e^{-{S}_{\rm pert}(Q,b_*)-S_{\rm NP}^{\rm SIDIS}(Q,b)}
\widetilde{F}_{UU}(b_*)\ ,  \label{fuucss}
\end{eqnarray}
with non-perturbative form factor decomposed into the distribution and fragmentation 
contributions,
\bea
S_{\rm NP}^{\rm SIDIS}(Q,b) = S_{\rm NP}^{f_1}(Q,b) + S_{\rm NP}^{D_1}(Q,b)\ ,
\eea
which should be determined from the global fit to the SIDIS, $e^+e^-$, and Drell-Yan data. 
In the standard CSS resummaton which we will follow in this paper, together with
the hard factor in the TMD factorization of Eq.~(\ref{eq:fuu_b}), the
functions $\widetilde{F}_q$ and $\widetilde{D}_q$ are absorbed into 
the $C$-functions by applying the renormalization group equation for
the running coupling constant in these two factors~\cite{Catani:2000vq}. With that,  we can 
write down $\widetilde{F}_{UU}(b_*)$ as,
\bea
\widetilde{F}_{UU}(b_*) =
\sum_{q} e_q^2 \, \left( C_{q\gets i}^{\rm (SIDIS)}\otimes f_{1}^{i}(x_B,\mu_b) \right) \, \left(\hat{C}_{j\gets q}^{\rm (SIDIS)}\otimes D_{h/j}(z_h,\mu_b)\right)  \ ,
\label{eq:fuu1}
\eea
where $\sum_q$ runs over both quark and anti-quark flavors, $f_{1}^{i}(x_B,\mu_b)$ and $D_{h/j}(z_h,\mu_b)$ are the usual unpolarized collinear parton distribution function and fragmentation function at the scale $\mu_b = c_0/b_*$.
We emphasize that the above $C$-coefficients are the same for all TMD schemes if hard factor $H$ and $\widetilde{F}_q$, and $\widetilde{\cal D}_q$ are absorbed in their definition. In 
particular, in the Ji-Ma-Yuan scheme, the $\rho$ dependence in $H$ of Eq.~(\ref{eq:fuu_b}),
$\widetilde{F}_q$ in Eq.~(\ref{tmdf}), and $\widetilde{\cal D}_q$ in Eq.~(\ref{tmdd})
are cancelled out. In the Collins-11 scheme when the hard factor $H$ is absorbed in the definition of $C$ functions, $C$ functions become process dependent and equal to those of the standard CSS scheme. The final expressions for $C^{\rm SIDIS}$ and $\hat{C}^{\rm SIDIS}$
do not depend on $\rho$, and they are the same in the Collins-11 scheme, which are also the
same as those used in the CSS literature~\cite{Nadolsky:1999kb,Koike:2006fn},
\bea
C_{q\gets q'}^{\rm (SIDIS)}(x,\mu_b) &=\delta_{q'q} \left[\delta(1-x)+\frac{\alpha_s}{\pi}\left(\frac{C_F}{2}(1-x) -2C_F\delta(1-x)  \right)\right]\; , \label{eq:cf_css}
\\
C_{q\gets g}^{\rm (SIDIS)}(x,\mu_b) &= \frac{\alpha_s}{\pi} {T_R} \, x (1-x)\; ,  \label{eq:cf1_css}
\\
\hat C_{q'\gets q}^{\rm (SIDIS)}(z,\mu_b) &=\delta_{q'q} \left[\delta(1-z)+\frac{\alpha_s}{\pi}\left(\frac{C_F}{2}(1-z) -2C_F\delta(1-z) + P_{q\gets q}(z)\, \ln z\right) \right]\; ,  \label{eq:cd_css}
\\
\hat C_{g\gets q}^{\rm (SIDIS)}(z,\mu_b) &= \frac{\alpha_s}{\pi} \left( \frac{C_F}{2} z\; +  P_{g\gets q}(z)\, \ln z \right) .  \label{eq:cd1_css}
\eea
Of course, there is a freedom to have a separate hard factor in Eq.~(\ref{fuucss}), so that the above $C$-coefficients
will be modified accordingly, compare to Eqs.~(\ref{eq:cf},\ref{eq:cf1},\ref{eq:cd},\ref{eq:cd1}). This is referred to as scheme dependence~\cite{Catani:2000vq} in the CSS resummation. 

For the non-perturbative form factors, we will follow   the parameterization of~\cite{Su:2014wpa},
\bea
S_{\rm NP}^{\rm SIDIS}(Q,b)&= g_2 \ln\left(\frac{b}{b_*}\right)\ln\left(\frac{Q}{Q_0}\right)+\left({g_q}+\frac{g_h}{z_h^2}\right)b^2 \ ,
\label{eq:sud_np}
\eea
where  $Q_0^2$ = 2.4 GeV$^2$, for the spin-averaged   contribution.
In the above parameterization, the parameters $g_q = g_1/2=0.106$, $g_2=0.84$, $g_h=0.042$  (GeV$^2$) have been determined
from the analysis of SIDIS and Drell-Yan processes in Ref.~\cite{Su:2014wpa}.  
In the fit of Ref.~\cite{Su:2014wpa}, it was found that the non-perturbative form factors do not
depend on $x$. We will use the non-perturbative factor of Eq.~\eqref{eq:sud_np} in this paper.

%
%
\subsubsection{Collins structure function with evolution}

Now, we turn to the Collins effects contribution to the single transverse spin 
asymmetry in SIDIS.   
We start again from the factorized TMD expression in the $b$ space \cite{Idilbi:2004vb,Boer:2011xd}:
\begin{equation}
\widetilde{F}_{\rm collins}^\alpha(Q;b) =\sum_q e_q^2\,  \tilde  h_1^{q\; (sub)}(x_B,b;\rho,\zeta,\mu)\, \widetilde H_{1\, h/q}^{\perp\alpha \; (sub)}(z_h,b;\rho,\hat\zeta,\mu)H(\rho,Q/\mu) \ ,
\label{eq:fut_b}
\end{equation}
where $\tilde h_1^q$ is the TMD quark transversity distribution, $\widetilde{H}_{1\, h/q}^\perp$ is the Collins
fragmentation function in the
$b$ space and is defined (omitting scale dependence) as,
\begin{equation}
\tilde H_{1\, h/q}^{\perp\alpha}(z_h,b)=\int d^2p_\perp e^{-ip_\perp\cdot b} p_\perp^\alpha H_{1\, h/q}^\perp(z_h,p_\perp) \ .
\end{equation}
Here $H_{1\, h/q}^\perp(z_h,p_\perp)$ is the quark Collins function as defined in \cite{Yuan:2009dw}, which differs by a factor of $\left(-1/{z_h}\right)$ from the so-called ``Trento convention''~\cite{Bacchetta:2004jz},
\bea
H_{1\, h/j}^\perp(z_h,p_\perp)= -\frac{1}{z_h}H_{1\, h/j}^\perp(z_h,p_\perp)|_{\rm Trento},
\eea
with $p_\perp$ the transverse component of the hadron with respect to the fragmenting quark momentum.

The following model independent relation of  Collins fragmentation  function $H_{1\, h/q}^\perp(z_h,p_\perp)$ and a twist-3 fragmentation function of quark flavor $q$ to hadron $h$, $\hat H_{h/q}^{(3)}(z_h)$,  can be obtained~\cite{Yuan:2009dw} :
\bea
\hat{H}_{h/j}^{(3)}(z_h)=\int d^2p_\perp \frac{|p_\perp^2|}{M_h} H_{1\, h/j}^\perp(z_h,p_\perp)\; .
\eea
One often defines the following so-called {\em first} moment of Collins fragmentation function
\bea
H_{1\, h/j}^{\perp (1)}(z_h)|_{\rm Trento} &\equiv \int d^2 {p}_{\perp} \ \frac{|p_\perp|^2}{2z_h^2M_h^2}  H_{1\, h/j}^\perp(z_h,p_\perp)|_{\rm Trento}\; .
\label{moments_our_main}
\eea
We thus find that
\bea
\hat{H}_{h/j}^{(3)}(z_h) = - {2 z M_h} H_{1\, h/j}^{\perp (1)}(z_h)|_{\rm Trento}\, .
\label{eq:relation_to_trento}
\eea
It is straightforward to show that $\hat{H}_{h/j}^{(3)}(z_h)$ can be written as  
\begin{eqnarray}
\hat{H}_{h/j}^{(3)}(z_h)&=&{n^+}{z_h^2}\int\frac{d\xi^-}{2\pi}e^{ik^+\xi^-}\frac{1}{2}  \Big\{
{\rm Tr}\sigma^{\alpha +}\langle 0|\left[ iD_\perp^\alpha\nonumber+\int_{\xi^-}^{+\infty} d\zeta^- gF^{\alpha+}(\zeta^-)\right]
\psi(\xi)|P_hX\rangle
\nonumber\\
&&~~\times \langle P_hX|\bar\psi(0)|0\rangle+h.c. \Big\} \ ,\label{hz}
\end{eqnarray}
where we have chosen the gauge link in Eq.~(\ref{hz}) going to $+\infty$, and
$F^{\mu\nu}$ is the gluon field strength tensor and we have suppressed
the gauge links between different fields and other indices for simplicity.
Since the Collins function is the same under different gauge links~\cite{Metz:2002iz,Gamberg:2008yt,{Meissner:2008yf}},
we shall obtain the same result if we replace $+\infty$
by $-\infty$ in the above equation. 
 
The TMD evolution for the quark transversity and Collins fragmentation 
functions have been derived in the literature~\cite{Collins:1984kg,Kang:2011mr,Ji:2004wu,Idilbi:2004vb,Echevarria:2014rua}.
When expressed in terms of the collinear transversity distribution $h_{1}^{q}(x_B)$
and the twist-three fragmentation function $\hat H^{(3)}_{h/q}(z_h)$, they
can be written as
\bea
\tilde h_{1}^{q\;(sub)}(x_B,b,\rho;Q^2,Q)&=   e^{-\frac{1}{2}{S}_{\rm pert}(Q,b_*)-S_{\rm NP}^{h_1}(Q,b)} \, \widetilde{{\cal H}}_{1q}(\alpha_s(Q))\,  \delta C_{q\gets q'}\otimes h_{1}^{q'}(x_B,\mu_b)   ,\label{tmdh}\\
\widetilde H_{1\, h/q}^{(sub)\perp\alpha}(z_h,b,\rho;Q^2,Q)&=\left(\frac{-i b^\alpha}{2 z_h}\right) e^{-\frac{1}{2}{S}_{\rm pert}(Q,b_*)-S_{\rm NP}^{D_1}(Q,b)}  \, \widetilde{{\cal H}}_c(\alpha_s(Q))\, \delta \hat{C}_{q'\gets q}\otimes \hat H_{h/q'}^{(3)}(z_h,\mu_b)  \label{tmdc}\ ,
\eea
where again, the scheme dependence are in the functions $\widetilde{{\cal H}}_{1q}(\alpha_s(Q))$
and $ \widetilde{{\cal H}}_c(\alpha_s(Q))$. 
They equal to 1 up to one-loop order in Collins-11 scheme. 
 The $C$-coefficient functions are found to be
\bea
\delta C_{q\gets q'} (x,\mu_b) &=\delta_{q'q} \left[\delta(1-x)+{\cal O}(\alpha_s^2) \right]\; \label{eq:ch1},\\
\delta \hat C_{q'\gets q}^{\rm (SIDIS)}(z,\mu_b) &=\delta_{q'q} \left[\delta(1-z)+\frac{\alpha_s}{\pi}\left(\hat P_{q\gets q}^{c}(z) \ln z\right) \right]\; \label{eq:ch1perp},
\eea
 where the function $\hat P_{q\gets q}^{c}(z)$ has the following form, see Eq.~\eqref{eq:collins_splitting},
\bea
\hat P_{q\gets q}^{c}(z) = C_F \left[ \frac{2 z}{(1- z)_+} +\frac{3}{2}\delta(1-z)\right].
\label{P_qqc}
\eea
Substituting the above results into the factorization formula, we obtain the final
result for $\widetilde {F}_{\rm collins}^\alpha$ as~\cite{Collins:1984kg,Kang:2011mr,Ji:2004wu,Idilbi:2004vb,Echevarria:2014rua} 
\begin{eqnarray}
\widetilde {F}_{\rm collins}^\alpha(Q;b)&=&\left(\frac{-i b^\alpha}{2 z_h}\right)e^{-{S}_{\rm pert}(Q,b_*)-S_{\rm NP\, collins}^{\rm SIDIS}(Q,b)}
\, \widetilde{F}_{\rm collins}(b_*)\ ,
\end{eqnarray}
with $\widetilde{F}_{\rm collins}(b_*)$ given by
\bea
\widetilde{F}_{\rm collins}(b_*) =
\sum_{q} e_q^2 \, \left( \delta C_{q\gets i}\otimes h_{1}^{i}(x_B,\mu_b) \right) \, \left(\delta \hat{C}_{j\gets q}^{\rm (SIDIS)}\otimes \hat H_{h/j}^{(3)}(z_h,\mu_b)\right)\, .
\label{eq:fcollins}
\eea
The convolutions are defined in Eqs.~(\ref{convx},\ref{convz}) and
the relevant coefficient functions up to the first order in $\alpha_s$ (compare to Eq.~\eqref{eq:ch1perp} to determine relation to hard factor $H$) are given by~\cite{Koike:2006fn,Bacchetta:2013pqa,Yuan:2009dw,Echevarria:2014rua}:
\bea
\delta C_{q\gets q'}^{\rm (SIDIS)}(z,\mu_b)  (x,\mu_b) &=\delta_{q'q} \left[\delta(1-x)+\frac{\alpha_s}{\pi}\left(-2C_F\delta(1-x)\right) \right]\; \label{eq:ch1_css},\\
\delta \hat C_{q'\gets q}^{\rm (SIDIS)}(z,\mu_b) &=\delta_{q'q} \left[\delta(1-z)+\frac{\alpha_s}{\pi}\left(\hat P_{q\gets q}^{c}(z) \ln z -2C_F\delta(1-z)\right) \right]\; \label{eq:ch1perp_css},
\eea
where again, the above $C$-coefficients contain the contributions from the 
hard factors in the TMD factorization. The hard factor is given in Eq.~(\ref{hqjmy}) for Ji-Ma-Yuan scheme and in Eq.~\eqref{hqjcc} for Collins-11 scheme.

To achieve the evolution at the NLL order, we have to evaluate both transversity $h_{1}^{q}(x_B,\mu_b)$ and twist-3 fragmentation function $H_{h/q}^{(3)}(z, \mu_b)$ up to the scale $\mu_b=c_0/b_*$. The evolution
for the quark transversity is well-known \cite{Artru:1989zv,Baldracchini:1980uq,Blumlein:2001ca,Stratmann:2001pt}, and we will use the leading order result 
\bea
\frac{\partial}{\partial \ln\mu^2}h_{1}^q(x_B,\mu)=\frac{\alpha_s}{2\pi} \int_{x_B}^1 \frac{d\hat x}{\hat x} P_{q\to q}^{h_1}\left(\hat x\right)
 h_1^q(x_B/\hat x, \mu)  \ ,
\label{eq:transversity}
\eea
where the splitting kernel 
\bea
{P}_{q\to q}^{h_1}(\hat x)= C_F \left[ \frac{2\hat x}{(1-\hat x)_+} +\frac{3}{2}\delta(1-\hat x)\right].
\eea
Note that since gluon transversity distribution for nucleons does not exist \cite{Barone:2001sp}, the quark transversity $h_{1}^q$ does not mix with gluons in its evolution and it evolves as a non-singlet quantity. On the other hand, the evolution equation for $\hat H^{(3)}_{h/j}$ was derived in \cite{Yuan:2009dw,Kang:2010xv} and has a more complicated form. However, if we
keep only the homogenous term, we can
write down the evolution equation as~\cite{Yuan:2009dw,Kang:2010xv}, 
\begin{equation}
\frac{\partial}{\partial \ln\mu^2}\hat H^{(3)}_{h/q}(z_h,\mu)=\frac{\alpha_s}{2\pi} \int_{z_h}^1 \frac{d\hat z}{\hat z}
\hat P_{q\gets q}^{\rm c}(\hat z)\,
\hat H^{(3)}_{h/q}(z_h/\hat z, \mu) \ ,
\label{eq:collins}
\end{equation}
where the splitting kernel $\hat P_{q\gets q}^{\rm c}$ of the homogenous term is given in Eqs.~(\ref{eq:collins_splitting},\ref{P_qqc}), and is the same as that for the evolution of the quark transversity function, as pointed out in \cite{Kang:2010xv}. We will take this approximation in our numerical studies below. In order to differentiate from the complete NLL accuracy we will call it NLL$'$ or approximate NLL.

For the non-perturbative form factors,
we follow the parameterizations of~\cite{Su:2014wpa},
\bea
S_{\rm NP\, collins}^{\rm SIDIS}(Q,b)&=g_2 \ln\left(\frac{b}{b_*}\right)\ln\left(\frac{Q}{Q_0}\right)+\left({g_q}+\frac{g_h-g_c}{z_h^2}\right)b^2 \ ,
\eea
where  we assume that the quark transversity follows the same
parameterization as unpolarized TMD, but introduce an additional parameter to constrain the
$p_\perp$-dependence in the Collins fragmentation. Therefore, $g_c$
will be a free parameter in the fit. It is also worthwhile to emphasize that the $\ln Q/Q_0$-dependent part (i.e.~$g_2 \ln\left(b/b_*\right)$ in our formalism above) is universal for all processes in the initial CSS formalism~\cite{Collins:1981uk,Collins:1984kg} as well as in the recent TMD formalism of \cite{Collins:2011zzd}. The other contributions in non-perturbative Sudakov form factor are $Q$-independent and can be 
associated with corresponding TMD distribution and fragmentation functions at an initial scale, see e.g. Ref.~\cite{Collins:2011zzd,Echevarria:2014xaa}.  

Finally performing Fourier transforms in Eqs.~\eqref{fuu} and \eqref{fut}, we obtain the expressions for both spin-averaged and spin-dependent structure functions in the transverse momentum space as
\bea
F_{UU}(Q;P_{h\perp}) &= \frac{1}{z_h^2}\int_0^\infty \frac{db \, b}{(2\pi)} J_0( {P}_{h\perp} b/z_h  )\, 
e^{-{S}_{\rm pert}(Q,b_*)-S_{\rm NP}^{\rm SIDIS}(Q,b)} \widetilde{F}_{UU}(b_*)
\; ,
\label{fuu1}
\\
F_{UT}^{\sin\left(\phi_h +\phi_s\right)}(Q;P_{h\perp}) &= \frac{1}{z_h^2}\left(-\frac{1}{2 z_h}\right)\int_0^\infty \frac{db \, b^2}{(2\pi)} J_1( {P}_{h\perp} b/z_h  )\, 
e^{-{S}_{\rm pert}(Q,b_*)-S_{\rm NP\, collins}^{\rm SIDIS}(Q,b)} \widetilde{F}_{\rm collins}(b_*)\, , 
\label{fut1}
\eea 
with $J_{0,1}$ the usual Bessel functions. 

Let us comment at this point about the usage of relations to collinear distributions 
in the structure functions $\widetilde F$ in Eqs.~(\ref{eq:fuu1},\ref{eq:fcollins}). 
One could in 
principle solve evolution equations starting at a particular scale $Q_0$ instead 
of introducing dynamical scale $\mu_b \propto 1/b$ and try to extract unknown 
functions, such as Collins fragmentation function or transversity, directly from 
the data without relying on collinear or twist-3 functions. However such a 
method has certain difficulties, both theoretically and phenomenologically. 
Theoretical difficulty consists in the fact that if one starts from a fixed 
scale $Q_0$ then the $\tilde F$ function will have potentially large 
logarithms of the type $\ln(b Q_0)$ which are obviously not present in $\mu_b$ 
method due to the choice of $\mu_b\sim 1/b$. 
Phenomenologically it might also be difficult to model
the unique $x, z$ and $b$ dependence as contained in the collinear function $f(x, \mu_b)$,
which further builds in some dependence on the collision energy~\cite{Qiu:2000ga}.
 Presently there are no successful descriptions of experimental observables simultaneously 
at both low and high energies that use the  method with fixed starting scale $Q_0$.
The method with the fixed starting scale can be applied for 
processes where the measured scale $Q$ is similar to $Q_0$, 
namely for processes where the most important contribution in cross section comes 
from $b\sim 1/Q \sim 1/Q_0$. An example of such a description is a fit of Sivers 
functions in Ref.~\cite{Anselmino:2012aa,Sun:2013hua}. 
 In our case, the characteristic scales of SIDIS, 
$Q^2\sim 2.4$ GeV$^2$, and $e^+e^-$, $Q^2\sim 110$ GeV$^2$, are 
substantially different. It means that the regions of $b$ explored are 
different and one needs to accurately take into account correct $b$ 
dependence of TMDs. 
That is why in this extraction we will use relations 
to collinear distributions, fragmentation functions, and twist-3 functions.   

 By applying the CSS formalism, we utilize the well-established framework of the 
collinear parton distribution and fragmentation functions to parameterize 
the TMDs at the input scale. For the unpolarized case, this is an obvious 
advantage because of the existing global fits for the integrated PDFs.
For the Collins fragmentation function case, it is also easier to parameterize TMDs
in terms of collinear twist-three function, for which the usual DGLAP
evolution can be applied. Another important point we want to emphasize is that
there are DGLAP-type logarithms in the TMD formalism when $b$ is small. 
The CSS formalism is the best way to resum these logarithms, by applying the relevant scales ($\mu_b$)
in  the associated integrated parton distribution and fragmentation functions. 
This is an important step to help the theory convergence in the perturbative calculations.

%
%
\subsection{Collins Azimuthal Asymmetries in $e^+e^-$ \label{subsectionII}}

In this section we present the formulas for the Collins azimuthal asymmetries in back-to-back di-hadron productions in $e^+e^-$ annihilations, 
\bea
e^+ + e^-\to h_1+h_2+X \ ,
\eea
with center of mass energy $S=Q^2=(P_{e^+}+P_{e^-})^2$, and the final state
two hadrons with momenta $P_{h1}$ and $P_{h2}$, respectively. We further identify the
longitudinal momentum fractions: $z_{hi}=2|P_{hi}|/Q$. Therefore, $z_{hi}$
represent the momentum fractions in the fragmentation functions which describe
the fragmentation processes. Ideally, at leading order these two hadrons are produced
in a back-to-back configuration. However, the gluon radiation and transverse
momentum dependence in the fragmentation processes will generate
non-zero imbalance between the two hadrons.

To describe the near-back-to-back
imbalance between the two hadrons in $e^+e^-$ annihilations,
the TMD factorization can be used to calculate the differential cross
sections. In particular, the Collins fragmentation function will
lead to a $\cos2\phi$ azimuthal angular asymmetries between these
two hadrons. In the literature, there are two proposed
experimental methods to investigate the Collins effects in this process: (1)
one is to define a thrust axis in $e^+e^-$ annihilation and measure
the relative azimuthal angular correlation between the two hadrons in
the back-to-back two jets, which is referred as $A_{12}$ asymmetries; (2)
one is to use one hadron as reference to define the azimuthal angle of
another hadron (in the back-to-back configuration), which is referred as
$A_0$ asymmetries. In the former case, we will have to measure two azimuthal
angles $\phi_1$ and $\phi_2$, and the Collins effects lead to an azimuthal
asymmetry proportional to $\cos(\phi_1+\phi_2)$, whereas in the latter case
only one azimuthal angle  $\phi_0$ is measured and the Collins asymmetry
appears as $\cos (2\phi_0)$. In the naive TMD factorization (Born level),
both asymmetries can be formulated in terms of the Collins fragmentation
functions for the hadrons. However, only for the second case, we can
immediately generalize a QCD factorization in terms of the TMDs. For the
first case, a certain modification has to be made to have a QCD factorization formula.
The reason of this complication is that, in order to describe the case of (1),
one has to define the jet direction, which
is beyond the usual situation of the TMD factorization such as the TMD factorization
in SIDIS and Drell-Yan lepton pair production.


\begin{figure}[tbh]
\includegraphics[width=8cm]{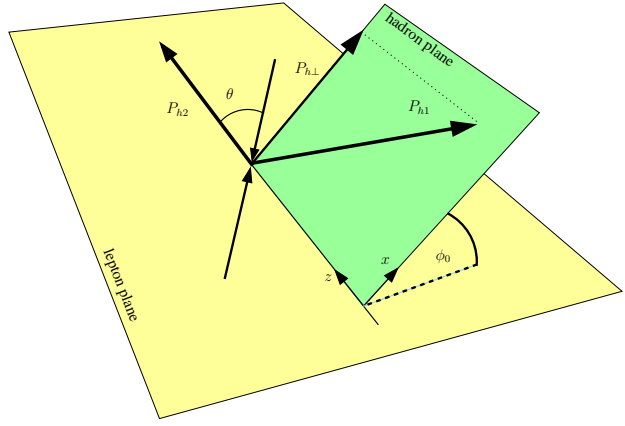}
\caption{$e^++e^-\to h_1+h_2+X$ process in the frame of method (2).}
\label{fig:epem}
\end{figure}

In this paper, as a first step, we only
consider the second case for the Collins asymmetries in $e^+e^-$ annihilations.
In this measurement, see Fig.~\ref{fig:epem}, the transverse momentum dependence is measured
for the hadron ($h_1$) relative to the direction of hadron ($h_2$). The total transverse
momentum dependence comes from the TMD fragmentation functions for hadron $h_1$
and hadron $h_2$, plus the soft factor generated from the soft gluon
radiation. Again, we focus on the low transverse momentum region,
where TMD factorization is appropriate and reads \cite{Boer:2008fr,Pitonyak:2013dsu}
\bea
\frac{d^5\sigma^{e^+e^-\to h_1 h_2 + X}}{dz_{h1}dz_{h2}d^2P_{h\perp}d\cos \theta}
=\frac{N_c \pi \alpha_{\rm em}^2}{2 Q^2}\left[\left(1+\cos^2\theta\right)Z_{uu}^{h_1h_2}
+\sin^2\theta \cos(2 \phi_0) Z_{\rm collins}^{h_1h_2}\right] \ ,
\label{e+e-}
\eea
where $\theta$ is the polar angle between the hadron $h_2$ and the beam of
$e^+e^-$, $\phi_0$ is defined as the azimuthal
angle of hadron $h_1$ relative to that of hadron $h_2$, i.e. of the plane containing hadrons $h_1$ and $h_2$ relative to the plane containing hadron $h_2$  and the lepton pair (see Fig.~\ref{fig:epem}), and $P_{h\perp}$ is the transverse momentum of 
hadron $h_1$ in this frame. We can rewrite the contribution corresponding to $Z_{\rm collins}^{h_1h_2}$ in Eq.~\eqref{e+e-} in the following form:
\bea
\sin^2\theta \cos(2 \phi_0) Z_{\rm collins}^{h_1h_2}  = \sin^2\theta \left(2\hat e_x^\alpha\hat e_x^\beta-g_\perp^{\alpha\beta}\right)Z_{\rm collins}^{h_1h_2\; \alpha\beta}  \ , 
\label{e+e-1}
\eea
where the unit vector $\hat e_x$   represents the transverse direction of the hadron in
the hadron frame and is defined in Fig.~\ref{fig:epem}. The tensor structure of this term leads to a $\cos 2\phi_0$ azimuthal
asymmetries between the two hadrons. 

The structure functions $Z_{uu}^{h_1h_2}$ and $Z_{\rm collins}^{h_1h_2}$ have the following form ,
\begin{eqnarray}
Z_{uu}^{h_1h_2}(Q;P_{h\perp})&=&\frac{1}{z_{h1}^2}\int \frac{d^2b}{(2\pi)^{2}} e^{i \vec{P}_{h\perp}\cdot \vec{b}/z_{h1}} \widetilde
{Z}_{UU}^{h_1h_2}(Q;b) +Y_{uu}(Q;P_{h\perp}) \ , \\
Z_{\rm collins}^{h_1h_2\; \alpha\beta}(Q;P_{h\perp})&=&\frac{1}{z_{h1}^2}\int \frac{d^2b}{(2\pi)^{2}} e^{i \vec{P}_{h\perp}\cdot \vec{b}/z_{h1}} \widetilde
{Z}_{\rm collins}^{h_1h_2\; \alpha\beta}(Q;b) +Y_{\rm collins}^{\alpha\beta}(Q;P_{h\perp}) \ ,
\end{eqnarray}
where the first term depends on the TMD fragmentation functions
for the two hadrons and dominates in $P_{h\perp}/z_{h1}\ll Q$ region, and the
second term dominates in the region of $P_{h\perp}/z_{h1}\gtrsim Q$.
For $\cos 2\phi_0$ asymmetries, we have an additional contribution from gluon
radiation \cite{Boer:2008fr} associated with spin-averaged fragmentation functions. This contribution
does not depend on the Collins fragmentation function, and
is proportional to $P_{h\perp}^2/Q^2$. It will become important at relatively 
large transverse momentum, and should be included in the above $Y$ terms.
However, in the following, we only consider the low transverse momentum
region $P_{h\perp}\ll Q$, where this contribution is power suppressed as
compared to the Collins contributions. In addition, in the experimental
measurements, the double ratio of the $\cos 2\phi$ asymmetries are
reported for di-hadron correlations in $e^+e^-$ annihilations, where this
contribution is cancelled out. Therefore, in the following analysis, we will
not include neither the contribution from gluon radiation independent of Collins FF nor the $Y$ term.

%
%
\subsubsection{Experimentally measured Collins azimuthal asymmetries in $e^+e^-$}

Let us now discuss the definitions of the asymmetries associated with Collins fragmentation functions in the actual experimental measurements. Collins function generates $\cos 2 \phi_0$ modulation in the $e^+e^-$ cross-section, let us rewrite Eq.~\eqref{e+e-} as follows:
\bea
\frac{d^5\sigma^{e^+e^-\to h_1 h_2 + X}}{dz_{h1}dz_{h2}d^2P_{h\perp}d\cos \theta} &=  \frac{\pi N_c\alpha_{em}^2}{2Q^2} (1+\cos^2\theta) Z_{uu}^{h_1h_2} \cdot R^{h_1h_2}(z_{h1},z_{h2},\theta,P_{h\perp}) \; ,  
\\ 
R^{h_1h_2}(z_{h1},z_{h2},\theta,P_{h\perp}) &\equiv  1+\cos(2 \phi_0) \frac{ \sin^2\theta }{ 1+\cos^2\theta  } \frac{Z_{\rm collins}^{h_1h_2}}{Z_{uu}^{h_1h_2}} \ .
\eea
One could also define analogously the $P_{h\perp}$-integrated modulation
\bea
R^{h_1h_2}(z_{h1},z_{h2},\theta) \equiv 
1+\cos(2 \phi_0) \frac{ \sin^2\theta }{  1+\cos^2\theta  } \frac{\int d P_{h\perp} P_{h\perp}\, Z_{\rm collins}^{h_1h_2}}{\int d P_{h\perp} P_{h\perp}\, Z_{uu}^{h_1h_2}} \ .
\eea
In order to eliminate false asymmetries BELLE and {\em BABAR} consider the ratios of unlike-sign ``U'' ($\pi^+\pi^- + \pi^-\pi^+$) over
like-sign ``L'' ($\pi^+\pi^+ + \pi^-\pi^-$) or charged ``C'' ($\pi^+\pi^+ + \pi^-\pi^- + \pi^+\pi^- + \pi^-\pi^+$) pion pairs. In our formalism, they can be written as follows:
\bea
\frac{R^U(z_{h1},z_{h2},\theta, P_{h\perp})}{R^{L}(z_{h1},z_{h2},\theta,P_{h\perp})} & \simeq
1+\cos(2 \phi_0) \frac{\langle\sin^2\theta\rangle}{\langle 1+\cos^2\theta \rangle} \left( \frac{Z_{\rm collins}^U}{Z_{uu}^U} - \frac{Z_{\rm collins}^{L}}{Z_{uu}^{L}}\right) \ ,
\\
\frac{R^U(z_{h1},z_{h2},\theta,P_{h\perp})}{R^{C}(z_{h1},z_{h2},\theta,P_{h\perp}) } & \simeq
1+\cos(2 \phi_0) \frac{\langle\sin^2\theta\rangle}{\langle 1+\cos^2\theta \rangle} \left( \frac{Z_{\rm collins}^U}{Z_{uu}^U} - \frac{Z_{\rm collins}^{C}}{Z_{uu}^{C}}\right) \ ,
\eea
and likewise for the $P_{h\perp}$-integrated modula:
\bea
\frac{R^U(z_{h1},z_{h2} ,\theta)}{R^{L}(z_{h1},z_{h2},\theta)} & \simeq
1+\cos(2 \phi_0) \frac{\langle\sin^2\theta\rangle}{\langle 1+\cos^2\theta \rangle} \left( \frac{\int d  P_{h\perp} P_{h\perp}\,Z_{\rm collins}^U}{\int d P_{h\perp} P_{h\perp}\, Z_{uu}^U} - \frac{\int d P_{h\perp} P_{h\perp}\, Z_{\rm collins}^{L}}{\int d P_{h\perp} P_{h\perp}\, Z_{uu}^{L}}\right) \ ,
\\
\frac{R^U(z_{h1},z_{h2},\theta )}{R^{C}(z_{h1},z_{h2},\theta)} & \simeq
1+\cos(2 \phi_0) \frac{\langle\sin^2\theta\rangle}{\langle 1+\cos^2\theta \rangle} \left( \frac{\int d P_{h\perp} P_{h\perp}\, Z_{\rm collins}^U}{\int d P_{h\perp} P_{h\perp}\,Z_{uu}^U} - \frac{\int d P_{h\perp} P_{h\perp}\,Z_{\rm collins}^{C}}{\int d P_{h\perp} P_{h\perp}\, Z_{uu}^{C}}\right) \ ,
\eea
where the relevant functions are given by
\bea
&Z_{uu}^U \equiv Z_{uu}^{\pi^+\pi^-} + Z_{uu}^{\pi^-\pi^+}  \; , 
\qquad
Z_{uu}^L \equiv Z_{uu}^{\pi^+\pi^+} + Z_{uu}^{\pi^-\pi^-} \; ,
\qquad
Z_{uu}^C \equiv Z_{uu}^{U} + Z_{uu}^{L}  \; ,
\\
&Z_{\rm collins}^U \equiv Z_{\rm collins}^{\pi^+\pi^-} + Z_{\rm collins}^{\pi^-\pi^+}  \; ,
\qquad
Z_{\rm collins}^L \equiv Z_{\rm collins}^{\pi^+\pi^+} + Z_{\rm collins}^{\pi^-\pi^-}   \; ,
\qquad
Z_{\rm collins}^C \equiv Z_{\rm collins}^{U} + Z_{\rm collins}^{L}  \; .
\eea
Experimentally measured asymmetries $A^{UL}_0$ and $A^{UC}_0$ are then given by
\bea
A^{UL}_0(z_{h1},z_{h2},\theta,P_{h\perp} ) &\equiv  \frac{\langle\sin^2\theta\rangle}{\langle 1+\cos^2\theta \rangle} \left( \frac{ Z_{\rm collins}^U}{ Z_{uu}^U} - \frac{ Z_{\rm collins}^{L}}{ Z_{uu}^{L}}\right) \ , 
\\
A^{UC}_0(z_{h1},z_{h2} ,\theta,P_{h\perp}) &\equiv
  \frac{\langle\sin^2\theta\rangle}{\langle 1+\cos^2\theta \rangle} \left( \frac{\  Z_{\rm collins}^U}{Z_{uu}^U} - \frac{ Z_{\rm collins}^{C}}{  Z_{uu}^{C}}\right) \ ,
\\
A^{UL}_0(z_{h1},z_{h2},\theta) &\equiv  \frac{\langle\sin^2\theta\rangle}{\langle 1+\cos^2\theta \rangle} \left( \frac{\int d P_{h\perp} P_{h\perp}\,Z_{\rm collins}^U}{\int d P_{h\perp} P_{h\perp}\, Z_{uu}^U} - \frac{\int d P_{h\perp} P_{h\perp}\,Z_{\rm collins}^{L}}{\int d P_{h\perp} P_{h\perp}\, Z_{uu}^{L}}\right) \ , 
\\
A^{UC}_0(z_{h1},z_{h2},\theta) &\equiv
  \frac{\langle\sin^2\theta\rangle}{\langle 1+\cos^2\theta \rangle} \left( \frac{\int d P_{h\perp} P_{h\perp}\, Z_{\rm collins}^U}{\int d P_{h\perp} P_{h\perp}\,Z_{uu}^U} - \frac{\int d P_{h\perp} P_{h\perp}\,Z_{\rm collins}^{C}}{\int d P_{h\perp} P_{h\perp}\, Z_{uu}^{C}}\right) \ .
\eea

%
%
\subsubsection{Structure functions in $e^+e^-$ with QCD evolution}

Corresponding structure functions $Z_{uu}^{h_1h_2}$ and $Z_{\rm collins}^{h_1h_2}$ are defined as Fourier transforms of structure functions in 
$b$-space,
\bea
Z_{uu}^{h_1h_2}(Q;P_{h\perp})&=
\frac{1}{z_{h1}^2}\int \frac{d^2b}{(2\pi)^{2}} e^{i \vec{P}_{h\perp}\cdot \vec{b}/z_{h1}} \widetilde
{Z}_{uu}^{h_1h_2}(Q;b)  \label{zuu} \ , 
\\
Z_{\rm collins}^{h_1h_2}(Q;P_{h\perp})&=
\frac{1}{z_{h1}^2}\int \frac{d^2b}{(2\pi)^{2}} e^{i \vec{P}_{h\perp}\cdot \vec{b}/z_{h1}}
\left(2\hat P^\alpha_{h\perp} \hat P^\beta_{h\perp}-g^{\alpha\beta}_\perp\right) \widetilde
{Z}_{\rm collins}^{h_1h_2 \; \alpha\beta}(Q;b) \label{zut} \ ,
\eea
where we only keep the term dominant in the low transverse momentum region.
According to the TMD
factorization, we can write down
\begin{eqnarray}
\widetilde{Z}_{uu}^{h_1h_2}(Q; b)&=& \sum_{q} \, e_q^2 \, D_{h_1/q}^{(sub)}(z_{h1},b,\zeta_1;\mu)\,  D_{h_2/\bar q}^{(sub)}(z_{h2},b,\zeta_2;\mu)\,  H_{uu}^{e^+e^-}(Q;\mu) \label{zuu0}\ ,\\
\widetilde{Z}_{\rm collins}^{h_1h_2 \; \alpha\beta}(Q; b)&=& \sum_{q} e_q^2 \, \widetilde{H}_{1\, h_1/q}^{\perp\alpha\; (sub)}(z_{h1},b,\zeta_1;\mu)\, \widetilde{H}_{1\, h_2/\bar q}^{\perp\beta\; (sub)} (z_{h2},b,\zeta_2;\mu)
\,  H_{\rm collins}^{e^+e^-}(Q;\mu) \ , \label{zcollins}
\end{eqnarray}
where again $\zeta_i$ and $\rho$ are parameters to regulate the light-cone
singularities in the TMD fragmentation functions: $\zeta_i^2=(2v_i\cdot P_{hi})^2/v_i^2$
and $\rho^2=(2v_1\cdot v_2)^2/v_1^2v_2^2$.
One-loop calculations can be performed for the above observables, and
the relevant hard factors shall follow those in SIDIS calculations.
In particular, for the $Z_{uu}$ term, the hard factor is the same as that for
Drell-Yan lepton pair production, which differs from SIDIS. This happens because
in $e^+e^-$ annihilation, the virtual photon is time-like $q^2 > 0$, the same
as that in Drell-Yan process, whereas in SIDIS, the virtual photon
is space-like $q^2 < 0$. Because of spin-independence of hard interaction
in perturbative QCD, the hard factor $Z_{\rm collins}$ will be the same
as $Z_{uu}$,
\begin{equation}
H_{\rm collins}^{e^+e^-(JMY)}(Q;\mu)=H_{uu}^{e^+e^-(JMY)}(Q;\mu)=1+
\frac{\alpha_s(\mu)}{2\pi}C_F\left[\ln\frac{Q^2}{\mu^2}+\ln\rho^2\ln\frac{Q^2}{\mu^2}-\ln\rho^2+\ln^2\rho+2\pi^2-4\right] \ ,
\end{equation}
in Ji-Ma-Yuan scheme.
For Collins-11 scheme, we will obtain similar results, 
\begin{equation}
H_{\rm collins}^{e^+e^-(JCC)}(Q;\mu)=H_{uu}^{e^+e^-(JCC)}(Q;\mu)=1+
\frac{\alpha_s(\mu)}{2\pi}C_F\left[3\ln\frac{Q^2}{\mu^2}-\ln^2\left(\frac{Q^2}{\mu^2}\right)+\pi^2-8\right] \ .
\label{eq:hsidisjcc}
\end{equation}
Following the previous section, we first derive the evolution 
results for the TMD unpolarized and Collins fragmentation functions and   substituting the results 
into the above factorization formulas of Eqs.~(\ref{zuu0},\ref{zcollins}) and obtain,
\bea
\widetilde {Z}_{uu}^{h_1h_2}(Q;b)&= e^{-{S}_{\rm pert}(Q, b_*)-S_{\rm NP}^{e^+e^-}(Q,b)}\,  \widetilde {Z}_{uu}^{h_1h_2}(b_*)\ ,
\\
\widetilde {Z}_{\rm collins}^{h_1h_2 \; \alpha\beta}(Q;b)&=\left(\frac{-i b^\alpha}{2z_{h1}}\right)\left(\frac{-i b^\beta}{2z_{h2}}\right)
e^{-{S}_{\rm pert}(Q, b_*)-S_{\rm NP\, collins}^{e^+e^-}(Q,b)}\,  \widetilde {Z}_{\rm collins}^{h_1h_2}(b_*)\ .
\eea
Again, the energy evolution effects are explicit in exponential factors, and 
\bea
\widetilde {Z}_{uu}^{h_1h_2}(b_*)&= \sum_{q} e_q^2 \, \left( \hat{C}_{i\gets q}^{(e^+e^-)}\otimes D_{h_1/i}(z_{h1},\mu_b )\right) \, \left(\hat{C}_{j\gets \bar q}^{(e^+e^-)}\otimes D_{h_2/j}(z_{h2},\mu_b ) \right)\ ,
\\
\widetilde {Z}_{\rm collins}^{h_1h_2}(b_*)&= \sum_{q} e_q^2 \, \left( \delta \hat C_{i\gets q}^{(e^+e^-)}\otimes \hat H^{(3)}_{h_1/i}(z_{h1},\mu_b )\right)
\, \left(\delta \hat{C}_{j\gets \bar q}^{(e^+e^-)}\otimes \hat H_{h_2/j}^{(3)}(z_{h2},\mu_b )\right) \ ,
\eea
where the convolution  is defined in Eq.~(\ref{convz}) and the coefficient functions read:
\bea
\hat{C}_{q'\gets q}^{(e^+e^-)}(z,\mu_b) &= \delta_{q'q} \left[\delta(1-z)+\frac{\alpha_s}{\pi}\left(\frac{C_F}{2}\left(1-z \right)+ P_{q \gets q}(z) \ln z +\frac{C_F}{4}\left(\pi^2 -8\right)\delta(1-z)\right)\right]\; , 
\\
\hat C_{g\gets q}^{(e^+e^-)}(z,\mu_b) &= \frac{\alpha_s}{\pi} \left( \frac{C_F}{2} z\; +  P_{g\gets q}(z)\, \ln z \right) , 
\\
\delta \hat C_{q'\gets q}^{(e^+e^-)}(z,\mu_b) &=\delta_{q'q}\left[\delta(1-z)+\frac{\alpha_s}{\pi}\left(\hat P_{q\gets q}^{c}(z) \ln z+\frac{C_F}{4}\left(\pi^2 -8\right)\delta(1-z)\right)\right]\; ,
\eea
with the functions $P_{q \gets q}$, $P_{g\gets q}$, and $\hat P_{q\gets q}^{c}$ given in Eqs.~\eqref{P_qq}, \eqref{P_gq}, and \eqref{P_qqc}, respectively. Again, the above $C$-coefficient functions contain the 
contributions from the associated hard factors in the TMD factorization, similar to the case of SIDIS in 
the last section (see Appendix A for detailed derivations).
On the other hand, the non-perturbative form factors are parameterized as
\bea
S_{\rm NP}^{e^+e^-}(Q,b)&= g_2 \ln\left(\frac{b}{b_*}\right)\ln\left(\frac{Q}{Q_0}\right)+\left(\frac{g_h}{z_{h1}^2}+\frac{g_{h}}{z_{h2}^2}\right)b^2 \ ,
\\
S_{\rm NP\, collins}^{e^+e^-}(Q,b)&= g_2 \ln\left(\frac{b}{b_*}\right)\ln\left(\frac{Q}{Q_0}\right)+\left(\frac{g_h-g_c}{z_{h1}^2}+\frac{g_{h}-g_c}{z_{h2}^2}\right)b^2 \ ,
\eea
where we have utilized the universality arguments for these parameters. Performing Fourier transforms in Eqs.~\eqref{zuu} and \eqref{zut}, we have  
\bea
Z_{uu}^{h_1h_2}(Q;P_{h\perp})&= \frac{1}{z_{h1}^2}\int_0^\infty \frac{db\, b}{(2\pi)} J_0(  {P}_{h\perp} {b}/z_{h1}) \; 
e^{-{S}_{\rm pert}(Q, b_*)-S_{\rm NP}^{e^+e^-}(Q,b)}\widetilde {Z}_{uu}^{h_1h_2}(b_*)  \label{zuu1} \ ,
\\
Z_{\rm collins}^{h_1h_2}(Q;P_{h\perp})&= \frac{1}{z_{h1}^2}\frac{1}{4z_{h1}z_{h2}}\int_0^\infty \frac{db\, b^3}{(2\pi)} J_2(  {P}_{h\perp} {b}/z_{h1})  \;
e^{-{S}_{\rm pert}(Q, b_*)-S_{\rm NP\, collins}^{e^+e^-}(Q,b)}\widetilde {Z}_{\rm collins}^{h_1h_2}(b_*) \label{zut1}\ ,
\eea
with $J_2$ the associated Bessel function.

%
%
\section{Global Fit with TMD Evolution}
\label{sec:phenomenology}

%
%
\subsection{Parametrizations}

As we have seen in previous sections, we have two unknown functions to be extracted from experimental data: collinear 
transversity distribution $h_1^q$ and collinear twist-3 fragmentation function $\hat{H}_{h/q}^{(3)}$. The QCD evolution of both functions is known, $x$-dependence of $h_1^q$ and $z$-dependence of $\hat{H}_{h/q}^{(3)}$ at the initial scale $Q_0$ should be parametrized.

In the global fit, we parameterize the quark transversity distributions
as
\begin{equation}
h_1^{q}(x,Q_0)=N_{q}^h x^{a_{q}}(1-x)^{b_{q}} \frac{(a_{q} + b_{q})^{a_{q} + b_{q}}}
{a_{q}^{a_{q}} b_{q}^{b_{q}}}\frac{1}{2}\left (f_{1}^q(x,Q_0) + g_{1}^q(x,Q_0)  \right ) \ ,
\end{equation}
at the initial scale $Q_0$, for up and down quarks $q=u,d$, respectively, where $f_{1}^q$ are the unpolarized
CT10 NLO quark distributions~\cite{Lai:2010vv} and $g_{1}^q$ are the NLO DSSV quark helicity distributions~\cite{deFlorian:2009vb}.
In our parametrization we enforce the so-called Soffer positivity bound~\cite{Soffer:1995ww} of transversity distribution at the initial scale.
This bound is known to be valid~\cite{Vogelsang:1997ak} up to NLO order in perturbative QCD. Soffer bound was criticized in Ref.~\cite{Ralston:2008sm} and was predicted to fail so it is very interesting to determine phenomenologically if there are signs of such violation in experimental data. Many extractions of transversity, for instance those of Refs.~\cite{Anselmino:2013vqa,Radici:2015mwa}, indeed show saturation of Soffer bound for $d$-quark transversity.

In this study, we assume that all the sea quark transversity distributions
are negligible. With more data available in the future, we hope we can constrain
the sea quark as well, in particular, with the Electron-Ion Collider. We leave estimates on possible
non-zero sea quarks transversity distributions for future publications.

Similarly, we parameterize 
the twist-3 Collins fragmentation functions in terms of the unpolarized fragmentation functions,
\begin{align}
\hat{H}_{fav}^{(3)}(z,Q_0)&= N_{u}^c z^{\alpha_{u}}(1-z)^{\beta_{u}} D_{\pi^+/u}(z,Q_0) \ , \\
\hat{H}_{unf}^{(3)}(z,Q_0)&= N_{d}^c z^{\alpha_{d}}(1-z)^{\beta_{d}} D_{\pi^+/d}(z,Q_0) \ , 
\end{align}
which correspond to the favored and unfavored Collins
fragmentation functions, respectively. 
  We also utilize the newest NLO extraction of fragmentation functions~\cite{deFlorian:2014xna}. The new DSS FF set
 is capable of describing pion multiplicities measured by COMPASS and HEMRES collaborations. In fact it is the {\em only}
 set of fragmentation functions that accurately describes COMPASS and HERMES data. The quality of the global fit improved from $\chi^2/{d.o.f.}\simeq 2.2$ for previous DSS NLO~\cite{deFlorian:2007aj} to $\chi^2/{d.o.f.}\simeq 1.2$ for the new NLO fit~\cite{deFlorian:2014xna}. Extractions of LO FFs~\cite{deFlorian:2007aj} have yielded a much less satisfactory description of the available pion data thus NLO sets ought to be used in extractions of TMDs. NLL accuracy allows to utilize this set at NLO. We have verified that results presented here are in complete agreement with previously published extraction of Ref.~\cite{Kang:2014zza}.

The rest can be obtained by applying the isospin relations.
We also neglect possible difference of favoured/unfavoured fragmentation function of $\bar u, \bar d$ and $u, d$: 
\begin{align} 
\hat{H}_{\pi^+/\bar u}^{(3)}(z,Q_0) = \hat{H}_{\pi^-/u}^{(3)}(z,Q_0) = \hat{H}_{\pi^-/\bar d}^{(3)}(z,Q_0) = \hat{H}_{unf}(z,Q_0)\ , \\
\hat{H}_{\pi^+/\bar d}^{(3)}(z,Q_0) =  \hat{H}_{\pi^-/d}^{(3)}(z,Q_0) = \hat{H}_{\pi^-/\bar u}^{(3)}(z,Q_0) = \hat{H}_{fav}(z,Q_0)\ . 
\end{align}
Strange quark fragmentation deserves an additional 
attention. Fragmentation of strange quarks to hadrons is different from just ``unfavored'' fragmentation functions,
such as $D_{\pi^+/\bar u}$ and in order to take this into account we will parametrize strange quark ``unfavored'' Collins fragmenation function
as
\begin{eqnarray}
\hat{H}_{unf s}^{(3)}(z,Q_0) \equiv \hat{H}_{\pi^\pm/s,\bar s}^{(3)}(z,Q_0) = N_{d}^c z^{\alpha_{d}}(1-z)^{\beta_{d}} D_{\pi^+/s, \bar s}(z,Q_0) \ . 
\end{eqnarray}

We would
like to emphasize that in the fit, we will solve the DGLAP evolution equations for both transversity and Collins FF to the scale $\mu_b=c_0/b_*$, in order to be complete at the NLL$'$ order. Numerical solution of DGLAP equations is performed
in $x$-space by {\tt HOPPET} evolution package \cite{Salam:2008qg}. Original code of {\tt HOPPET} is modified by us 
so that transversity splitting functions are included, the initial scale for the evolution is chosen to be $Q_0^2 = 2.4$ GeV$^2$. and the {\tt HOPPET} code is executed using $\alpha_s(Q_0) = 0.327$. 
In our numerical calculations we consistently use 2-loop order result for $\alpha_s(\mu)$ with $n_f = 5$ effective quark flavors and
$\Lambda_{QCD} = 0.225$ GeV such that $\alpha_s(M_Z) = 0.118$. 

For the non-perturbative form factors, we use the following parameters from 
 Ref.~\cite{Su:2014wpa}: $g_q =g_1/2=0.106$, $g_2=0.84$, $g_h=0.042$  (GeV$^2$). The NLL formula has a large negative contribution coming from $C^{(1)}\propto -2C_F\delta(1-x)$, see Eq.~\eqref{eq:cf_css}, and $H^{(1)}\propto -8 C_F$, see Eq.~\eqref{eq:hardfactor}, this makes the need for 
potentially large $K$ factors in description of the data. We assume that $K$ factors will be largest in lowest $Q^2$ region where $\alpha_s$ is relatively large.
In fact the fit of SIDIS and Drell-Yan data of Ref.~\cite{Su:2014wpa} revealed $K_{SIDIS} \sim 2$ for COMPASS and HERMES and $K_{DY} \sim 1$ for Drell-Yan data. In asymmetry $K$ factors cancel, so we will not use them in present analysis.

The existing experimental data does not allow to determine precisely shapes of all polarised distributions in
coordinate space, we make a simplifying assumption and allow for the Collins fragmentation function to modify
its shape with respect to unpolarised fragmentation distributions and have $g_c$ as a free parameter 
in the fit. 

Therefore, we have total of 13 parameters in our global fit:
$N_u^h$, $N_d^h$, $a_u$, $a_d$, $b_u$, $b_d$, $N_u^c$, $N_d^c$, $\alpha_u$, $\alpha_d$, $\beta_d$, $\beta_u$, $g_c$ (GeV$^2$). 

%
%
\subsection{Experimental data}

Let us also discuss available experimental data. In this paper we extract Collins fragmentation functions for pions and transversity distributions for $u$, $d$ and favoured/unfavoured Collins fragmentation functions for pions. Thus we will select the data involving pion production only.

The HERMES Collaboration measured Collins asymmetries
in electron proton scattering at the laboratory electron beam energy 27.5 GeV in production of $\pi^+$, $\pi^-$, and $\pi^0$ \cite{Airapetian:2009ae}. The data are presented in bins of $x_B$, $z_h$, and $P_{h\perp}$ respectively. Clear non-zero asymmetries were found for both $\pi^+$ and $\pi^-$. Large negative asymmetry for $\pi^-$ suggest that unfavoured Collins fragmentation function is big and not suppressed with respect to the favoured one.

The COMPASS Collaboration uses muon beam of energy 160 GeV and have measured Collins asymmetries on both NH$_3$
(proton) \cite{Adolph:2014zba} and LiD (deuterium) \cite{Alekseev:2008aa} targets. The data are presented as function of $x_B$, $z_h$, and $P_{h\perp}$. Results on the proton target are compatible with HERMES findings and asymmetries are found to be compatible with zero on the deuteron target. The beam energy of COMPASS is higher than the energy of HERMES and thus COMPASS reaches lower values of $x_B\sim 10^{-3}$. For each point in $x_B$ the scale $Q^2$ is higher at COMPASS as one has $Q^2 = s x y$. Both experiments consider $Q^2 >1 $ GeV$^2$ in order to be in the perturbative region and the energy of $\gamma^* p$, $W^2 > 10$ GeV$^2$ for HERMES and $W^2 > 25$ GeV$^2$ for COMPASS in order to be outside of the resonance region. The COMPASS Collaboration considers $z_h >0.2$ region and the HERMES Collaboration uses $0.2 < z_h < 0.7$ in order to minimize both target fragmentation effects and exclusive reaction contribution. All other experimental cuts are described in Refs.~\cite{Airapetian:2009ae,Alekseev:2008aa,Adolph:2014zba}.

Jefferson Lab's HALL A published data
of $\pi^\pm$ pion production in 5.9 GeV electron scattering on $^3$He (effective neutron) target \cite{Qian:2011py}.
Jefferson Lab operates at relatively low energy and reaches higher values of $x_B \sim 0.35$.

Information on Collins fragmentation functions is contained in $e^+e^-$ at the energy $\sqrt{s} \simeq 10.6$ GeV data of the BELLE \cite{Seidl:2008xc} and the {\em BABAR} \cite{TheBABAR:2013yha} Collaborations. Note that usual feature of TMD evolution is widening of 
distributions with increase of the hard scale. Thus it is very important to check our knowledge against available data on $P_{h\perp}$ distributions
and corresponting $P_{h\perp}$ dependencies of asymmetries. For this reason, we include {\em BABAR} \cite{TheBABAR:2013yha} data on $P_{h\perp}$ dependence in our fit. We will also present predictions of $P_{h\perp}$ dependence of the unpolarised cross section that will be the ultimate test of the model. As we mentioned in Sec.~\ref{subsectionII} we will use $A_0$ data on Collins asymmetries in $e^+e^-$ in our fit. Both BELLE and {\em BABAR} Collaborations 
require the momentum of the virtual photon $P_{h\perp}/z_{h1} < 3.5$ GeV in order to remove contributions
from hadrons assigned to the wrong hemisphere and it also helps to remove contribution from gluon radiation.  
The analysis of BELLE is performed in ($z_{h1}$,$z_{h2}$) bins with boundaries
at $z_{hi}=$ $0.2$, $0.3$, $0.5$, $0.7$ and $1.0$. The {\em BABAR} Collaboration chooses 6 $z_{hi}$-bins: $[0.15-0.2]$, $[0.2-0.3]$,
$[0.3-0.4]$, $[0.4-0.5]$, $[0.5-0.7]$, $[0.7-0.9]$. A characteristic feature of the asymmetry is growth with $z_{hi}$ which is compatible with $z_{hi}$ dependence of theoretical formula, and the asymmetry should vanish in the limit  $z_{hi}\to 0$.

%
%
\subsection{Fitting procedure}

We proceed with a global fit of SIDIS and $e^+e^-$ data using {\rm MINUIT} package \cite{James:1975dr,James:2004xla} by minimizing the total $\chi^2$:
\begin{align} 
\chi^2(\{a\}) = \sum_{i=1}^{N} \sum_{j=1}^{N_i} \frac{(T_{j}(\{a\})-E_{j})^2}{\Delta E_{j}^2}\; ,
\end{align}
for $i=1,...,N$ data sets each containing $N_i$ data points. Experimental measurement of each point is $E_j$, experimental uncertainty is $\Delta E_j$ and theoretical estimate, $T_j$, is calculated for a given set of parameters $\{a\}$=$\{N_u^h$, $N_d^h$, $a_u$, $a_d$, $b_u$, $b_d$, $N_u^c$, $N_d^c$, $\alpha_u$, $\alpha_d$, $\beta_d$, $\beta_u$, $g_c\}$. We include both statistical and systematical  experimental uncertainty in quadrature. Normalization uncertainties are not included in this fit. We have in total $N=26$ sets of which $N_{SIDIS}=20$ sets for SIDIS and $N_{e^+e^-}=6$ sets in $e^+e^-$.
The formalism is valid for low values of $P_{h\perp}/z\ll Q$, so we include only 
SIDIS data for $P_{h\perp}$-dependence using  a conservative choice $P_{h\perp}<0.8$ GeV. We also limit $P_{h\perp}/z_{h1} < 3.5$ GeV from BELLE~\cite{Seidl:2008xc}
and {\em BABAR}~\cite{TheBABAR:2013yha} data following the experimental cuts. The number of points is $N_{total}^{SIDIS}=140$ and 
$N_{total}^{e^+e^-}=122$. The number of fitted parameters, $13$, is adequate for fitting the total number of data points, $N_{total}=262$. More flexible parametrizations will be explored in future publications. In the fit we use the average values of $\langle x_B\rangle$,$\langle z_h\rangle$,$\langle y\rangle$,$\langle  P_{h\perp}\rangle$ for each bin in SIDIS and $\langle z_{h1}\rangle$,$\langle z_{h2}\rangle$,$\langle P_{h\perp}\rangle$,  $\langle \sin^2\theta\rangle/\langle 1+cos^2\theta \rangle$ for each bin in $e^+e^-$.

We present an estimate at 90\% confidence level (C.L.) interval for  the
nucleon tensor charge contributions and estimate errors on our results using the strategy outlined in
Refs.~\cite{deFlorian:2014yva,Martin:2009iq}. The method consists of exploring 
the parameter space $\{a_i\}$ by exploring possible values of $\chi^2$ so that
\bea
\chi^2(\{a_i\}) \leq \chi^2_{min} + \Delta \chi^2 \, ,
\eea
where $\Delta \chi^2$ corresponds to the so-called fit {\em tolerance} $T \equiv \sqrt{\Delta \chi^2}$. In the ideal case of uncorrelated measurements without unknown sources of error and gaussian errors of the measured observables, the 68\% C.L. corresponds to $\Delta \chi^2 = 1$ and
90\% C.L to  $\Delta \chi^2 = 2.71$. In typical measurement of asymmetries or other observables one encounters either correlated measurements or some inconsistent data sets due to uncontrolled experimental and/or theoretical errors. In order to deal with those issues 
the tolerance is changed with respect to the standard values.

A very rough idea of a good fit of the data set that contains $N$ points is the resulting $\chi^2$ being in the range of $N\pm \sqrt{2 N}$.
A more precise quantification of the allowed tolerance or $\Delta \chi^2$ can be estimated by 
 assuming that calculated $\chi^2$ follows the $\chi^2$-distribution for $N$ degrees of freedom with the probability density function
\bea
 \frac{1}{2 \Gamma(N/2)}\left( \frac{\chi^2}{2} \right)^{N/2 -1} \exp \left[-{\frac{\chi^2}{2}}\right]
\eea
The most probable value is the 50\% percentile  $\xi_{50}$ (compare to the goodness of fit):
\bea
 \int_{0}^{\xi_{50}} d \chi^2 \frac{1}{2 \Gamma(N/2)}\left( \frac{\chi^2}{2} \right)^{N/2 -1} \exp \left[-{\frac{\chi^2}{2}}\right] = 0.5\, .
\eea
This percentile is of order of $N$.
The 90\% percentile, $\xi_{90}$, is accordingly
\bea
 \int_{0}^{\xi_{90}} d \chi^2 \frac{1}{2 \Gamma(N/2)}\left( \frac{\chi^2}{2} \right)^{N/2 -1} \exp \left[-{\frac{\chi^2}{2}}\right] = 0.9\, .
\eea
The $\Delta \chi^2$ is defined then as
\bea
\Delta \chi^2 \equiv \xi_{90} - \xi_{50}\, .
\label{eq:delta}
\eea
Analogously we can define
\bea
\Delta \chi^2_{68} \equiv \xi_{68} - \xi_{50}\, ,
\label{eq:delta68}
\eea
for 68\% C.L.
In our particular case with 13 fitting parameters we have $N = N_{total} - 13 = 249$, $\xi_{50} = 248.3$,  and $\xi_{90} = 278.0$, and thus
$\Delta \chi^2 = 29.7$. It is comparable to $\sqrt{2 N} =22.3$.
We also have $\Delta \chi^2_{68} = 10.6$.

For each set of experimental data $i$, the 90\% C.L. is defined as in Ref.~\cite{Martin:2009iq}
\bea
\chi^2_i \leq \left(\frac{\chi^2_{i\, min}}{\xi_{50}}\right)\xi_{90} \, ,
\label{eq:deltachi2}
\eea
note that the value of $\xi_{90}$ is renormalized by ${\chi^2_{i\, min}}/\xi_{50}$ due to the fact that in the total global minimum $\chi^2_{min} = \sum_i \chi^2_{i\, min}$
the value of $\chi^2_{i\, min}$ may be away from possible minimal value.

In order to estimate errors on parameters and on calculation of asymmetries we will utilize a Monte Carlo sampling method
explained in Ref.~\cite{Anselmino:2008sga}. We are going to generate samples of parameters $\{a_i\}$ in the vicinity
of the minimum found by MINUIT $\{a_0\}$ that defines the minimal value of total  $\chi^2_{min}$. In order to account for 
correlations in parameters and improve numerical performance we will generate {\em correlated} parameter samples using CERNLIB's\footnote{\rm http://cernlib.web.cern.ch/cernlib/} Monte Carlo generators CORSET and CORGEN utilizing correlation matrix found by MINUIT. 
We generate $135$ sets of parameters $\{a_i\}$ that satisfy
\bea
\chi^2(\{a_i\}) \leq \chi^2_{min} + \Delta \chi^2\, ,
\label{condition}
\eea
with $\Delta \chi^2$ from Eq.~\eqref{eq:delta}.
 By definition these sets correspond to the   hyper volume in parameter space that defines 90\% C.L.   region. Any observable then will be calculated using these sets and the maximum and minimal value found will define our 90\% C.L. estimate. This Monte Carlo method underestimates the errors due to the limited number of generated parameter sets ($135$). The errors on asymmetries and functions that we quote are thus estimates and we will use a more robust method to estimate the errors on tensor charge. Errors on the tensor charge will be calculated using the evaluation of the $\chi^2$-profile by varying parameters of the model and careful analysis of the possible values.

%
%
\subsection{Results}

The resulting parameters after minimization procedure are presented in Table~\ref{parameters}. Only relative sign of transversity can be determined and we present here a solution with positive $u$ quark transversity as in Refs.~\cite{Anselmino:2007fs,Anselmino:2008jk,Anselmino:2013vqa,Bacchetta:2012ty,Bacchetta:2011ip}. Indeed transversity and helicity distributions can be related via boost and rotation of correspondding operators, however boost and rotation do not commute in quantum theory, thus these two distributions are independent and in principle different. It is unlikely that they differ by sign, thus we choose the same sign for $u$-quark transversity and $u$-quark helicity distribution~\cite{deFlorian:2009vb} which is positive.
Transversity of $d$ quark is negative. Favored and unfavored Collins FFs are of opposite signs, indeed $N_u^c<0$, $N_d^c>0$ and of approximately the same magnitudes. It means that {\em favored } Collins fragmentation function is positive and  {\em unfavored } Collins fragmentation function is negative, see Eq.~\eqref{eq:relation_to_trento}.  
The corresponding sum rule \cite{Schafer:1999kn,Meissner:2010cc} for Collins fragmentation functions read:
\bea
\sum_h \sum_{S_h} \int_0^1 d z_h z_h H_{1\, h/j}^{\perp (1)}(z_h)|_{\rm Trento} = 0 \, ,
\label{sumrule}
\eea
which suggests the compensation of  {\em favored } and  {\em unfavored }  Collins fragmentation functions.

We observe that parameters that define $z$ dependence of Collins FF $\alpha_u$ and $\alpha_d$ are different, thus
the $z$-shapes of favorite and unfavorite Collins FFs are different. The same is true for transversity distributions,
both large-$x$ region controlled by $b_u$ and $b_d$ and low-$x$ region controlled by $a_u$ and $a_d$ indicates that the $x$-shape
of transversity for $u$ and $d$ quarks is different.  It might be well possible that $k_\perp$-shape of transversity and Collins fragmentation functions is also flavor dependent, however the current experimental data does not allow to determine whether it is true or not.
\begin{table}[htb]
\begin{tabular}{l c l l c l l c l}
\hline
 & & & & & & & &\\
$N_u^h$ &=& $0.85\pm 0.09$ & $a_u$ &=& $ 0.69 \pm 0.04$ & $b_u$ &=& $ 0.05 \pm 0.04$ \\
$N_d^h$ &=& $-1.0\pm 0.13$ & $a_d$ &=& $ 1.79 \pm 0.32$ & $b_d$ &=& $ 7.00 \pm 2.65$  \\
$N_u^c$ &=& $-0.262\pm 0.025$ & $\alpha_u$ &=& $ 1.69 \pm 0.01$ & $\beta_u$ &=& $ 0.00 \pm 0.54$ \\
$N_d^c$ &=& $0.195\pm 0.007$ & $\alpha_d$ &=& $ 0.32 \pm 0.04$ & $\beta_d$ &=& $ 0.00 \pm 0.79$ \\
$g_c$ &=& $0.0236\pm 0.0007$&\multicolumn{3}{l}{(GeV$^2$)}\\
& & & & & & & &\\
\hline
& & & & & & & &\\
\multicolumn{3}{l}{$\chi_{min}^2 =  218.407$} & \multicolumn{3}{l}{$\chi^2_{min}/{n.d.o.f}=0.88$}  \\
\end{tabular}
\caption{Fitted parameters of the transversity quark distributions for $u$ and $d$ and Collins fragmentation functions.
The fit is performed by using MINUIT minimization package. Quoted errors correspond to MINUIT estimate.}
\label{parameters}
\end{table}

The total $\chi^2_{min} = 218.407$ and $\chi^2/{n_{d.o.f.}} = 218.407/249=0.88$.  We calculate the goodness of fit using the well known formula \cite{Press:1992zz}:
\bea
P(\chi^2_{min},n_{d.o.f.}) = 1 - \int_{0}^{\chi^2_{min}} d \chi^2 \frac{1}{2 \Gamma(n_{d.o.f.}/2)}\left( \frac{\chi^2}{2} \right)^{n_{d.o.f.}/2 -1} \exp \left[-{\frac{\chi^2}{2}}\right]\, .
\eea
The goodness of fit  describes how well it fits a set of observables. In principle if the model adequately describes the data then one would expect
$\chi^2/{n_{d.o.f.}} \simeq 1$. In case $\chi^2/{n_{d.o.f.}} \gg 1$ the model fails to describe the data, $\chi^2/{n_{d.o.f.}} \ll 1$ means that the model
starts fitting the statistical noise in the data. Notice that in our fit we obtained $\chi^2/{n_{d.o.f.}} = 0.88$ which means that the number of parameters is adequately chosen. An attempt to extract more information from the data, such as flavor dependence etc, would lead to  $\chi^2/{n_{d.o.f.}} \ll 1$. One of course can estimate a number of different hypothesis and we leave those estimates for further publications.

We obtain that the probability that the fit indeed is the underlying mechanism for the measured asymmetries is $P(218.407,249) = 92\%$. This gives us full confidence in presented results. It is very important to note that we have used the data from two different processes implementing appropriate factorization and evolution. Thus we have also presented a phenomenological proof that these processes, SIDIS and $e^+e^-$, are consistent with TMD factorization.

The results of the fit including partial values of $\chi^2$ are presented in Table~\ref{results_sidis} for SIDIS experiments and in Table~\ref{results_epem} for $e^+e^-$ experiments. One observes that $\chi^2$ values are quite satisfactory and {\em homogeneous} for both SIDIS, $\chi^2_{SIDIS}/{N_{total}^{SIDIS}} = 0.93$, and $e^+e^-$, $\chi^2_{e^+e^-}/{N_{total}^{e^+e^-}} = 0.72$. TMD factorization approach is describing data of both SIDIS and $e^+e^-$ adequately. 

Description of SIDIS data is very good. The data span the energy range
starting from $P_{lab}^{\rm JLAB} = 5.9$ to $P_{lab}^{\rm HERMES} = 27.5$, and to $P_{lab}^{\rm COMPASS} = 160$ GeV. 
The resolution scale changes also in a relatively wide region $1 \lesssim \langle Q^2 \rangle \lesssim 6$ (GeV$^2$) for HERMES and $1 \lesssim \langle Q^2 \rangle \lesssim  21$ (GeV$^2$) for COMPASS.
One can see from Table~\ref{results_sidis} that description of the individual subsets is also very satisfactory.

As we mentioned in Sec.~\ref{introduction}, it is very important to include appropriate QCD evolution in order to be able 
to have a controlled accuracy and adequate description of  $e^+e^-$ data that is measured at $Q^2 \simeq 110$ GeV$^2$. One can see from Table~\ref{results_epem}
that both BELLE \cite{Seidl:2008xc} and {\em BABAR} \cite{TheBABAR:2013yha} data sets on $A_0$ are described well. Both 
methods $UC$ and $UL$ from BELLE \cite{Seidl:2008xc} and {\em BABAR} \cite{TheBABAR:2013yha} appear to be consistent with
our description and also
 $P_{h\perp}$ dependence of asymmetry is well described. We will give predictions for $P_{h\perp}$ dependent unpolarized cross sections in $e^+e^-$ in the following section.

\begin{table}[htb]
\begin{tabular}{l l l c c c c}
Experiment & hadron & Target &  dependence & \# ndata & $\chi^2$ &
$\chi^2/{ndata}$ \\
\hline
COMPASS \cite{Alekseev:2008aa} &$\pi^+$ &LiD& $x$  & 9 &   11.16 & 1.24 \\
COMPASS \cite{Alekseev:2008aa} &$\pi^-$ &LiD&  $x$  & 9 &  9.08  & 1.01 \\
COMPASS \cite{Alekseev:2008aa} &$\pi^+$ &LiD&  $z$  & 8 &  3.26  & 0.41 \\
COMPASS \cite{Alekseev:2008aa} &$\pi^-$ &LiD&  $z$  & 8 &  7.29  & 0.91 \\
COMPASS \cite{Alekseev:2008aa} &$\pi^+$ &LiD&  $P_{h \perp}$ & 6 & 4.19   &
0.70  \\
COMPASS \cite{Alekseev:2008aa} &$\pi^-$ &LiD&  $P_{h \perp}$ & 6 &  4.50   &
0.75  \\
\hline
COMPASS \cite{Adolph:2014zba} &$\pi^+$   &NH$_3$&$x$ & 9 &   21.46   &
2.38 \\
COMPASS \cite{Adolph:2014zba} &$\pi^-$   &NH$_3$&$x$ & 9 &  6.23   &
0.69 \\
COMPASS \cite{Adolph:2014zba} &$\pi^+$  &NH$_3$&$z$ & 8 &   7.80  &
0.98 \\
COMPASS \cite{Adolph:2014zba} &$\pi^-$   &NH$_3$&$z$ & 8 &  10.29  &
1.29 \\
COMPASS \cite{Adolph:2014zba} &$\pi^+$  &NH$_3$&$P_{h \perp}$  & 6 &
3.82 & 0.64  \\
COMPASS \cite{Adolph:2014zba} &$\pi^-$   &NH$_3$&$P_{h \perp}$  & 6 &
3.85  & 0.64  \\
\hline
HERMES \cite{Airapetian:2009ae} &$\pi^+$ &H&  $x$  & 7 & 5.37   & 0.77 \\
HERMES \cite{Airapetian:2009ae} &$\pi^-$ &H&  $x$  & 7 & 12.61   & 1.80 \\
HERMES \cite{Airapetian:2009ae} &$\pi^+$ &H&  $z$  & 7 & 3.04  & 0.43 \\
HERMES \cite{Airapetian:2009ae} &$\pi^-$ &H&  $z$  & 7 & 3.23   & 0.46 \\
HERMES \cite{Airapetian:2009ae} &$\pi^+$ &H&  $P_{h \perp}$  & 6 &    1.60 & 0.27  \\
HERMES \cite{Airapetian:2009ae} &$\pi^-$ &H&  $P_{h \perp}$  & 6 &    4.82  & 0.80  \\
  \hline
JLAB \cite{Qian:2011py} &$\pi^+$ &$^3$He&  $x$  & 4 &   3.90   & 0.98 \\
JLAB \cite{Qian:2011py} &$\pi^-$ &$^3$He&  $x$  & 4 &  3.11   & 0.78 \\
\hline
  & & & & 140 & 130.65 & 0.93 \\
\end{tabular}
\caption{Partial $\chi^2$ values of the global best fit for SIDIS experiments.}
\label{results_sidis}
\end{table}

\begin{table}[htb]
\begin{tabular}{l l  c c c c}
Experiment & Observable &  dependence & \# ndata & $\chi^2$ &
$\chi^2/{ndata}$ \\
\hline
& & & & & \\
BELLE \cite{Seidl:2008xc}  & $A_0^{UL}$ & $z$  & 16 &   13.02   & 0.81 \\
BELLE \cite{Seidl:2008xc}  & $A_0^{UC}$ & $z$  & 16 &   11.54  & 0.72 \\
{\em BABAR}\cite{TheBABAR:2013yha}  & $A_0^{UL}$ & $z$  & 36 &   34.61   & 0.96 \\
{\em BABAR}\cite{TheBABAR:2013yha}  & $A_0^{UC}$ & $z$  & 36 &   15.17  & 0.42 \\
{\em BABAR}\cite{TheBABAR:2013yha}  & $A_0^{UL}$ & $P_{h\perp}$  & 9 &   9.09   & 1.01 \\
{\em BABAR}\cite{TheBABAR:2013yha}  & $A_0^{UC}$ & $P_{h\perp}$  & 9 &   4.33  & 0.48 \\
& & & & & \\
\hline
  & &  & 122 & 87.76 & 0.72 \\
\end{tabular}
\caption{Partial $\chi^2$ values of the global best fit for $e^+e^-$ experiments.}
\label{results_epem}
\end{table}

%
%
\subsection{Transversity, Collins fragmentation functions and tensor charge}

\begin{figure}[tbp]
\centering
\includegraphics[width=8cm]{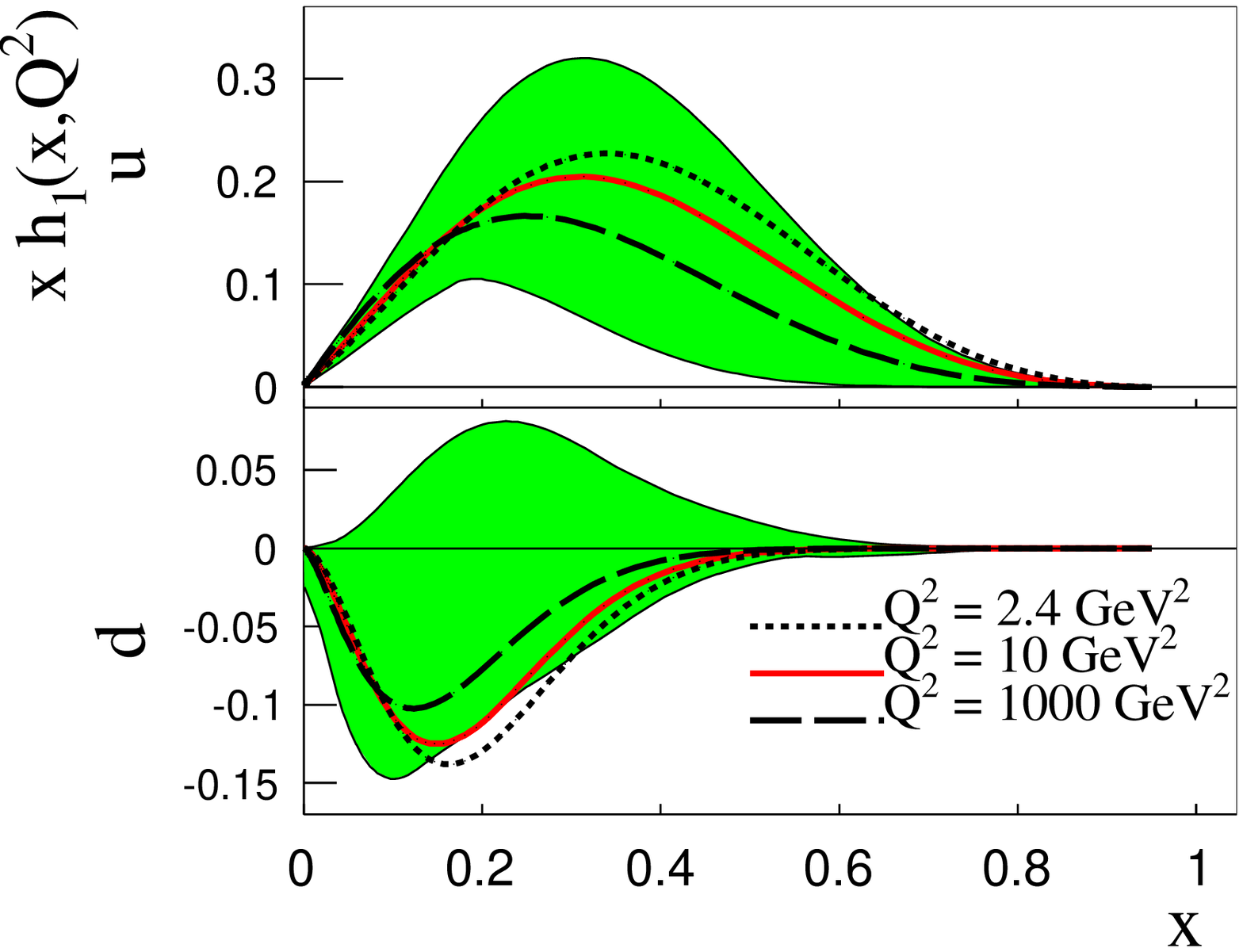}(a)
\includegraphics[width=8cm]{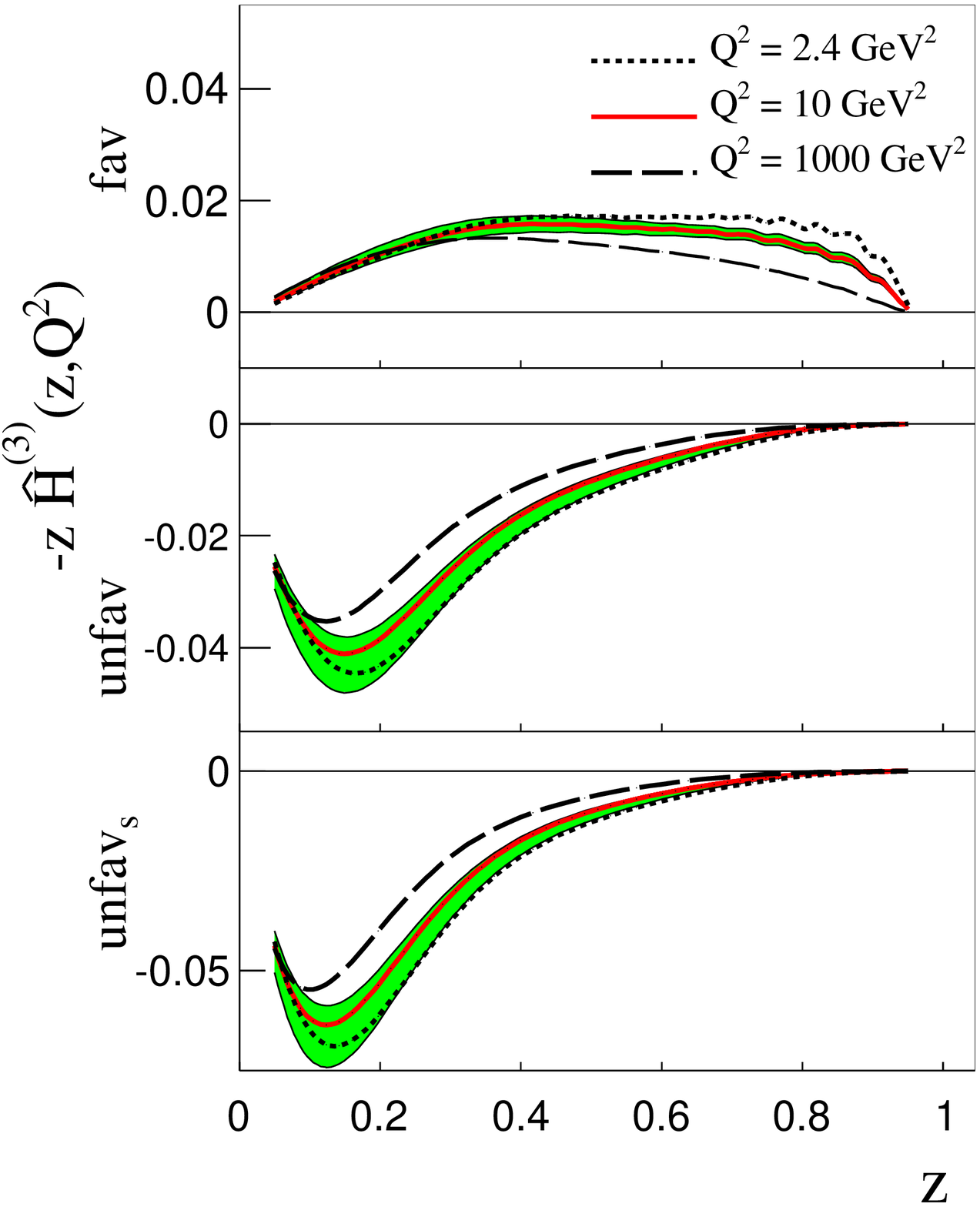}(b)
\caption{Extracted transversity distribution (a) and Collins regimentation function (b) at three different scales  $Q^2 = 2.4$ (dotted lines), $Q^2 = 10$ (solid lines) and $Q^2 = 1000$ (dashed lines) GeV$^2$. The shaded region corresponds to our estimate of 90\% C.L. error band at $Q^2=10$ GeV$^2$.}
\label{fig:functions}
\end{figure}

We plot transversity and the Collins fragmentation function in Fig.~\ref{fig:functions} at two different scales $Q^2 = 10$ and 1000 GeV$^2$. In order to evaluate functions we solve appropriate DGLAP equations for transversity Eq.~\eqref{eq:transversity} and twist-3 collins functions Eq.~\eqref{eq:collins}. Due to the fact that neither of the functions mixes with gluons, these distributions do not change drastically in low-$x$ region due to DGLAP evolution.

\begin{figure}[tbp]
\centering
\includegraphics[width=4.2cm,bb= 50 10 595 550]{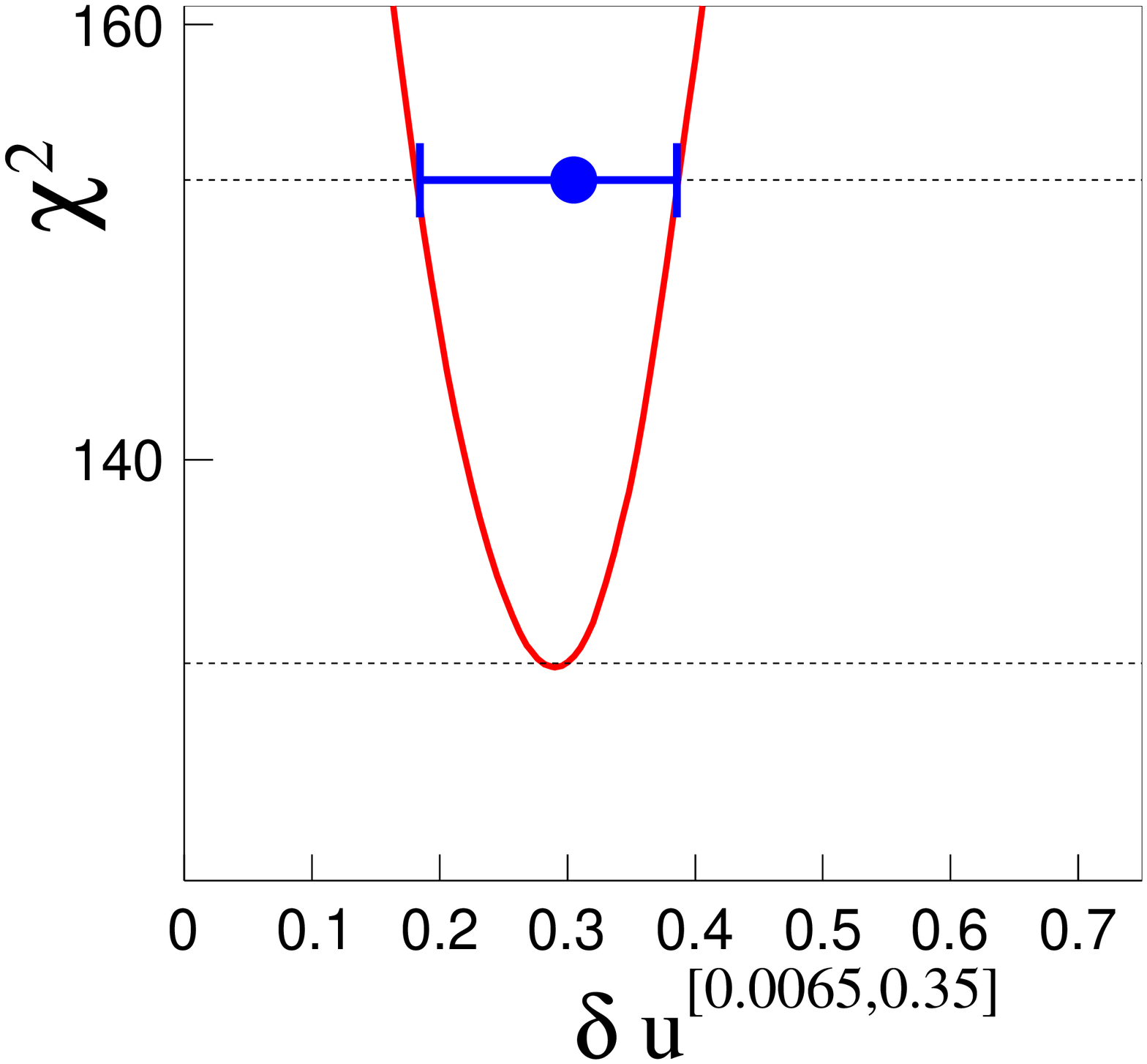}
\includegraphics[width=4.2cm,bb= 50 10 595 550]{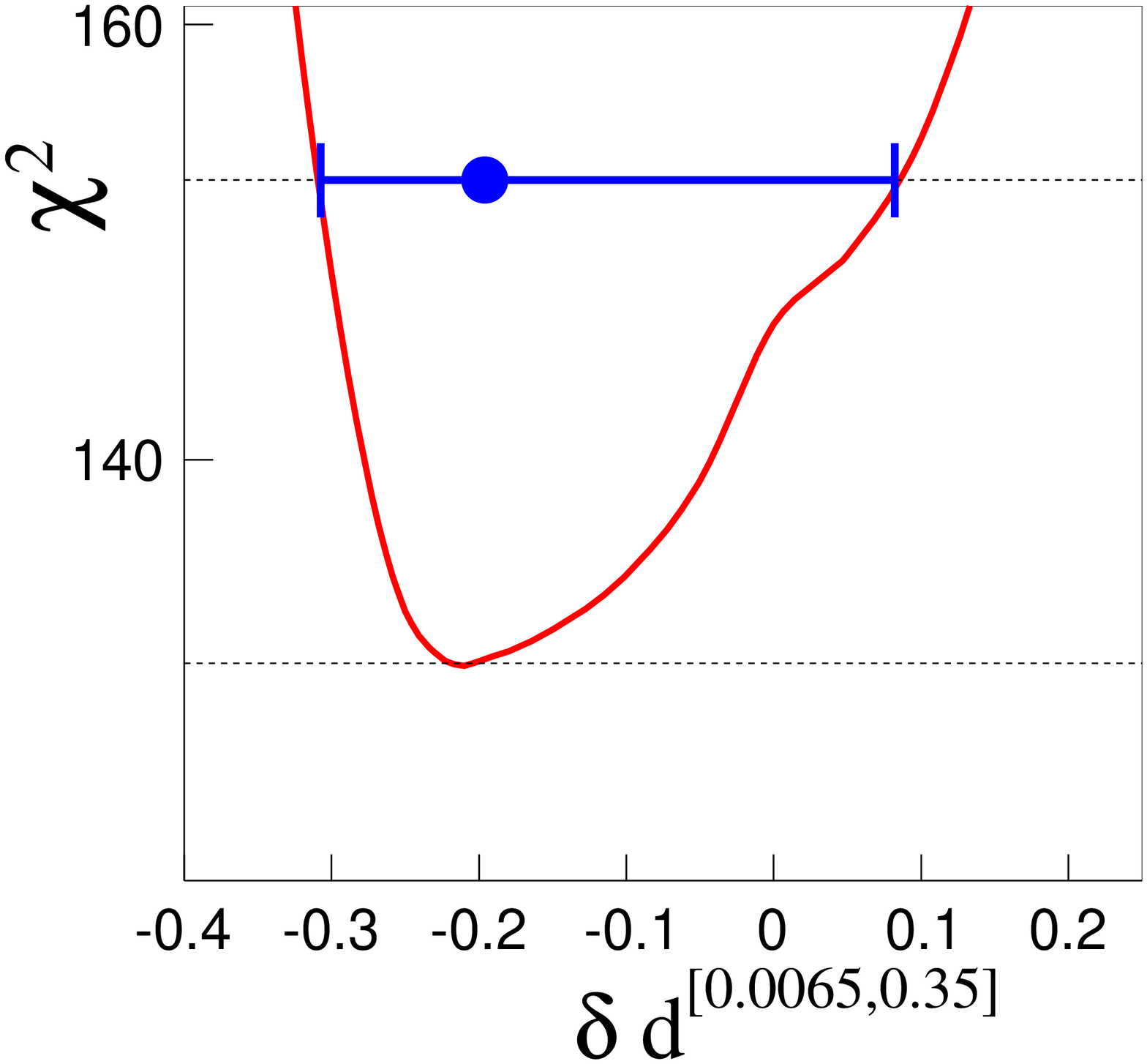}
\caption{$\chi^2$ profiles for up and down quark contributions to the tensor charge. The errors of points correspond to 90\% C.L. interval at $Q^2=10$ GeV$^2$.}
\label{fig:chi2}
\end{figure}

\begin{figure}[tbp]
\centering
\includegraphics[width=4.2cm,bb= 50 10 595 550]{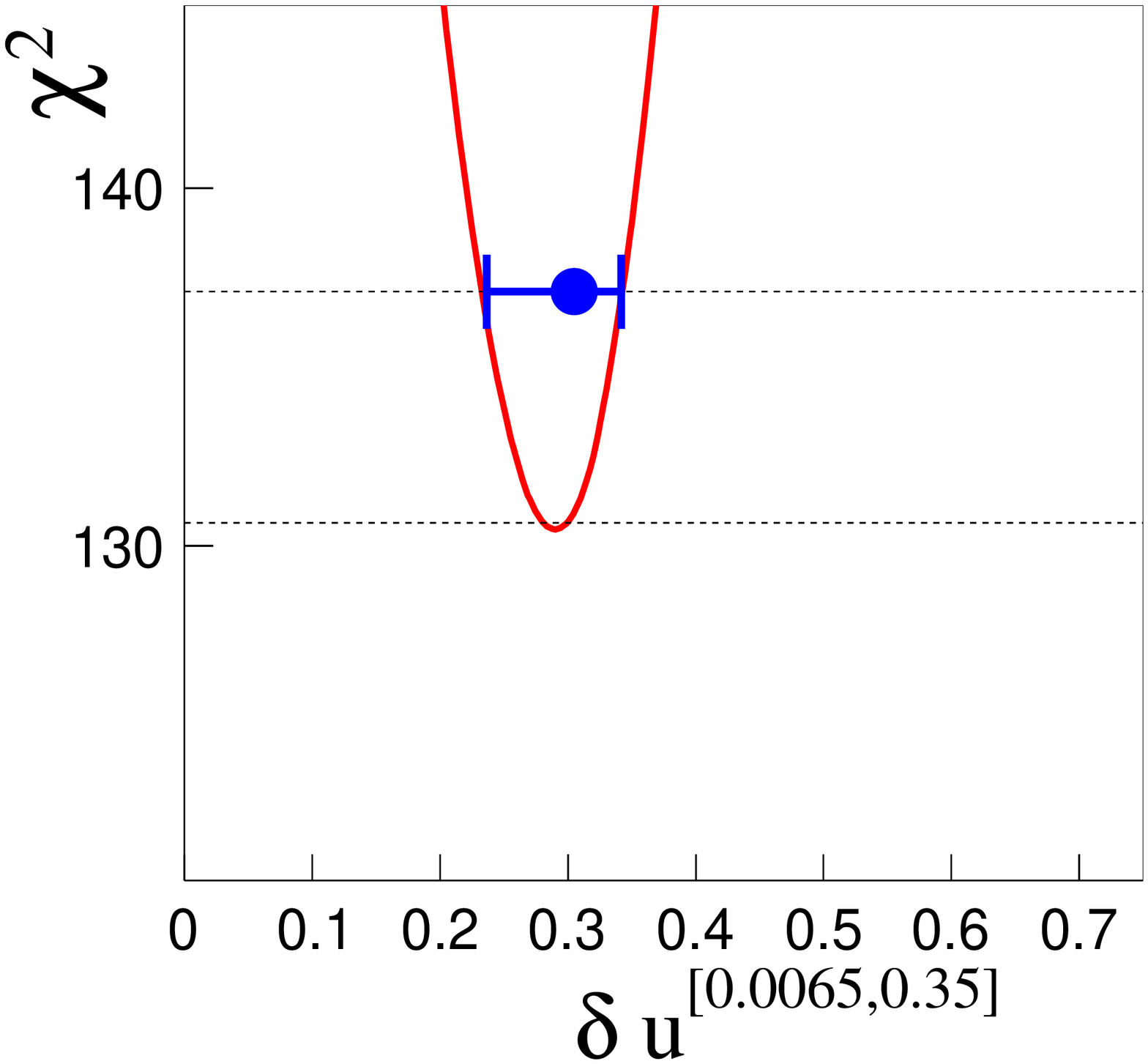}\hspace{1cm}
\includegraphics[width=4.2cm,bb= 50 10 595 550]{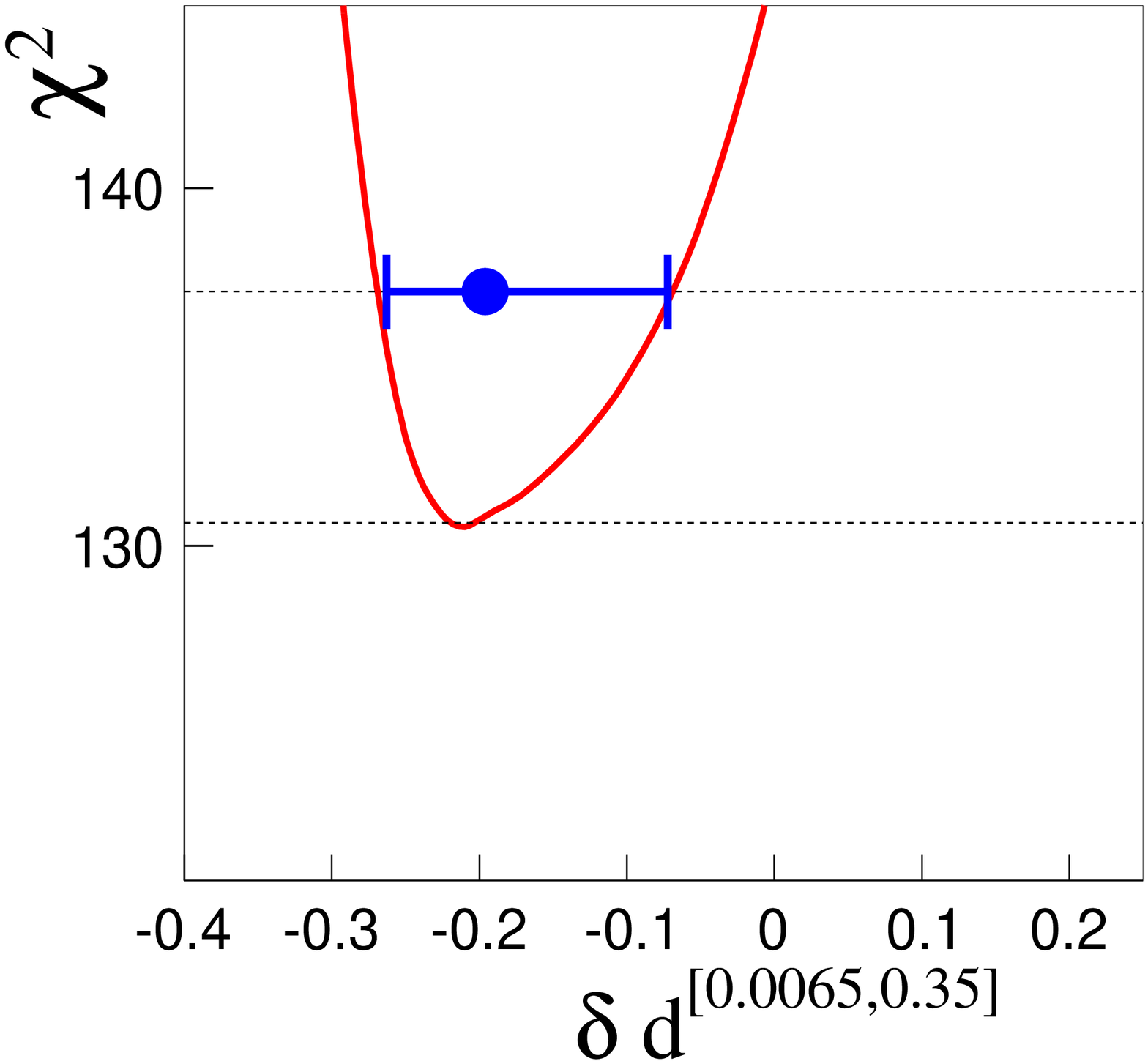}
\caption{$\chi^2$ profiles for up and down quark contributions to the tensor charge. The errors of points correspond to 68\% C.L. interval at $Q^2=10$ GeV$^2$.}
\label{fig:chi2_68}
\end{figure}

 Transversity enters directly in SIDIS
asymmetry and we find that  the main constraints come from
SIDIS data only, its correlations with errors of Collins FF turn out
to be numerically negligible. We thus vary only $\chi^2_{SIDIS}$ and use $\Delta \chi^2_{SIDIS} = 22.2$ for 90\% C.L. and $\Delta \chi^2_{SIDIS} = 6.4$ for 68\% C.L. calculated using Eq.~\eqref{eq:deltachi2}.
Since the experimental data has only probed the limited region $0.0065 < x_B < 0.35$,
we define the following partial contribution to the tensor charge
\begin{eqnarray}
\delta q^{[x_{\rm min},x_{\rm max}]}\left(Q^2\right) \equiv   \int_{x_{\rm min}}^{x_{\rm max}}dx \, h_1^q(x,Q^2) \ .
\end{eqnarray}
In Fig.~\ref{fig:chi2}, we plot the
$\chi^2$ Monte Carlo scanning of SIDIS data for the contribution to
the tensor charge from such a region, and find ~\cite{Kang:2014zza}
\begin{eqnarray}
\delta u^{[0.0065,0.35]} &=&  +0.30_{-0.12}^{+0.08} \ ,\\
\delta d^{[0.0065,0.35]} &=&  -0.20_{-0.11}^{+0.28}   \ ,
\end{eqnarray}
at 90\% C.L. at  $Q^2=10$ GeV$^2$.
Analogously in Fig.~\ref{fig:chi2_68}, we plot the
$\chi^2$ Monte Carlo scanning of SIDIS data at 68\% C.L. at  $Q^2=10$ GeV$^2$ and find
\begin{eqnarray}
\delta u^{[0.0065,0.35]} &=&  +0.30_{-0.07}^{+0.04} \ ,\\
\delta d^{[0.0065,0.35]} &=&  -0.20_{-0.07}^{+0.12}   \ . 
\end{eqnarray}
We notice that this result is comparable with previous TMD extractions
without evolution~\cite{Anselmino:2007fs,Anselmino:2008jk,Anselmino:2013vqa}
and di-hadron method~\cite{Bacchetta:2012ty,Bacchetta:2011ip}.

\begin{figure}[tbp]
\centering
\includegraphics[width=4.2cm,bb= 50 10 595 550]{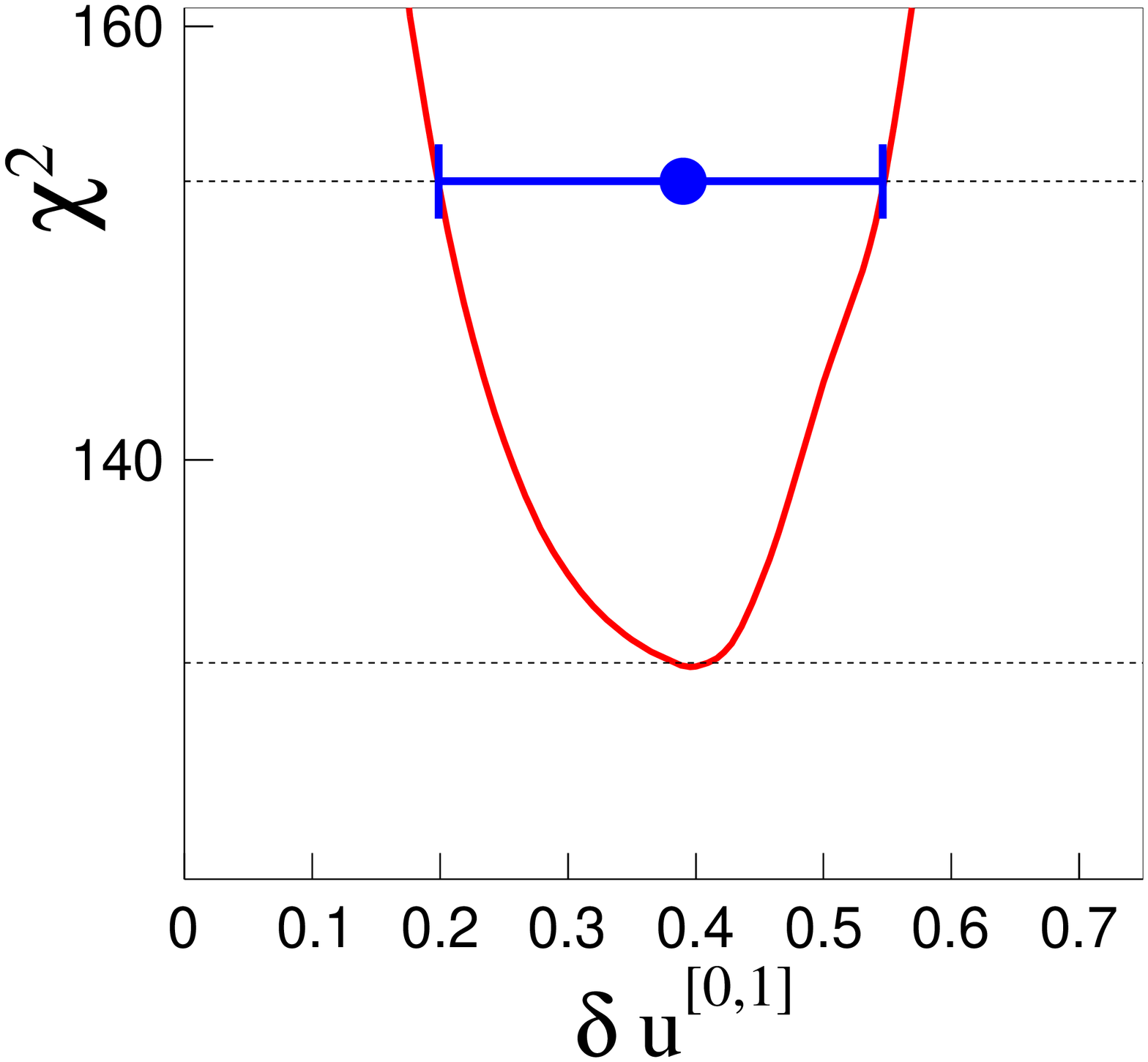}\hspace{1cm}
\includegraphics[width=4.2cm,bb= 50 10 595 550]{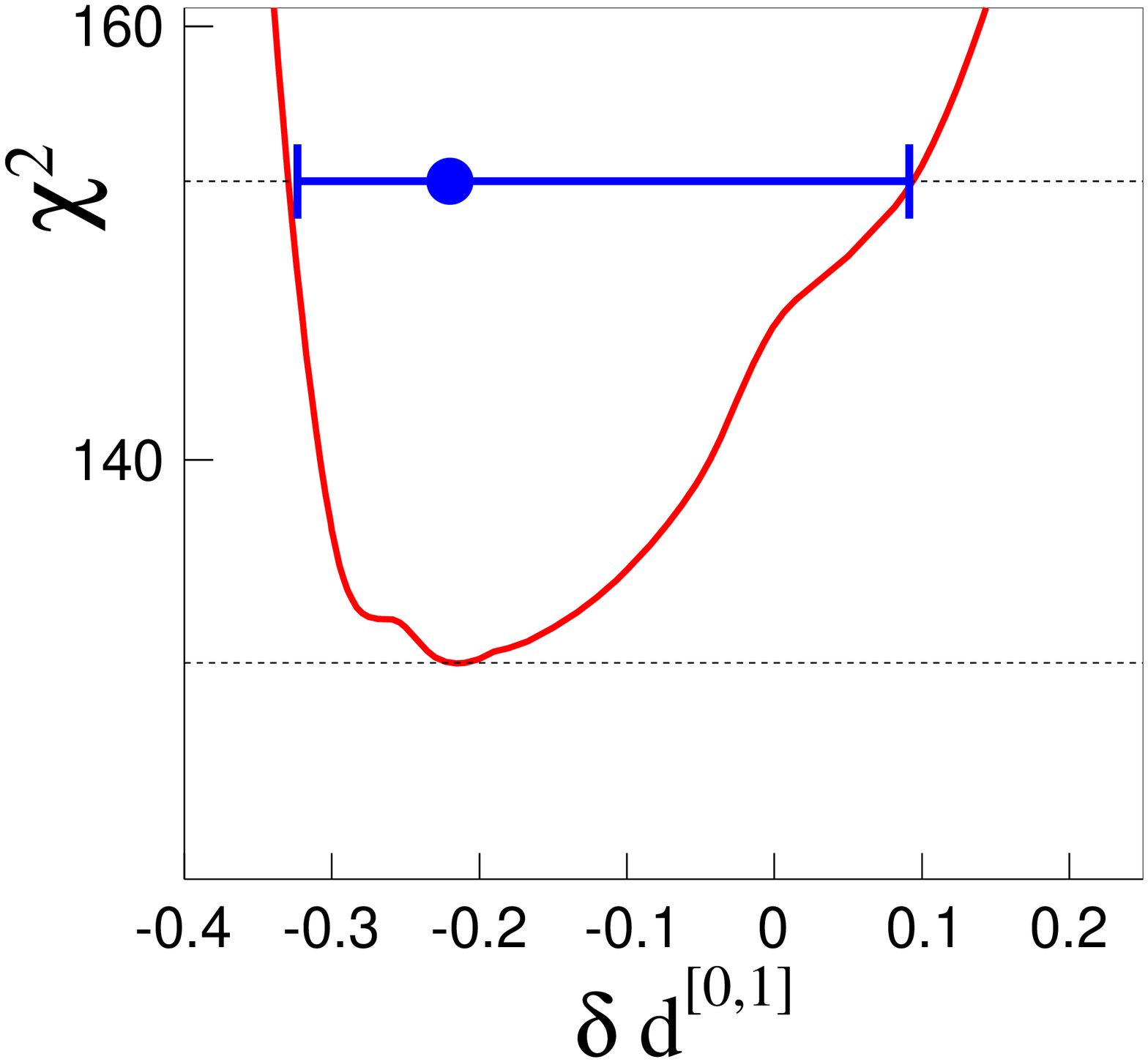}
\caption{$\chi^2$ profiles for up and down quark contributions to the tensor charge in the whole kinematical region. The errors of points correspond to 90\% C.L. interval.}
\label{fig:chi201}
\end{figure}

\begin{figure}[tbp]
\centering
\includegraphics[width=4.2cm,bb= 50 10 595 550]{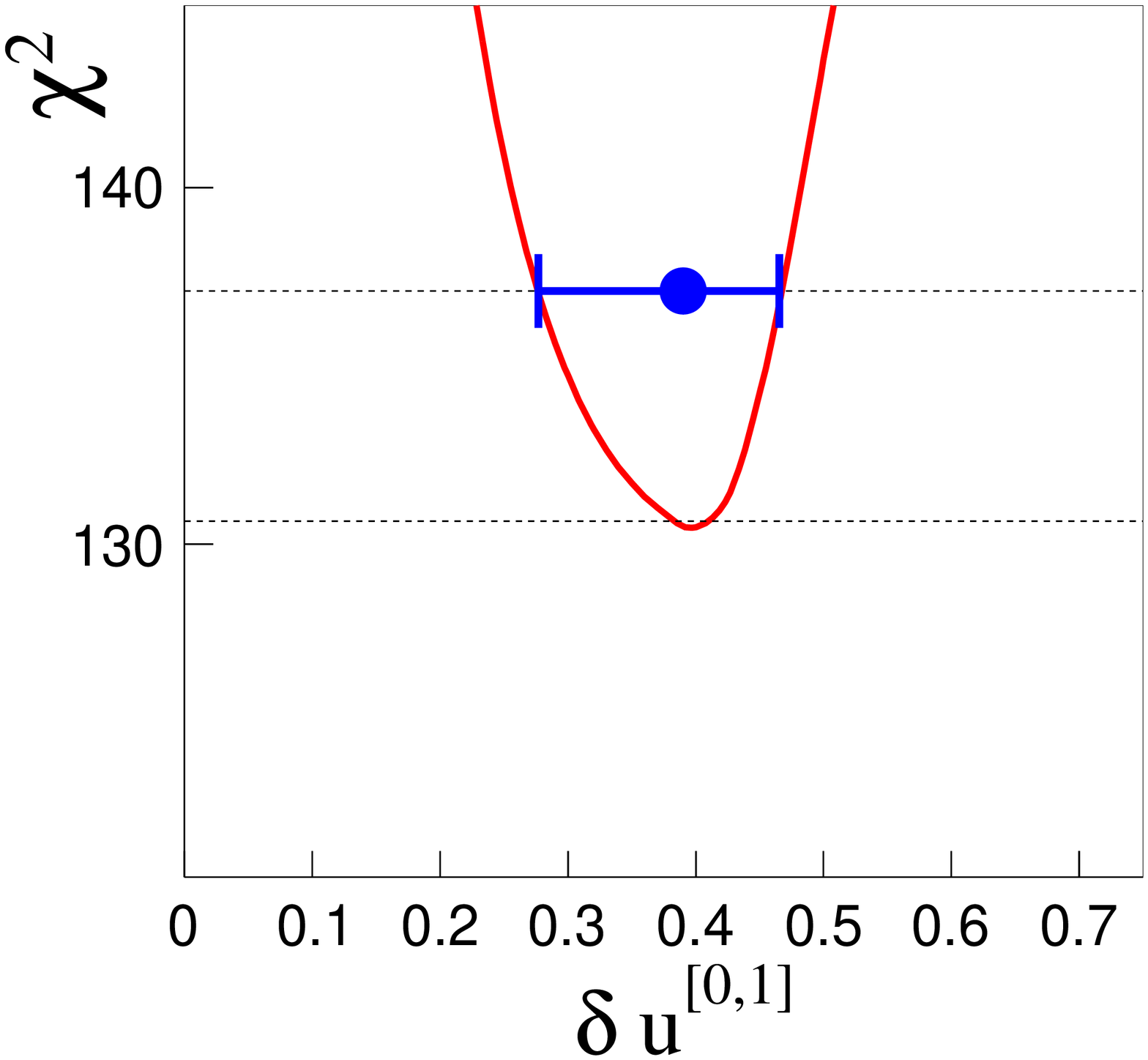}\hspace{1cm}
\includegraphics[width=4.2cm,bb= 50 10 595 550]{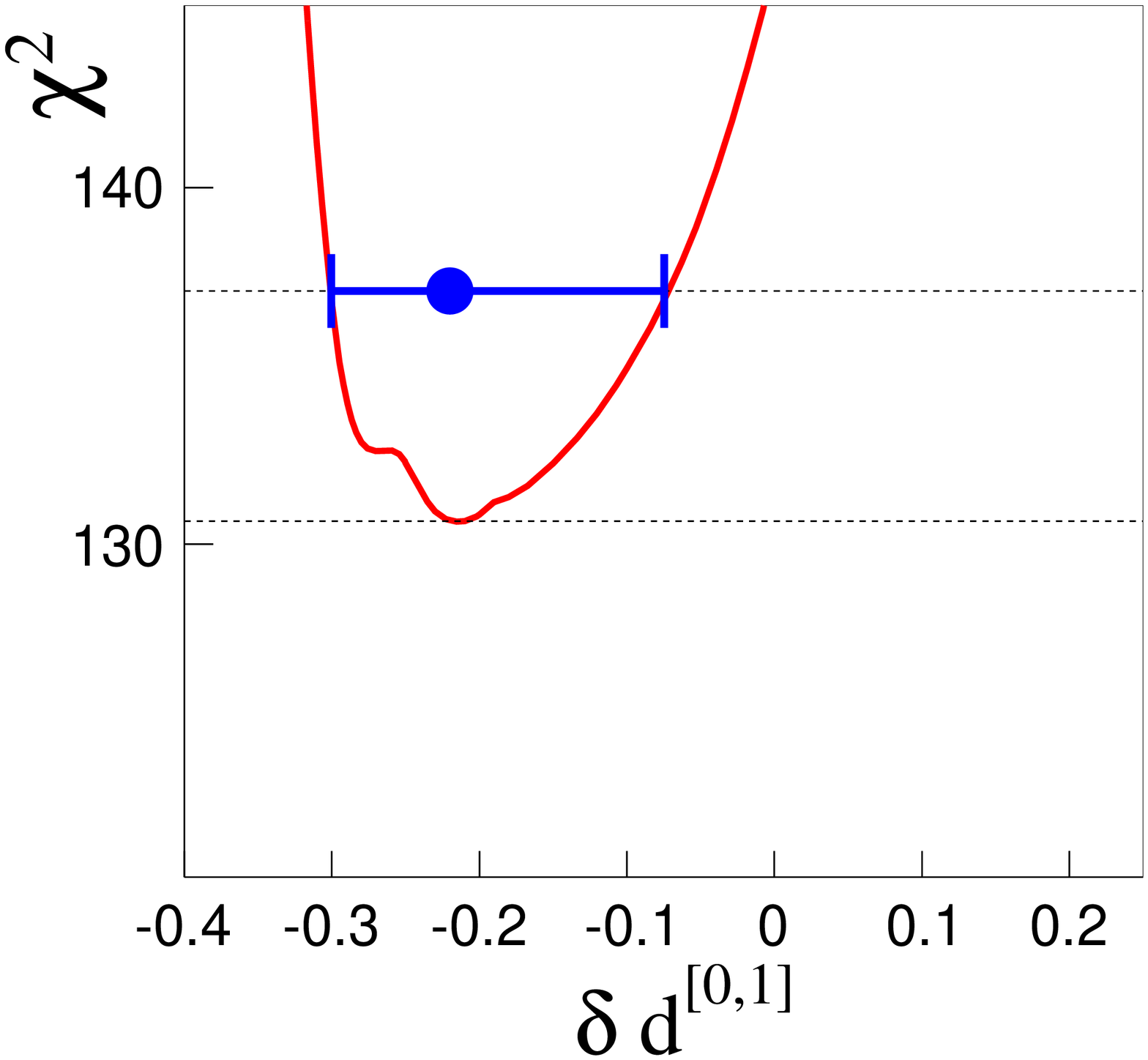}
\caption{$\chi^2$ profiles for up and down quark contributions to the tensor charge in the whole kinematical region. The errors of points correspond to 68\% C.L. interval.}
\label{fig:chi201_68}
\end{figure}

Existing experimental data covers a limited kinematic region,
thus a simple extension of our fitted parametrization to the whole range of $0<x_B<1$
will significantly underestimate the uncertainties, in particular, in the dominant large-$x_B$
regime.
It is extremely important to extend the experimental study of the quark transversity
distribution to both large and small $x_B$ to constrain the total tensor charge
contributions. This requires future experiments to provide measurements
at the Jefferson Lab 12 GeV upgrade~\cite{Dudek:2012vr} and the planned
Electron Ion Collider~\cite{Boer:2011fh,Accardi:2012qut,Aschenauer:2014twa}.
Nevertheless for completeness let us present our results on the tensor charge calculated over the whole kinematical region $\delta q^{[0,1]}$:
\begin{eqnarray}
\delta u^{[0,1]} &=&  +0.39_{-0.20}^{+0.16} \ ,\\
\delta d^{[0,1]} &=& -0.22_{-0.10}^{+0.31} \ ,
\end{eqnarray}
at 90\% C.L. and
\begin{eqnarray}
\delta u^{[0,1]} &=&  +0.39_{-0.11}^{+0.07} \ ,\\
\delta d^{[0,1]} &=& -0.22_{-0.08}^{+0.14} \ ,
\end{eqnarray}
at 68\% C.L. both at  $Q^2=10$ GeV$^2$, as shown in Figs.~\ref{fig:chi201} and \ref{fig:chi201_68}. The tensor charge for $u$-quark can have a bigger contribution with respect to $d$-quark from unexplored region of $x$ according to our estimates.

%
%
\subsection{TMD Interpretations of Our Results}
\label{sec:interpretation}

As we mentioned in the Introduction and elaborated in Sec.~II, there exists TMD
interpretation of CSS results. In particular the equations of the previous section that 
represent the solution of evolution equations are quite complicated. One might
formulate the solutions of TMD evolution equations for TMD functions directly,
in such a way that expressions will look very much like extension of a simple 
parton-like model, for instance used in Ref.~\cite{Anselmino:2011ch}. 

Let us start with writing the cross sections in terms of the individual TMDs:
\bea
F_{UU}(Q;P_{h\perp}) &= H_{\rm SIDIS}(Q, \mu=Q) \sum_q e_q^2  \int_{k_\perp, p_\perp} f_1^q(x_B, k_\perp^2; Q) D_{h/q}(z_h, p_\perp^2; Q) , \nonumber
\\
F_{UT}^{\sin\left(\phi_h +\phi_s\right)}(Q;P_{h\perp}) &=  - H_{\rm SIDIS}(Q, \mu=Q)\sum_q e_q^2\int_{k_\perp, p_\perp} h_1^q(x_B, k_\perp^2; Q) \nonumber
\frac{\hat P_{h\perp}\cdot p_\perp}{M_h}H_{1 h/q}^\perp(z_h, p_\perp^2; Q) ,
\\
Z_{uu}^{h_1h_2}(Q;P_{h\perp})&= H_{e^+e^-}(Q, \mu=Q) \sum_q e_q^2 \int_{p_{1\perp}, p_{2\perp}}  D_{h/q}(z_{h1}, p_{1\perp}^2; Q) D_{h/\bar q}(z_{h2}, p_{2\perp}^2; Q) , \nonumber
\\
Z_{\rm collins}^{h_1h_2}(Q;P_{h\perp}) & = H_{e^+e^-}(Q, \mu=Q)  \sum_q e_q^2 \int_{p_{1\perp}, p_{2\perp}} 
\left(2\hat P^\alpha_{h\perp} \hat P^\beta_{h\perp}-g^{\alpha\beta}_\perp\right)
\frac{p_{1\perp}^\alpha}{M_{h1}} H_{1\, h/q}^\perp(z_{h1}, p_{1\perp}^2; Q)\nonumber\\
&~~~~~~~~~~~~~~~~~\times  \frac{p_{2\perp}^\alpha}{M_{h2}}
H_{1\, h/\bar q}^\perp(z_{h2}, p_{2\perp}^2; Q) , \label{eq:tmd_inter}
\eea
where we have set the factorization scale $\mu=Q$, and the evolution effects have been fully taken into 
account in the TMDs, $f_1^q(x_B, k_\perp^2; Q)$, $h_1^q(x_B, k_\perp^2; Q)$, $D_{h/q}(z, p_{\perp}^2; Q)$, 
and $H_{1 h/\bar q}^\perp(z, p_{\perp}^2; Q)$ are the transverse momentum dependent unpolarized 
parton distribution function, quark transversity, unpolarized fragmentation function, and the Collins 
function at the scale $\mu=Q$ and $\zeta = Q^2$, respectively. These TMDs are also understood as the soft factor 
subtracted TMDs. 

The short-hand notations for the integrations have the following explicit forms:
\bea
\int_{k_\perp, p_\perp} &\equiv \int d^2k_\perp d^2p_\perp \delta^2\left(z_h \vec{k}_\perp + \vec{p}_\perp - \vec{P}_{h\perp} \right),
\\
\int_{p_{1\perp}, p_{2\perp}} & \equiv \int d^2p_{1\perp} d^2p_{2\perp} \delta^2\left(\vec{P}_{h\perp} - \vec{p}_{1\perp} - \vec{p}_{2\perp} \frac{z_{h1}}{z_{h2}}\right).
\eea
As discussed in Sec.~II, the TMDs and the associated hard factors depend on the scheme to regulate
the light-cone singularities. However, in the final results for the structure functions, this scheme dependence
cancels out between the TMDs and the hard factors. In the following, we present the results
in the   Collins-11
scheme~\cite{Collins:2011zzd}. The functions that encode scheme dependence from Eqs.~(\ref{tmdf},\ref{tmdd},\ref{tmdh},\ref{tmdc}) are $\widetilde{{\cal F}}_q(\alpha_s(Q)) = 1$, $\widetilde{{\cal D}}_q(\alpha_s(Q)) =1$,
$\widetilde{{\cal H}}_{1q}(\alpha_s(Q))=1$, $\widetilde{{\cal H}}_{c}(\alpha_s(Q))=1$ at one loop. For all other schemes, the results can be obtained accordingly. 

In the Collins-11 scheme, the associated hard factors can be written using Eqs.~(\ref{hqjcc},\ref{eq:hsidisjcc}) as
\bea
H_{\rm SIDIS}(Q, \mu=Q) &= 1+\frac{\alpha_s(Q)}{2\pi} C_F\left(- 8\right)\ , \nonumber
\\
H_{e^+ e^-}(Q, \mu=Q) &= 1+\frac{\alpha_s(Q)}{2\pi} C_F\left(\pi^2- 8\right)\ .
\label{eq:hardfactor}
\eea

The TMDs are Fourier transformations of the relevant expressions in $b$
space in Sec.~II, Eqs.~(\ref{tmdf},\ref{tmdd},\ref{tmdh},\ref{tmdc}),
\bea
f_1^q(x, k_\perp^2; Q) &= \int_0^{\infty} \frac{db\; b}{(2\pi)} J_0(k_\perp b)\, C_{q\gets i}^{f_1}\otimes f_1^i(x, \mu_b)\, e^{-\frac{1}{2} S_{\rm pert}(Q, b_*)  - S_{\rm NP}^{f_1}(Q, b)}, \label{eq:f1_tmd}
\\
h_1^q(x, k_\perp^2; Q) &= \int_0^{\infty} \frac{db\; b}{(2\pi)} J_0(k_\perp b)\, \delta C_{q\gets i}\otimes h_1^i(x, \mu_b)\, e^{-\frac{1}{2} S_{\rm pert}(Q, b_*)  - S_{\rm NP}^{h_1}(Q, b)}, \label{eq:h1_tmd}
\\
D_{h/q}(z, p_\perp^2; Q) &= \frac{1}{z^2}\int_0^{\infty} \frac{db\; b}{(2\pi)} J_0(p_\perp b/z)\, \hat C_{i\gets q}^{D_1}\otimes D_{h/i}(z, \mu_b) \, e^{-\frac{1}{2} S_{\rm pert}(Q, b_*)  - S_{\rm NP}^{D_1}(Q, b)}, \label{eq:D1_tmd}
\\
\frac{p_\perp}{M_h} H_{1\,h/q}^\perp(z, p_\perp^2; Q) & = \frac{1}{z^2}\int_0^{\infty} \frac{db\; b^2}{(4\pi z)} J_1(p_\perp b/z)\, \delta \hat C_{i\gets q}^{\rm collins} \otimes \hat H^{(3)}_{h/i}(z, \mu_b) \, e^{-\frac{1}{2} S_{\rm pert}(Q, b_*)  - S_{\rm NP}^{\rm collins}(Q, b)}\ ,
\eea
where the TMD evolution has been taken into account, and one-loop results
of $\widetilde{\cal F}_q$, $\widetilde{\cal D}_q$, $\widetilde{\cal H}_{1q}$, and $\widetilde{\cal H}_c$
equal to one in the Collins-11 scheme have been applied, and 
$C$-functions are given in Eqs.~(\ref{eq:cf},\ref{eq:cf1},\ref{eq:cd},\ref{eq:cd1},\ref{eq:ch1},\ref{eq:ch1perp}). Using relation to Trento conventions of Eq.~\eqref{eq:relation_to_trento} we
can write
\bea
\frac{p_\perp}{M_h} H_{1\,h/q}^\perp(z, p_\perp^2; Q )& = -\frac{1}{z^2}\int_0^{\infty} \frac{db\; b^2}{(2\pi)} J_1(p_\perp b/z)\, \delta \hat C_{i\gets q}^{\rm collins} \otimes \hat H_{1\, h/j}^{\perp (1)}(z)|_{\rm Trento}(z, \mu_b) \, e^{-\frac{1}{2} S_{\rm pert}(Q, b_*)  - S_{\rm NP}^{\rm collins}(Q, b)}\ , \\
\frac{p_\perp}{z M_h} H_{1\,h/q}^\perp(z, p_\perp^2; Q)|_{\rm Trento} & = \frac{1}{z^2}\int_0^{\infty} \frac{db\; b^2}{(2\pi)} J_1(p_\perp b/z)\, \delta \hat C_{i\gets q}^{\rm collins} \otimes \hat H_{1\, h/j}^{\perp (1)}(z)|_{\rm Trento}(z, \mu_b) \, e^{-\frac{1}{2} S_{\rm pert}(Q, b_*)  - S_{\rm NP}^{\rm collins}(Q, b)}\ .
\label{eq:collins_tmd}
\eea

We can also write explicitly the non-perturbative Sudakov form factor $S_{\rm NP}(Q, b)$ for all the TMDs discussed in our paper:
\bea
S_{\rm NP}^{f_1}(Q, b) &= S_{\rm NP}^{h_1}(Q, b) = \frac{g_2}{2} \ln\left(\frac{b}{b_*}\right)\ln\left(\frac{Q}{Q_0}\right)+{g_q}\, b^2\; ,
\\
S_{\rm NP}^{D_1}(Q, b) &=  \frac{g_2}{2} \ln\left(\frac{b}{b_*}\right)\ln\left(\frac{Q}{Q_0}\right) + \frac{g_h}{z^2}\, b^2 \; ,
\\
S_{\rm NP}^{\rm collins}(Q, b) &=  \frac{g_2}{2} \ln\left(\frac{b}{b_*}\right)\ln\left(\frac{Q}{Q_0}\right) + \frac{g_h - g_c}{z^2}\, b^2 \; ,
\eea
where we have assumed that the non-perturbative Sudakov form factors are the same for $f_1^q$ and $h_1^q$ as a first study following~\cite{Anselmino:2013vqa}. With the expressions for individual TMDs given in Eqs.~(\ref{eq:f1_tmd}, \ref{eq:h1_tmd}, \ref{eq:D1_tmd}, \ref{eq:collins_tmd}), and the fitted parameters in this section, we are now ready to present all these TMDs as a function of both longitudinal momentum fraction ($x$ or $z$) and the transverse component ($k_\perp$ or $p_\perp$). 

\begin{figure}[tbp]
\centering
\includegraphics[width=8cm]{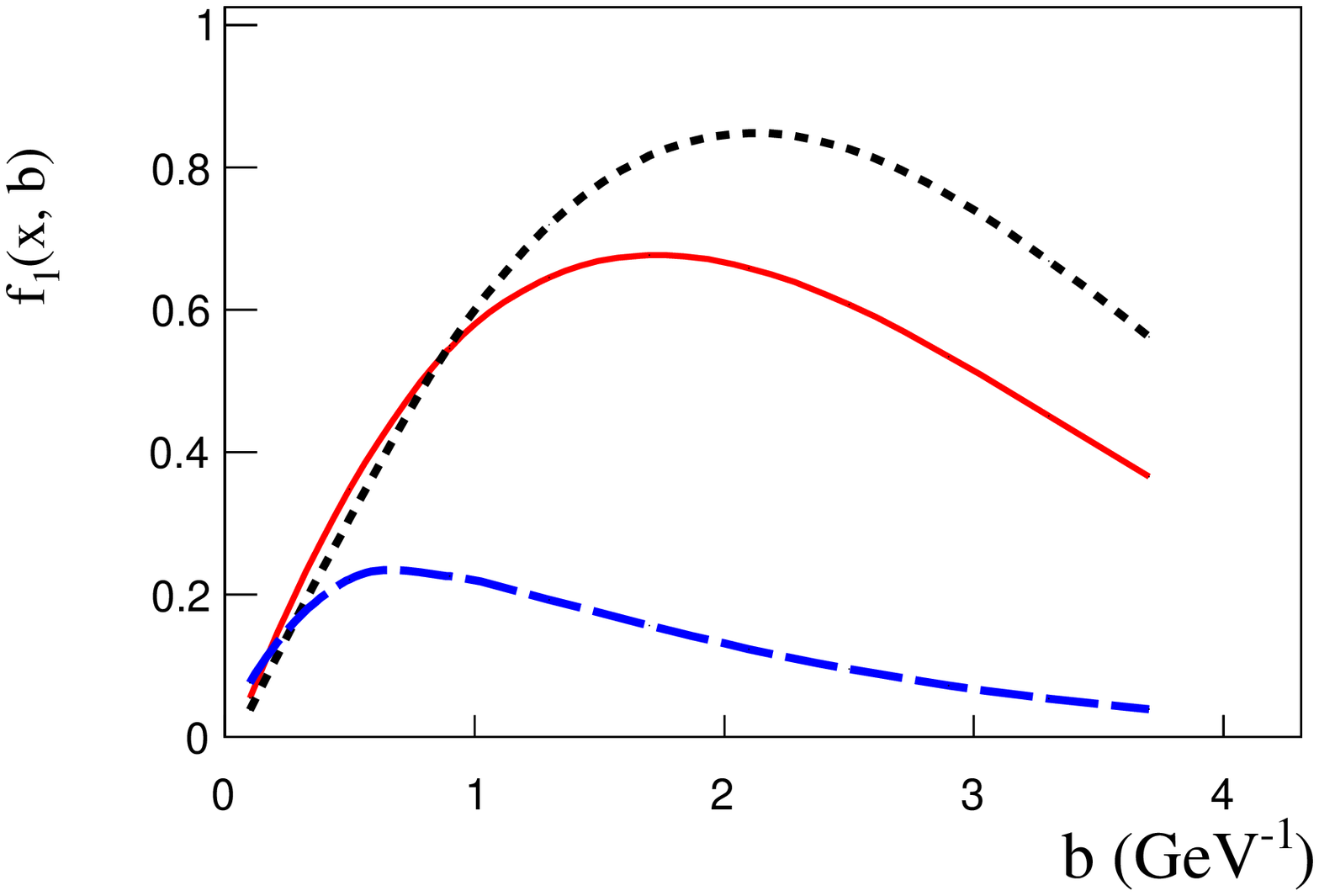}(a)
\includegraphics[width=8cm]{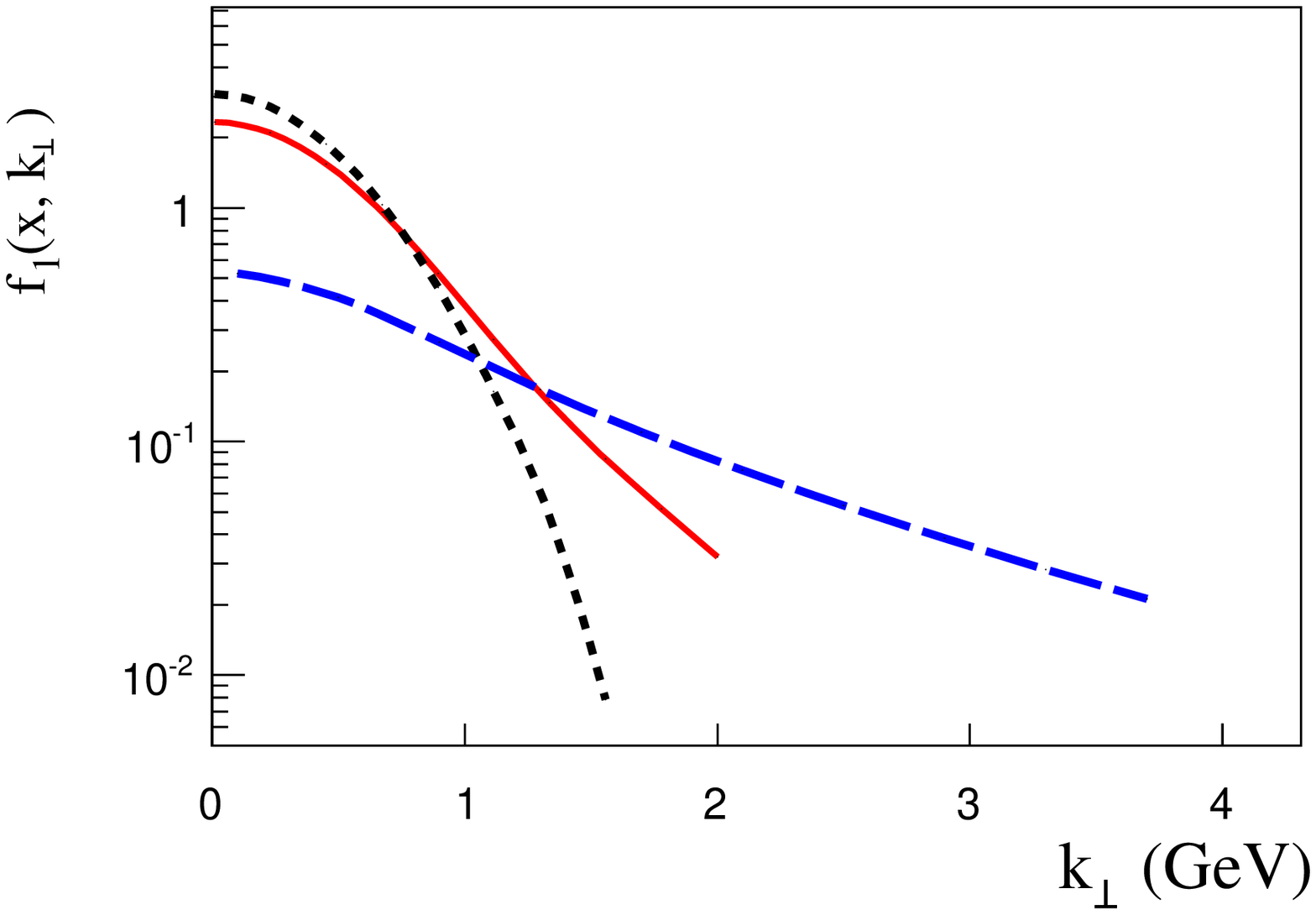}(b)
\caption{Unpolarised $u$-quark distribution   as function of $b$ (a) and as function of $k_\perp$ (b) at three different scales  $Q^2 = 2.4$ (dotted lines), $Q^2 = 10$ (solid lines) and $Q^2 = 1000$ (dashed lines) GeV$^2$.}
\label{fig:unpolarised_f1}
\end{figure}

In Fig.~\ref{fig:unpolarised_f1} we present  unpolarized $u$-quark distribution $f_1$ at $x=0.1$ as a function of $b$ (left) and $k_\perp$ (right).
We plot
\bea
f_1^q(x, b; Q) &\equiv   \frac{b}{(2\pi)} \, C_{q\gets i}^{f_1}\otimes f_1^i(x, \mu_b)\, e^{-\frac{1}{2} S_{\rm pert}(Q, b_*)  - S_{\rm NP}^{f_1}(Q, b)}, 
\eea
while $k_\perp$-dependence is defined in Eq.~\eqref{eq:f1_tmd}.
The distribution is calculated at three different scales: $Q^2 = 2.4$ (dotted lines), $Q^2 = 10$ (solid lines) and $Q^2 = 1000$ (dashed lines) GeV$^2$. As one can see, at large scale  $Q^2 = 1000$ GeV$^2$, the distribution is highly dominated by perturbative region of $b< b_{max}$ while at lower scales $Q^2 = 2.4$  and $10$ GeV$^2$ the distribution is shifted towards large values of $b\sim 2\div 3$  GeV$^{-1}$, in  this region of $b$ one needs to carefully account for non-perturbative effects of TMD evolution and intrinsic motion of quarks. The distribution in $k_\perp$ space is becoming wider with growth of $Q^2$ and has developed perturbative tail, while at low values of $Q^2$ it resembles gaussian-type parametrization used in tree level extractions, for instance that of Refs.~\cite{Anselmino:2011ch,Anselmino:2013vqa}.

\begin{figure}[tbp]
\centering
\includegraphics[width=8cm]{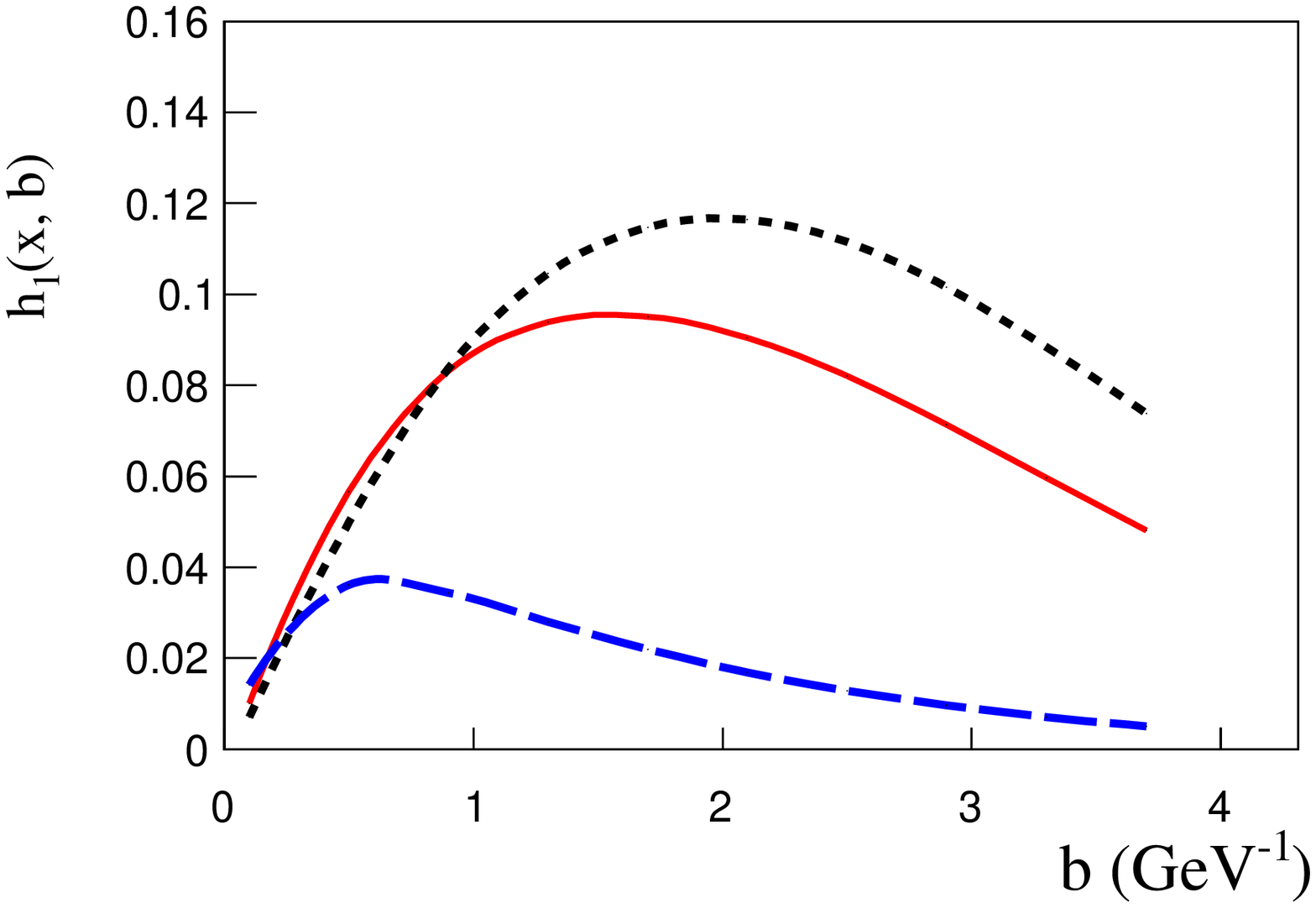}(a)
\includegraphics[width=8cm]{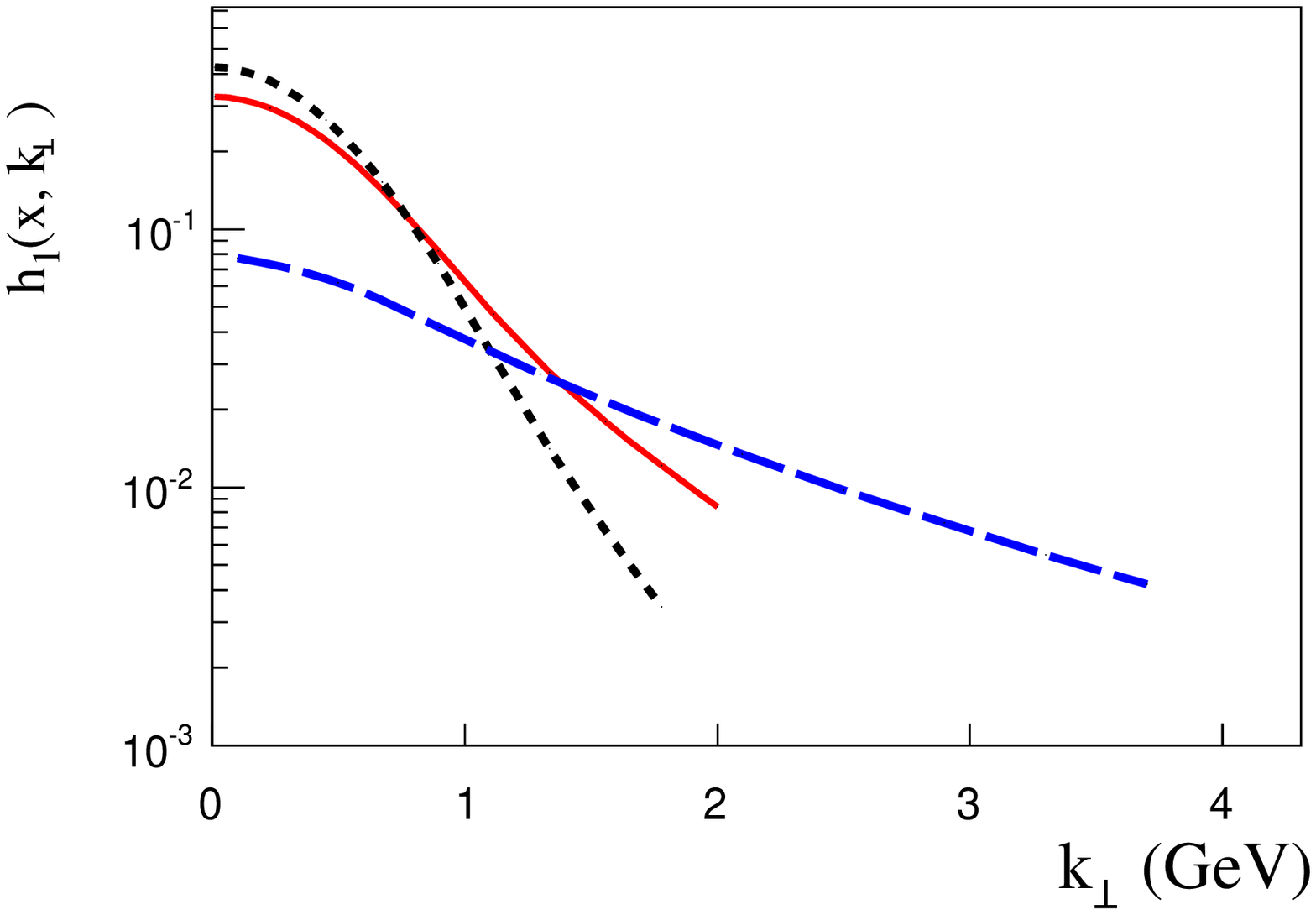}(b)
\caption{Transversity $u$-quark distribution   as function of $b$ (a) and as function of $k_\perp$ (b) at three different scales  $Q^2 = 2.4$ (dotted lines), $Q^2 = 10$ (solid lines) and $Q^2 = 1000$ (dashed lines) GeV$^2$.}
\label{fig:transversity_h1}
\end{figure}

The same observation is true for transversity distribution. We present  transversity  $u$-quark distribution $h_1$ at $x=0.1$ as function of $b$ and  $k_\perp$ in Fig.~\ref{fig:transversity_h1}. 
We plot
\bea
h_1^q(x, b; Q) &\equiv   \frac{b}{(2\pi)} \,\delta C_{q\gets i}\otimes h_1^i(x, \mu_b)\, e^{-\frac{1}{2} S_{\rm pert}(Q, b_*)  - S_{\rm NP}^{h_1}(Q, b)}, 
\eea
while $k_\perp$ distribution is defined in Eq.~\eqref{eq:h1_tmd}. Note that coefficient functions for transversity distribution $\delta C_{q\gets i}$ are different from those of unpolarized distribution. This difference affects the shape of distributions in $b$ and $k_\perp$ space. Moreover the width of transversity can be different from that of unpolarized distribution as well, however   features of TMD evolution are very similar in both cases. Generic results on transversity TMD evolution were also presented in Ref.~\cite{Bacchetta:2013pqa}.

\begin{figure}[tbp]
\centering
\includegraphics[width=8cm]{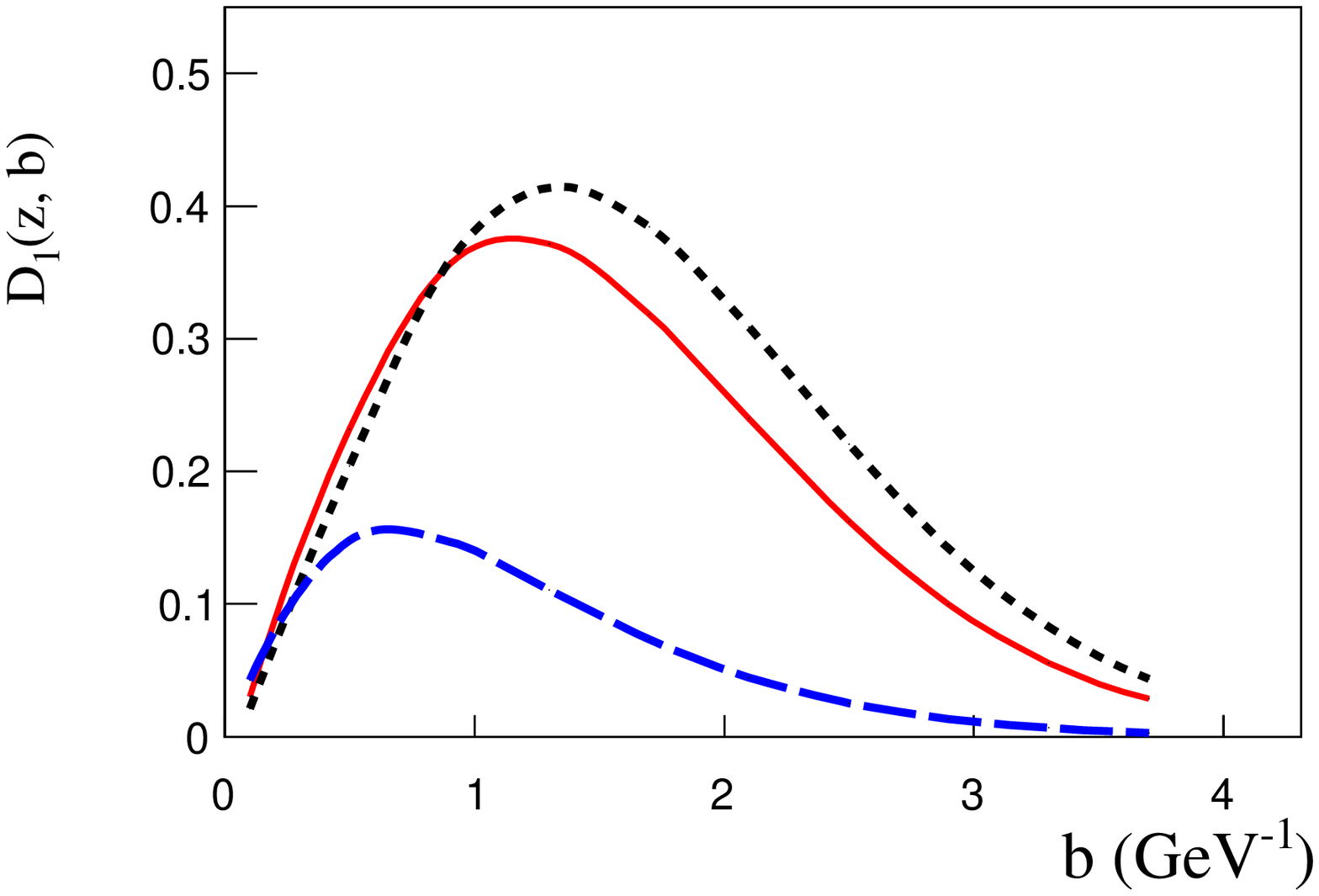}(a)
\includegraphics[width=8cm]{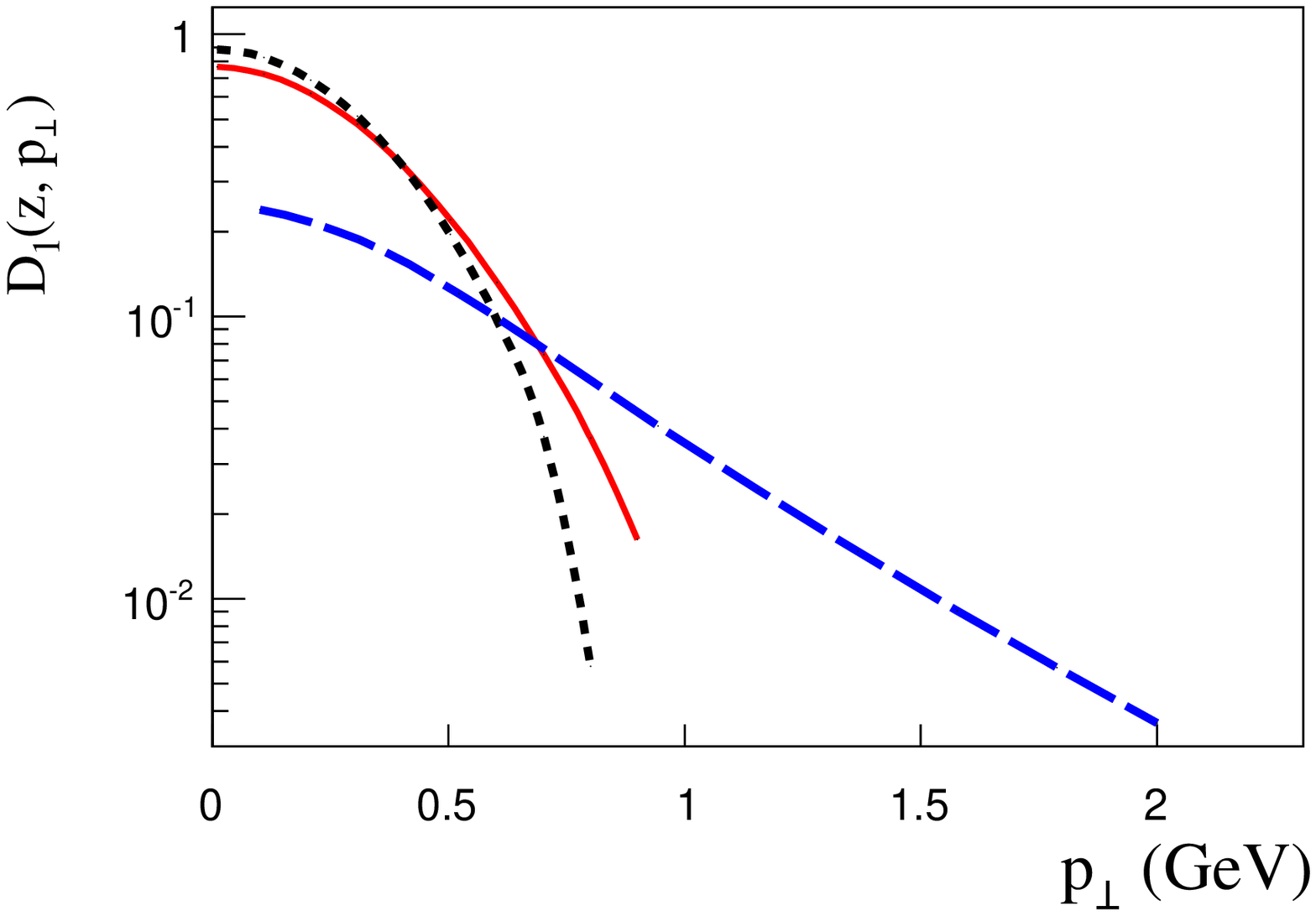}(b)
\caption{Unpolarised FF $u\to\pi^+$    as function of $b$ (a) and as function of $p_\perp$ (b) at three different scales  $Q^2 = 2.4$ (dotted lines), $Q^2 = 10$ (solid lines) and $Q^2 = 1000$ (dashed lines) GeV$^2$.}
\label{fig:unpolarised_D1}
\end{figure}

Unpolarized fragmentation TMD as function of $b$ is defined as
\bea
D_{h/q}(z, b; Q) &\equiv \frac{1}{z^2}  \frac{ b}{(2\pi)}  \, \hat C_{i\gets q}^{D_1}\otimes D_{h/i}(z, \mu_b) \, e^{-\frac{1}{2} S_{\rm pert}(Q, b_*)  - S_{\rm NP}^{D_1}(Q, b)}, 
\eea
and as function of $p_\perp$ it can be calculated using Eq.~\eqref{eq:D1_tmd}. In Fig.~\ref{fig:unpolarised_D1} we present unpolarized TMD FF at $z=0.4$ and at three different scales $Q^2 = 2.4$ (dotted lines), $Q^2 = 10$ (solid lines) and $Q^2 = 1000$ (dashed lines) GeV$^2$. Again as in case of other TMDs above one observes widening of distributions in $p_\perp$ and shift towards lower values $b$ of the maximum of the distribution with increase of $Q^2$ . In relatively low $Q^2$ region the effects of TMD evolution are quite moderate.

\begin{figure}[tbp]
\centering
\includegraphics[width=8cm]{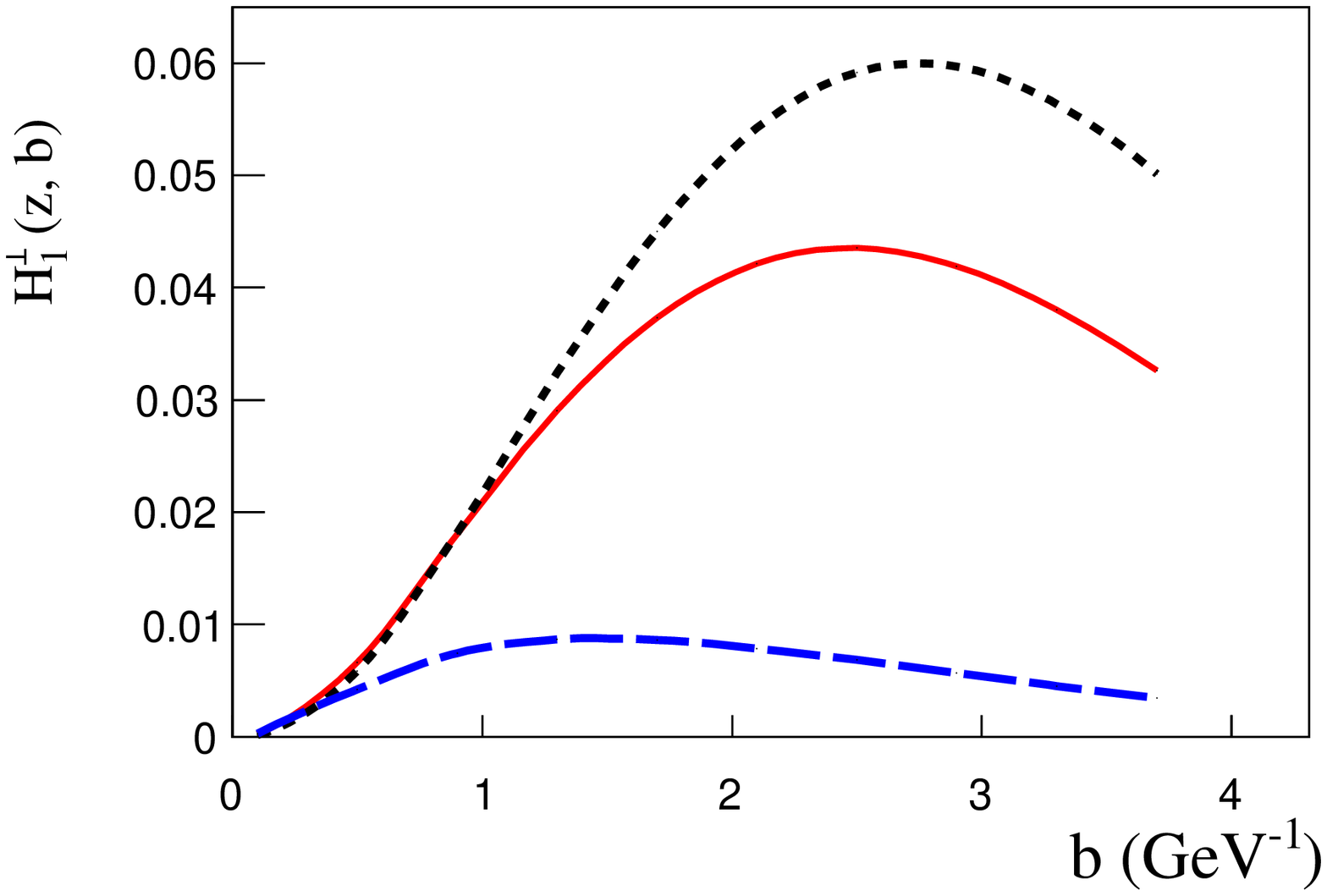}(a)
\includegraphics[width=8cm]{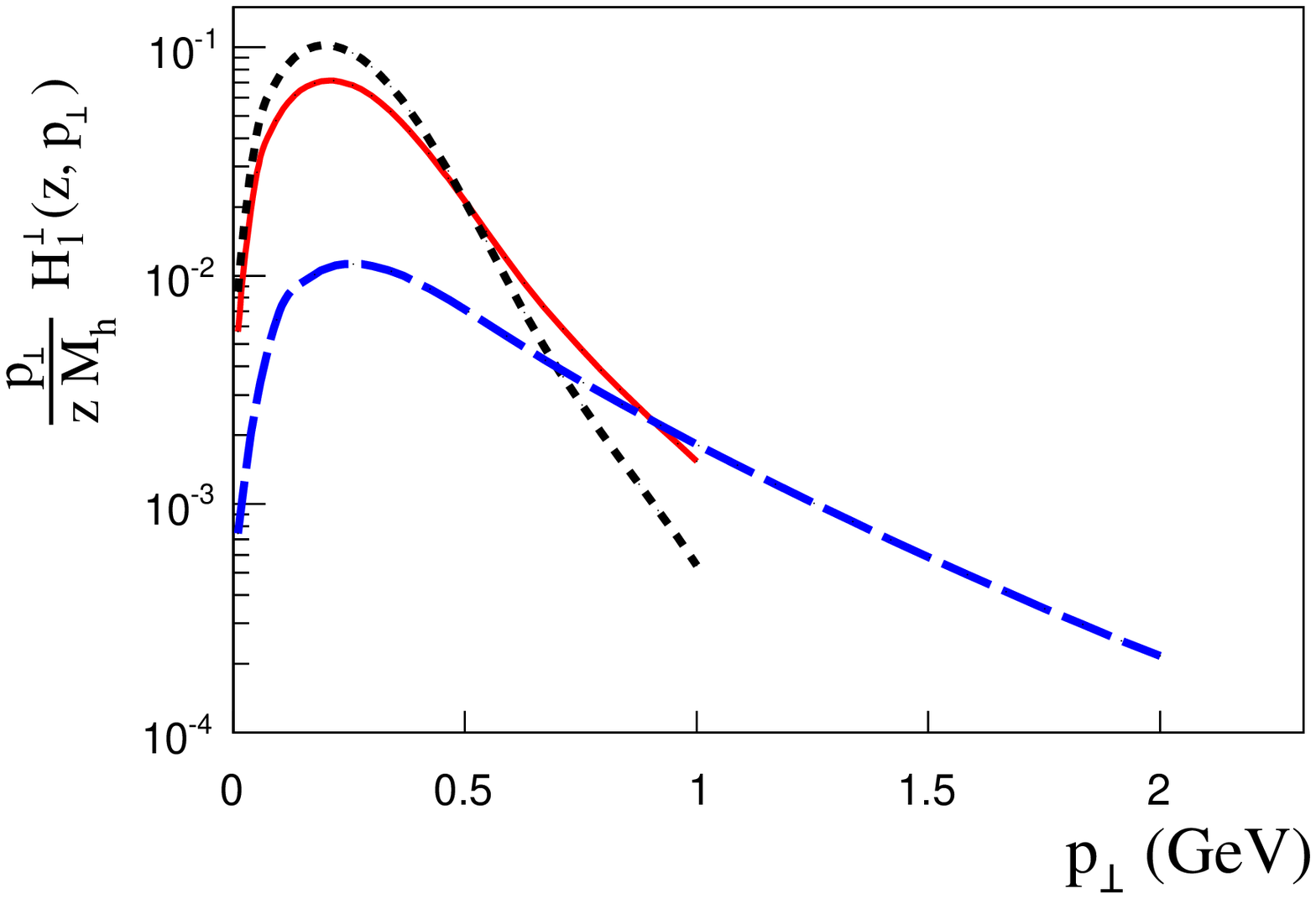}(b)
\caption{Collins FF $u\to\pi^+$    as function of $b$ (a) and as function of $p_\perp$ (b) at three different scales  $Q^2 = 2.4$ (dotted lines), $Q^2 = 10$ (solid lines) and $Q^2 = 1000$ (dashed lines) GeV$^2$.}
\label{fig:Collins_H1}
\end{figure}

Collins fragmentation function with evolution is presented for the first time in this paper. The $b$ dependent function can be defined as
\bea
 H_{1\,h/q}^\perp(z, b; Q)|_{\rm Trento} & \equiv \frac{1}{z^2}  \frac{  b^2}{(2\pi)}  \, \delta \hat C_{i\gets q}^{\rm collins} \otimes \hat H_{1\, h/j}^{\perp (1)}(z)|_{\rm Trento}(z, \mu_b) \, e^{-\frac{1}{2} S_{\rm pert}(Q, b_*)  - S_{\rm NP}^{\rm collins}(Q, b)}\ ,
\eea
and $p_\perp$ dependent function is in Eq.~\eqref{eq:collins_tmd}. In Fig.~\ref{fig:Collins_H1} we present TMD Collins FF at $z=0.4$ and at three different scales $Q^2 = 2.4$ (dotted lines), $Q^2 = 10$ (solid lines) and $Q^2 = 1000$ (dashed lines) GeV$^2$.  One observes widening of distributions in $p_\perp$ and shift towards lower values $b$ of the maximum of the distribution with increase of $Q^2$ . Note that TMD Collins FF has a kinematical zero due to the pre-factor ${p_\perp}/{z M_h}$.

%
%
\subsection{Description of the experimental data}
\begin{figure}[htbp]
\centering
\includegraphics[width=9cm]{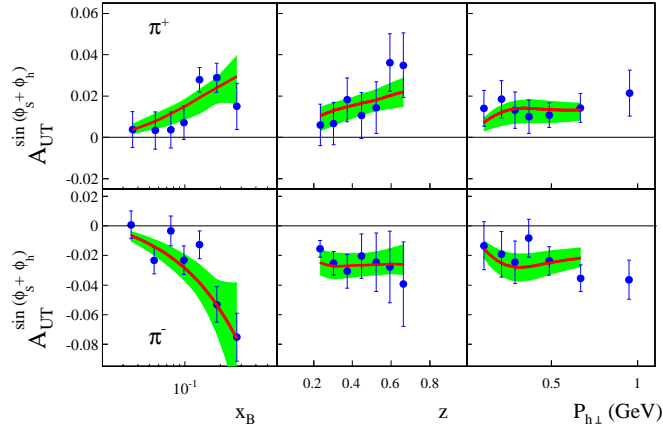}  
\caption{Description of Collins asymmetries measured by the HERMES Collaboration
 \cite{Airapetian:2009ae} as a function of $x_B$, $z$, $P_{h\perp}$ on  the proton target. The shaded region corresponds to our estimate of 90\% C.L. error band.}
\label{fig:hermes}
\end{figure}

The description of the HERMES data \cite{Airapetian:2009ae} is shown in Fig.~\ref{fig:hermes}. One can see 
that the description is good for all $x_B$, $z$, and $P_{h\perp}$ dependencies. The formalism that we use is appropriate 
in the region of low $P_{h\perp}$ and we limit our description by $P_{h\perp} < 0.8$ GeV. The data is in the region of $1 \lesssim \langle Q^2 \rangle \lesssim  6$ (GeV$^2$). The estimate of the error band is presented as the shaded region.

One can see from Fig.~\ref{fig:hermes} that both data and the model obey kinematical suppression of asymmetries at low $z_h$, and $P_{h\perp}$.  Additionally the data indicates that asymmetry
becomes smaller in the region of small-$x_B$ and thus transversity 
becomes small in the small-$x_B$ region as well as can be seen in Fig.~\ref{fig:functions} (a). Positive asymmetry of $\pi^+$ production implies that the product of $u$-quark transversity and the favored Collins fragmentation function is positive. We choose the solution with positive $u$-quark transversity (the same sign as $u$-quark helicity distribution) and obtain favored Collins fragmentation function is positive, see Fig.~\ref{fig:functions} (b).  Large negative asymmetry of $\pi^-$ production indicates that the so-called unfavored Collins fragmentation function is large and negative and indeed it is the case, see Fig.~\ref{fig:functions} (b). Measurements on proton targets are dominated by $u$-quark functions as far as 
$e_u^2/e_d^2 = 4$, thus we have better precision for the extraction of $u$-quark transversity and tensor charge $\delta u$.

\begin{figure}[tbp]
\centering
\includegraphics[width=8cm]{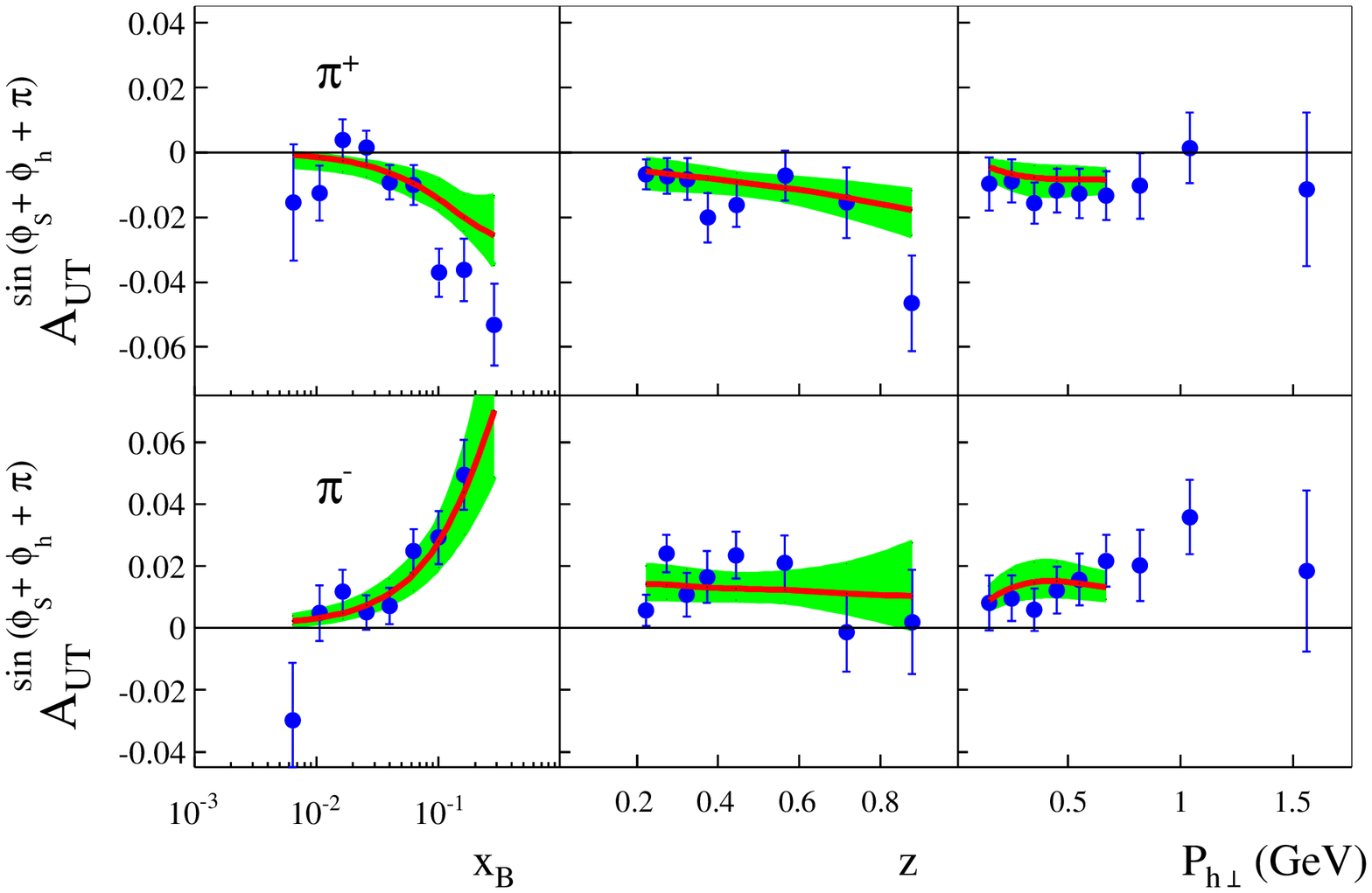} (a)
\includegraphics[width=8cm]{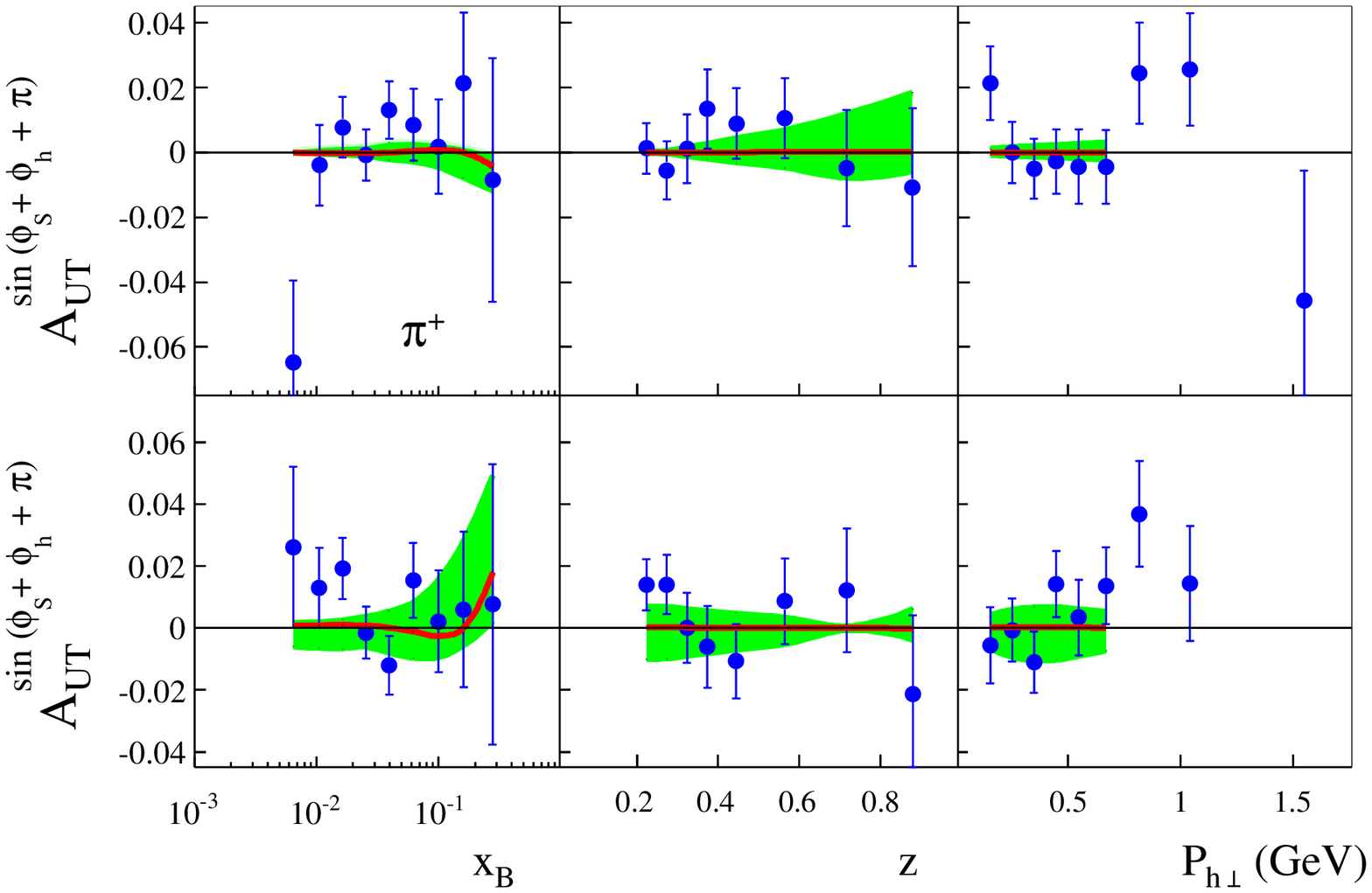}(b)
\caption{Description of Collins asymmetries measured by the COMPASS Collaboration
as a function of $x_B$, $z$, $P_{h\perp}$ on (a) NH$_3$ proton \cite{Adolph:2014zba} and (b) LiD deuterium \cite{Alekseev:2008aa} targets. The shaded region corresponds to our estimate of 90\% C.L. error band.}
\label{fig:compass}
\end{figure}

The COMPASS data \cite{Alekseev:2008aa,Adolph:2014zba} extend the region of resolution scale by a factor of three, $\langle Q^2 \rangle \lesssim  21$ (GeV$^2$). We present results of our description in Fig.~\ref{fig:compass}. Again we exclude the region
of $P_{h\perp} > 0.8$ GeV where relation $P_{h\perp}/\langle z \rangle < Q$ is not satisfied. The COMPASS data extends the region of $x_{B}$ up to 
$x_{B}\sim 10^{-2}$ and the measured asymmetry indicates that transversity is rather small in the small-$x$ region. Indeed the extracted transversity shown in Fig.~\ref{fig:functions} (a) becomes small in the small-$x$ region. The COMPASS data on effective deuterium target Fig.~\ref{fig:compass} (b) indicate that the sum of $u$-quark and $d$-quark transversities is small, and thus both functions are approximately of the same size, it can be seen in Fig.~\ref{fig:functions} (a). 

\begin{figure}[tbp]
\centering
\includegraphics[width=7cm]{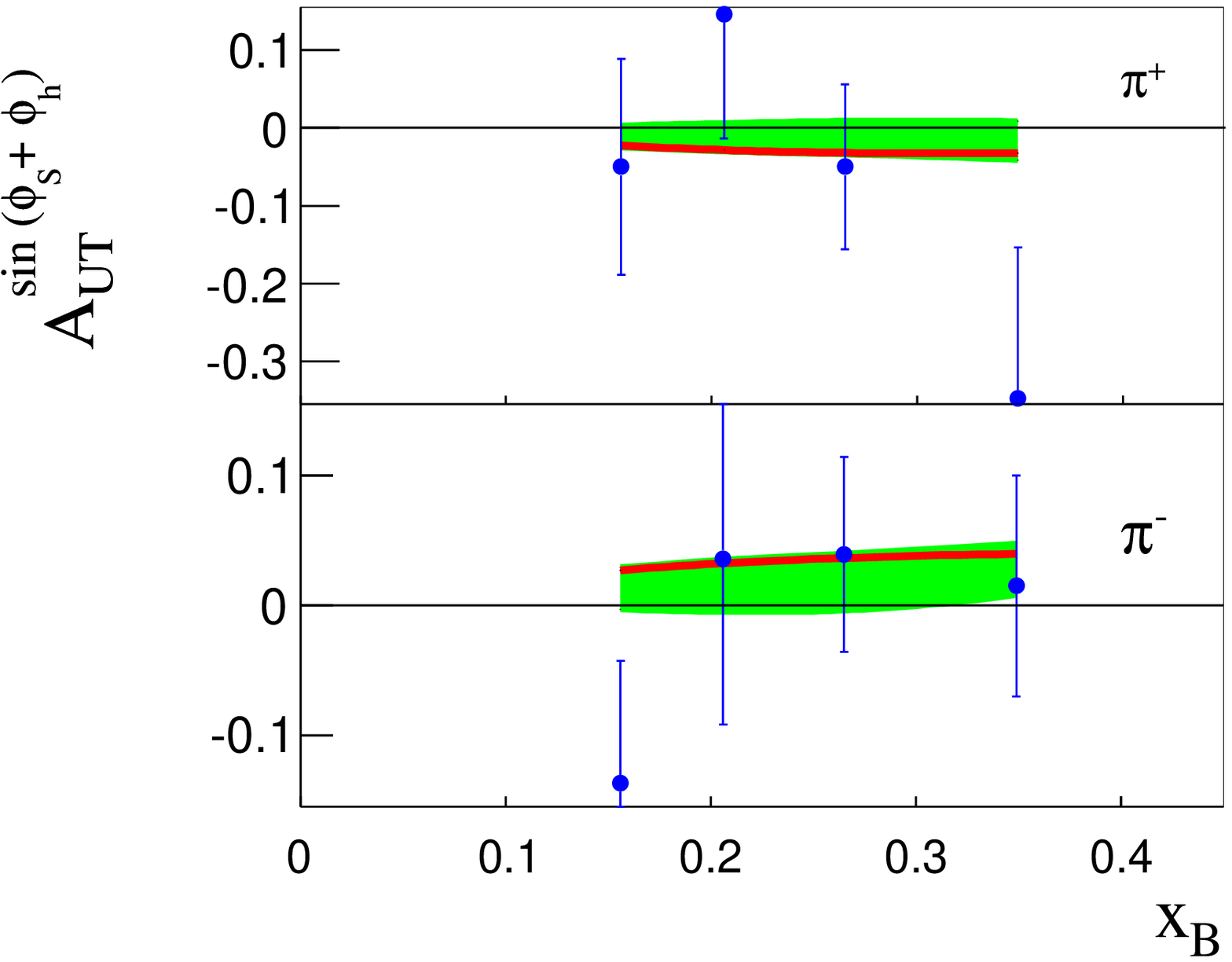}  
\caption{Description of Collins asymmetries measured by the JLab's HALL A \cite{Qian:2011py}
  as a function of $x_B$  on  the effective neutron target. The shaded region corresponds to our estimate of 90\% C.L. error band.}
\label{fig:jlab6}
\end{figure}

Description of JLab's HALL A data \cite{Qian:2011py} is shown in Fig.~\ref{fig:jlab6}. The data extend the region of $x_B$ toward large-$x$ and one can see that our fit is compatible with the data. The measurement on effective neutron target ($^3$He) is sensitive to $d$-quark functions, however the current experimental errors are too big to allow better extraction of $d$-quark transversity.

\begin{figure}[tbp]
\centering
\includegraphics[width=8cm]{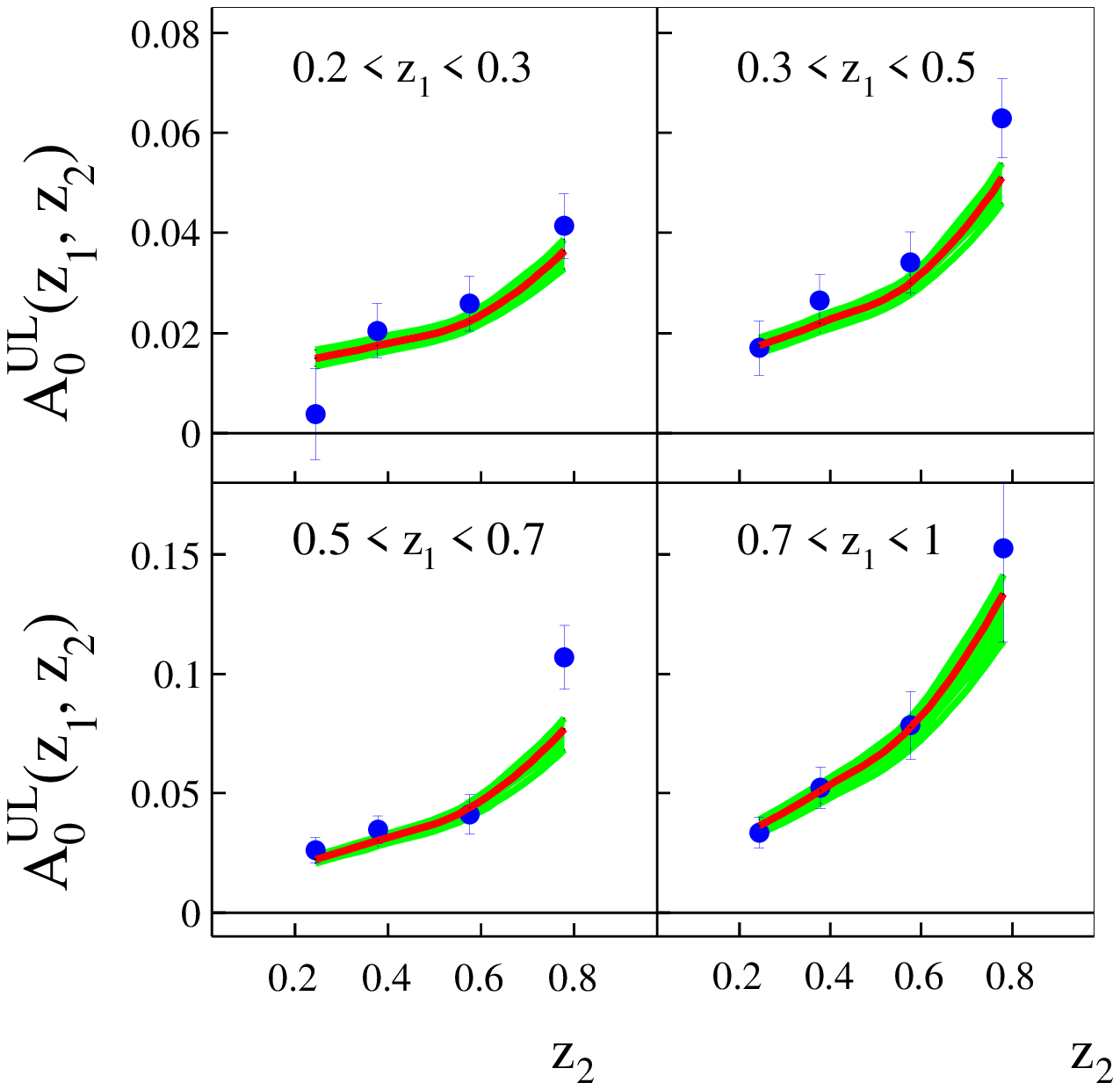}(a)
\includegraphics[width=8cm]{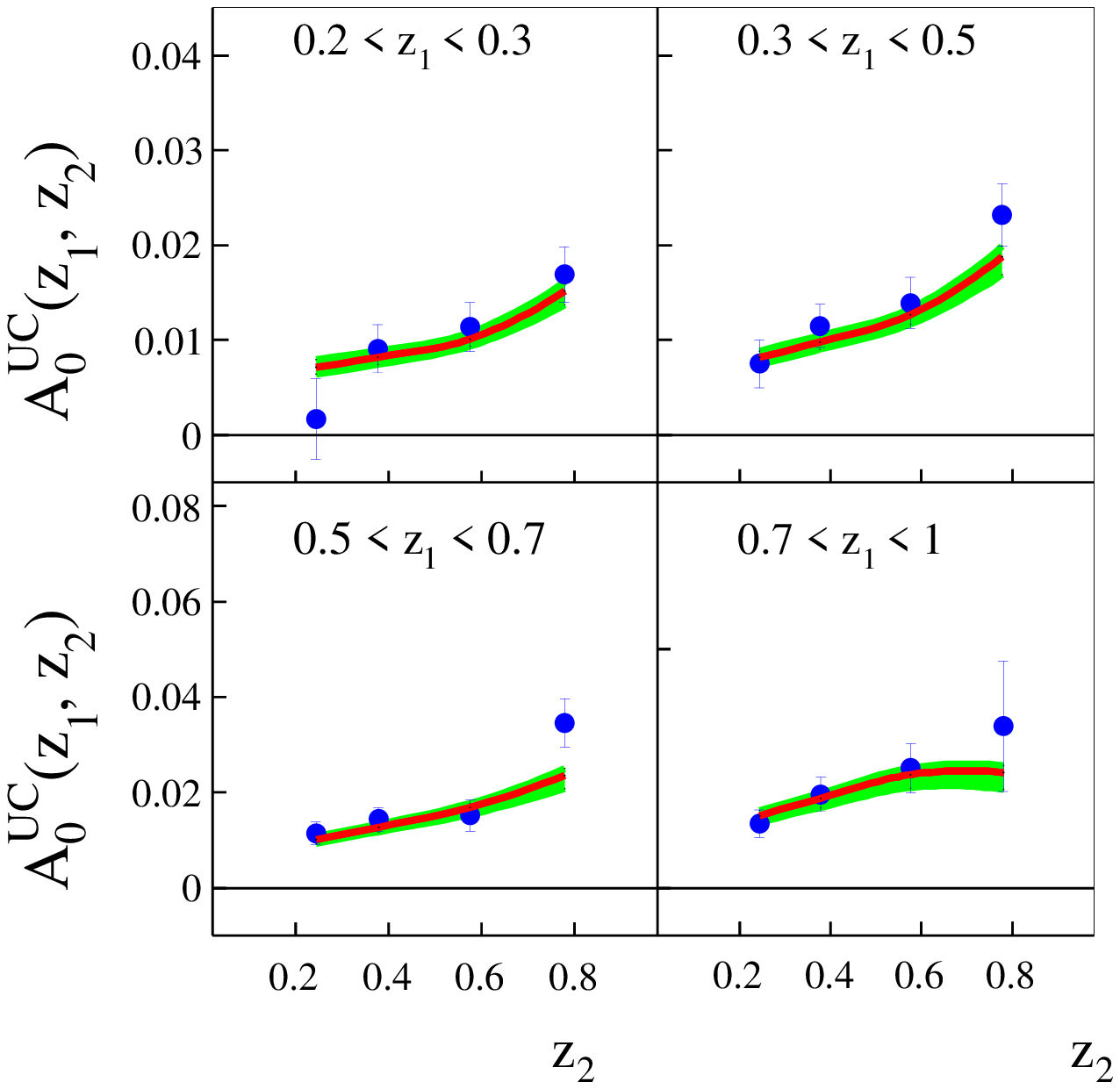}(b)
\caption{Collins asymmetries in $e^+e^-$ at $\sqrt{s}=10.6$ GeV measured by BELLE collaboration
\cite{Seidl:2008xc} as a function of $z_{h2}$ in different bins of $z_{h1}$ (a) UL and (b) UC. Calculations are performed with parameters from Table~\ref{parameters}. The shaded region corresponds to our estimate of 90\% C.L. error band.}
\label{fig:epem_belle}
\end{figure}

Both BELLE \cite{Seidl:2008xc} and {\em BABAR} \cite{TheBABAR:2013yha} collaborations measured the Collins asymmetries in $e^+e^-$ at $\sqrt{s}\simeq 10.6$ GeV.  Comparison of BELLE data \cite{Seidl:2008xc} on $A_0$ asymmetries for both $UL$ and 
$UC$ methods is presented in Fig.~\ref{fig:epem_belle}. The data are measured in four different bins of $z_{h1}, z_{h2}$
and one can see that the description of the data is very good. The asymmetry becomes small when  $z_{h1}, z_{h2}$ become small due to kinematical suppression and one can see from Fig.~\ref{fig:epem_belle} that our calculations are compatible with this behavior.

\begin{figure}[tbp]
\centering
\includegraphics[width=7cm]{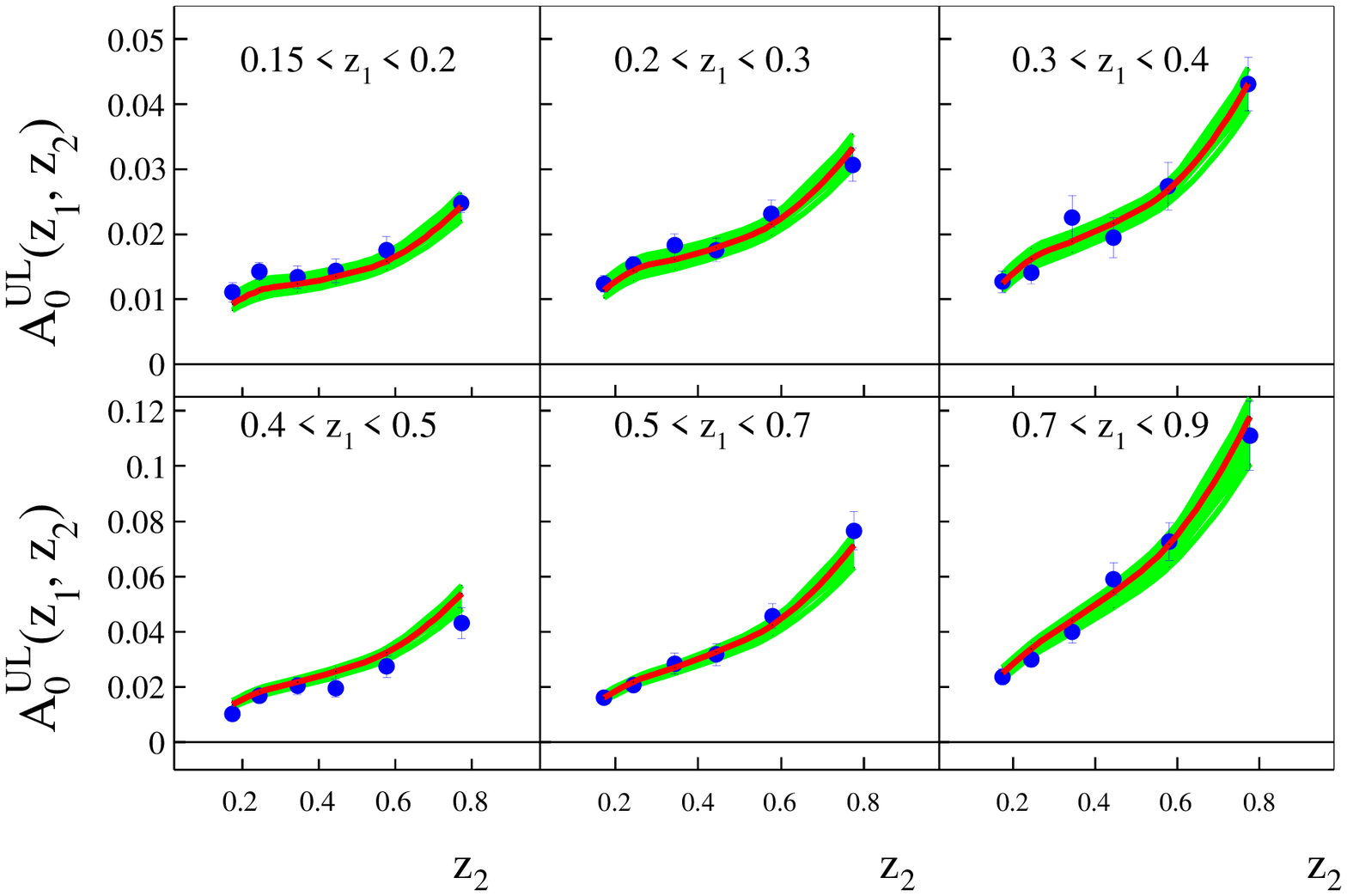}(a)
\includegraphics[width=7cm]{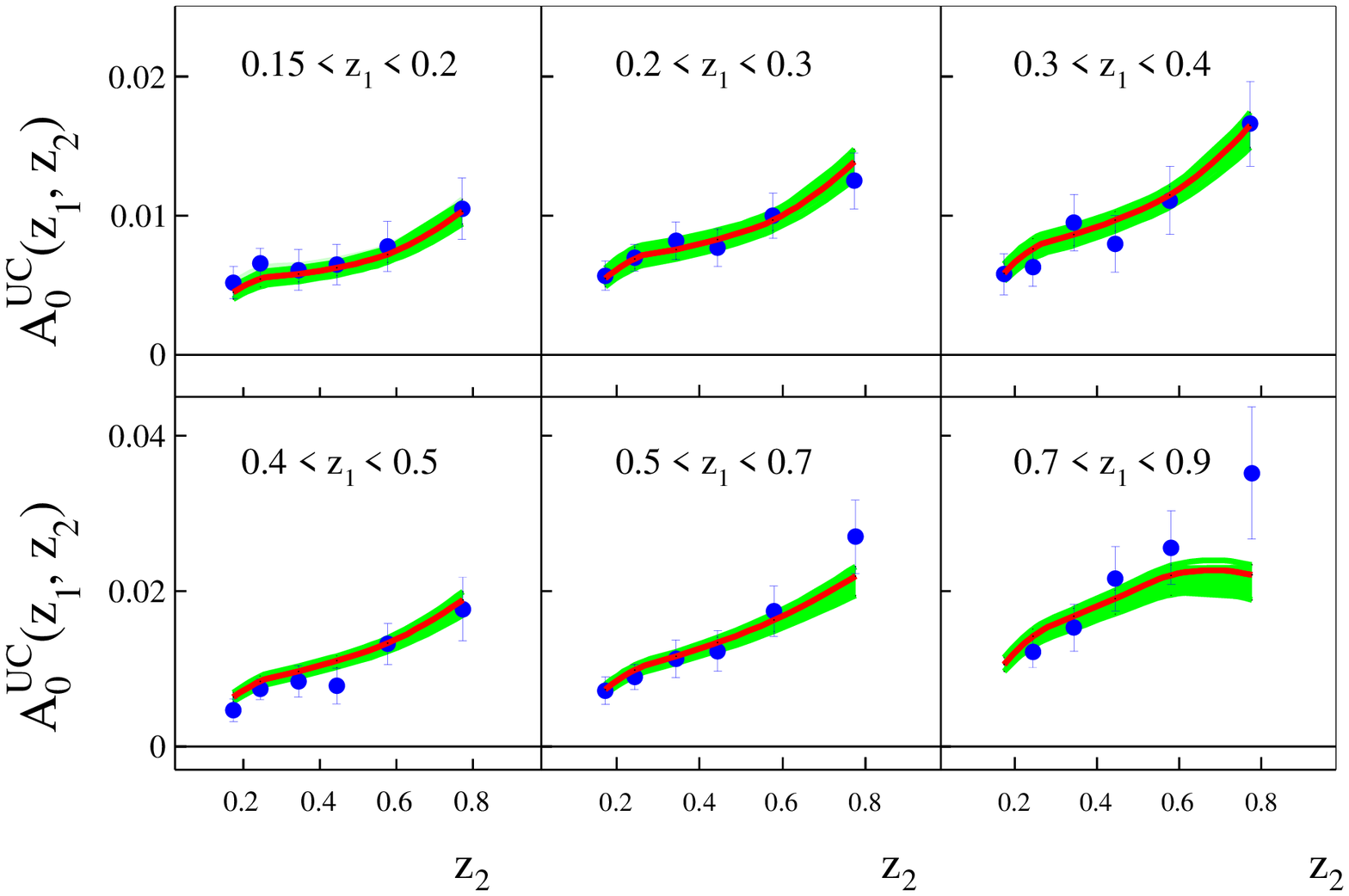}(b)
\caption{Collins asymmetries in $e^+e^-$ at $\sqrt{s}=10.6$ GeV measured by {\em BABAR} Collaboration
\cite{TheBABAR:2013yha}  as a function of $z_{h2}$ in different bins of $z_{h1}$ (a) UL and (b) UC. Calculations are performed with parameters from Table~\ref{parameters}. The shaded region corresponds to our estimate of 90\% C.L. error band.}
\label{fig:epem_babar}
\end{figure}

In Fig.~\ref{fig:epem_babar} we present description of {\em BABAR} data \cite{TheBABAR:2013yha} on $A_0$ asymmetries for both $UL$ and 
$UC$ methods. The data are in six bins of $z_{h1}, z_{h2}$ with six points in each bin. This allows for better extraction 
of the shape of Collins fragmentation functions. One can see that also in this case the description is very good. The large-$z$ region deserve a special comment. One expects that the formalism will become unreliable when $z_{h1}\to 1$ and/or $z_{h2}\to 1$ due to the influence of exclusive pion production. Indeed one can see from Figs.~\ref{fig:epem_belle} and \ref{fig:epem_babar} that in large-$z$ bins the quality of description deteriorates. Nevertheless both magnitude and the shape of the data are reproduced perfectly in the plot. It is achieved by allowing parameters that describe shape of favored and unfavored Collins fragmentation functions be different and independent of each other. Additionally the correct $Q^2$ evolution reproduces the shape much better compared to the case of absence of the evolution. Note that we have not attempted to fit the data without TMD evolution, thus our conclusion is valid only for 
comparison of results with and without evolution using parameters of NLL fit.

Even though a priori it is very difficult to expect perfect description of the data in the whole $z$ region, our fit indeed is capable of reproducing the data very well. Both $A_0^{UL}$ and $A_0^{UC}$ are described very well, we observe no tension between the measurements and it indicates the robustness of the method. $A_0^{UL}$ and $A_0^{UC}$ have slightly different sensitivity to 
different combinations of Collins fragmentation functions as can be seen from Eq.~\eqref{eq:fut_collins} and the usage of both measurements helps to constrain the functions better. We believe that favored Collins fragmentation functions are well determined and future experimental data could test our findings.

\begin{figure}[tbp]
\centering
\includegraphics[width=7cm]{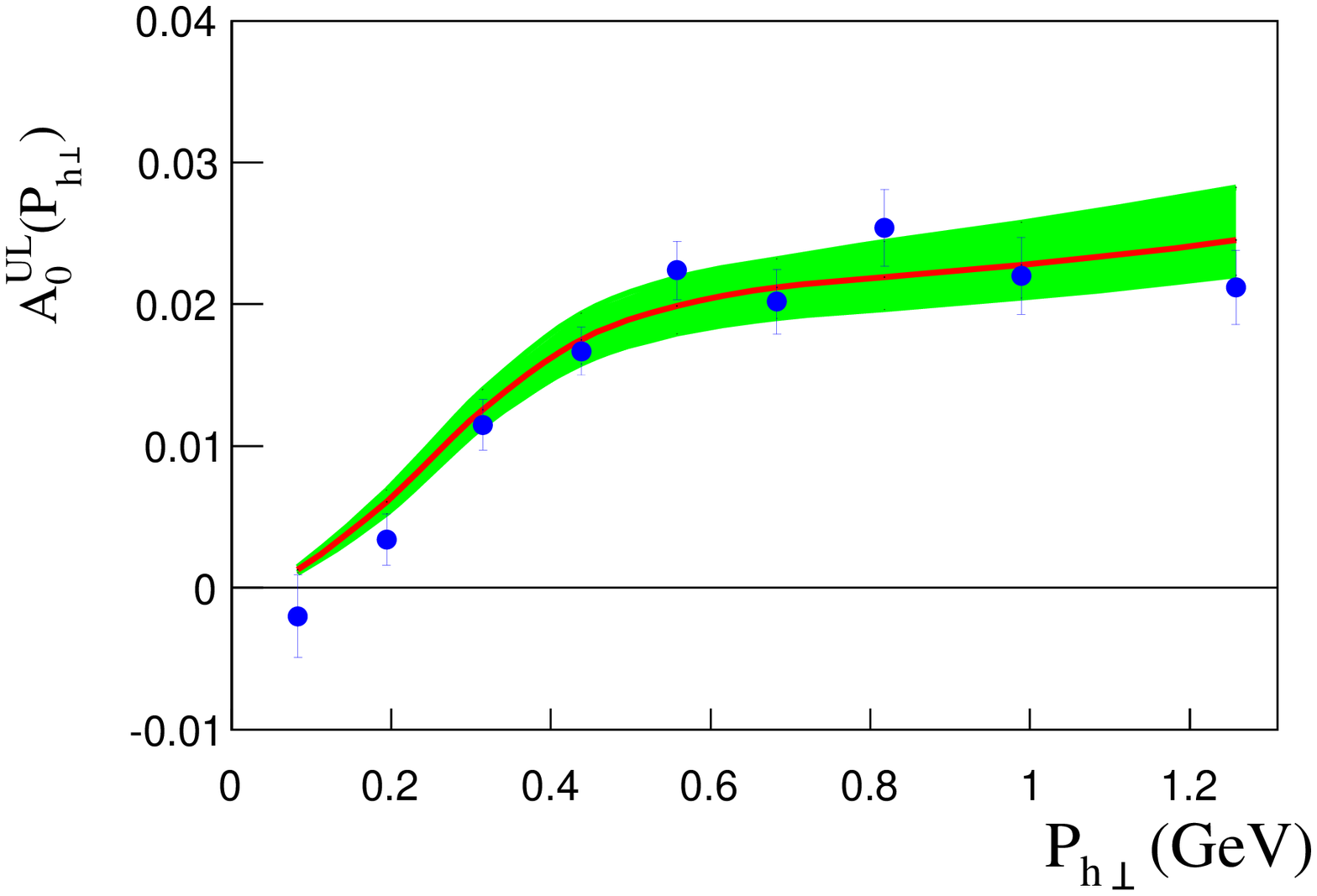}(a)
\includegraphics[width=7cm]{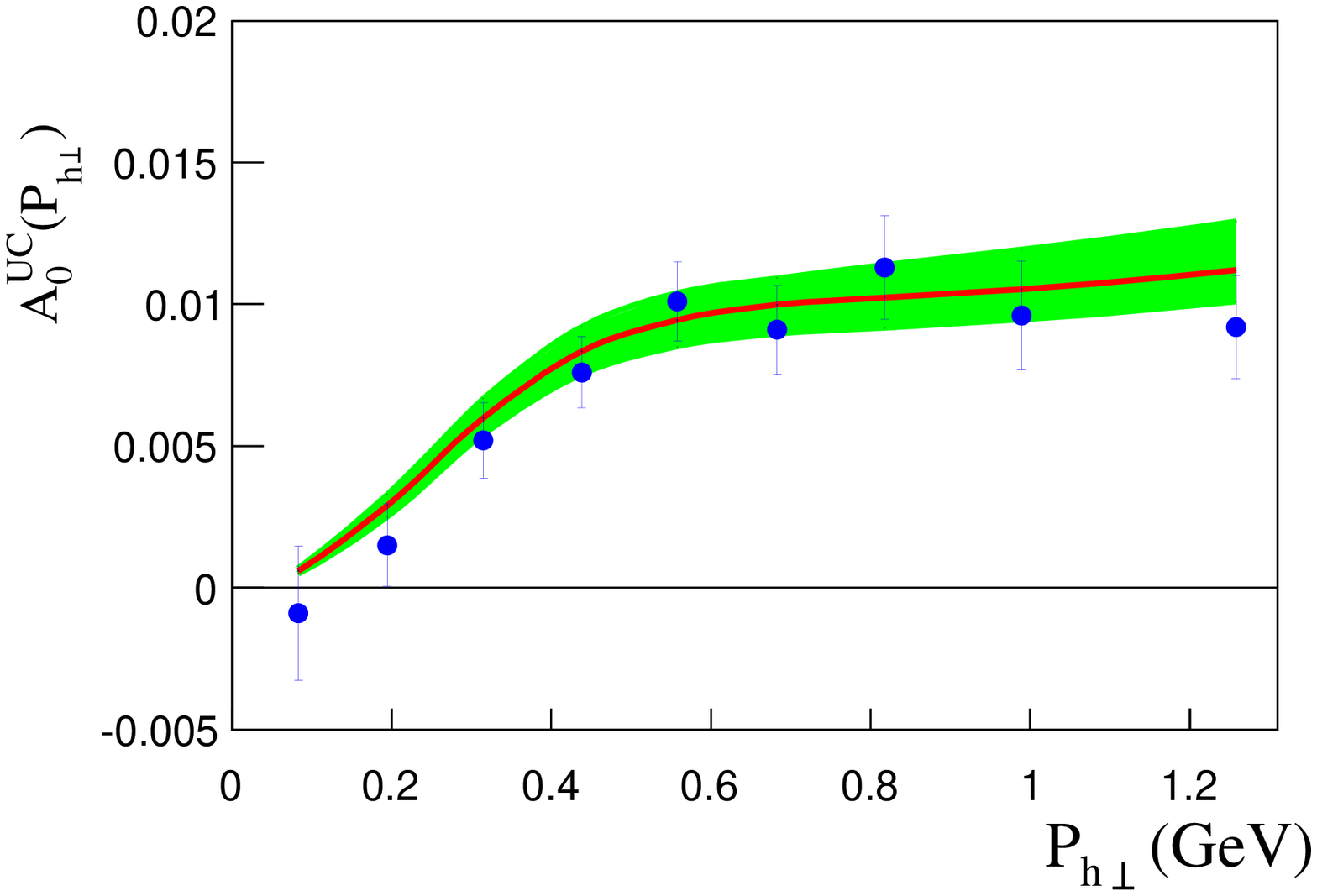}(b)
\caption{Collins asymmetries in $e^+e^-$ at $\sqrt{s}=10.6$ GeV measured by {\em BABAR} collaboration
\cite{TheBABAR:2013yha}  as a function of $P_{h\perp}$ (a) UL and (b) UC. Calculations are performed with parameters from Table~\ref{parameters}. The shaded region corresponds to our estimate of 90\% C.L. error band.}
\label{fig:epem_babar_pt}
\end{figure}

Finally we present comparison of our calculations with $P_{h\perp}$ dependence of $e^+e^-$ asymmetries in Fig.~\ref{fig:epem_babar_pt}.
Both $A_0^{UL}$ and $A_0^{UC}$ are described very well.

From the comparisons of the data and theoretical computations we can deduce that the TMD evolution at NLL$'$ can describe both $e^+e^-$ and SIDIS data adequately well. The highest resolution scale in our analysis is
quite big $Q^2 = s \simeq 110$ GeV$^2$ and we found that using appropriate QCD evolution was essential in order
to describe the data. It allows us to have a controlled theoretical precision of our computations. 
 Let us  study sensitivity of our results to the theoretical precision of computations. We will fix the parameters to the NLL$'$ fit results presented in Table.~\ref{parameters} and calculate asymmetries in different kinematical configurations using {\em tree level} approximation, i.e.  without TMD evolution,
Leading Logarithmic accuracy (LL), and Next-to-Leading Logarithmic accuracy  NLL$'$. As far as parameters are defined by fitting at NLL$'$ we expect that NLL$'$ will describe the data better than LL or tree approximation. We will not attempt to fit data at either tree approximation or LL, even though such fits can be well performed and may give reasonable descriptions of the data. By computing results with three different precisions with fixed parameters we will be able to answer two different questions: 
\begin{enumerate}
\item How big are effects of inclusion of higher orders in calculation of a particular asymmetry in a particular kinematical region?
\item How sensitive are experimental data to the inclusion of higher orders?
\end{enumerate}

\begin{figure}[tbp]
\centering
\includegraphics[width=5.5cm]{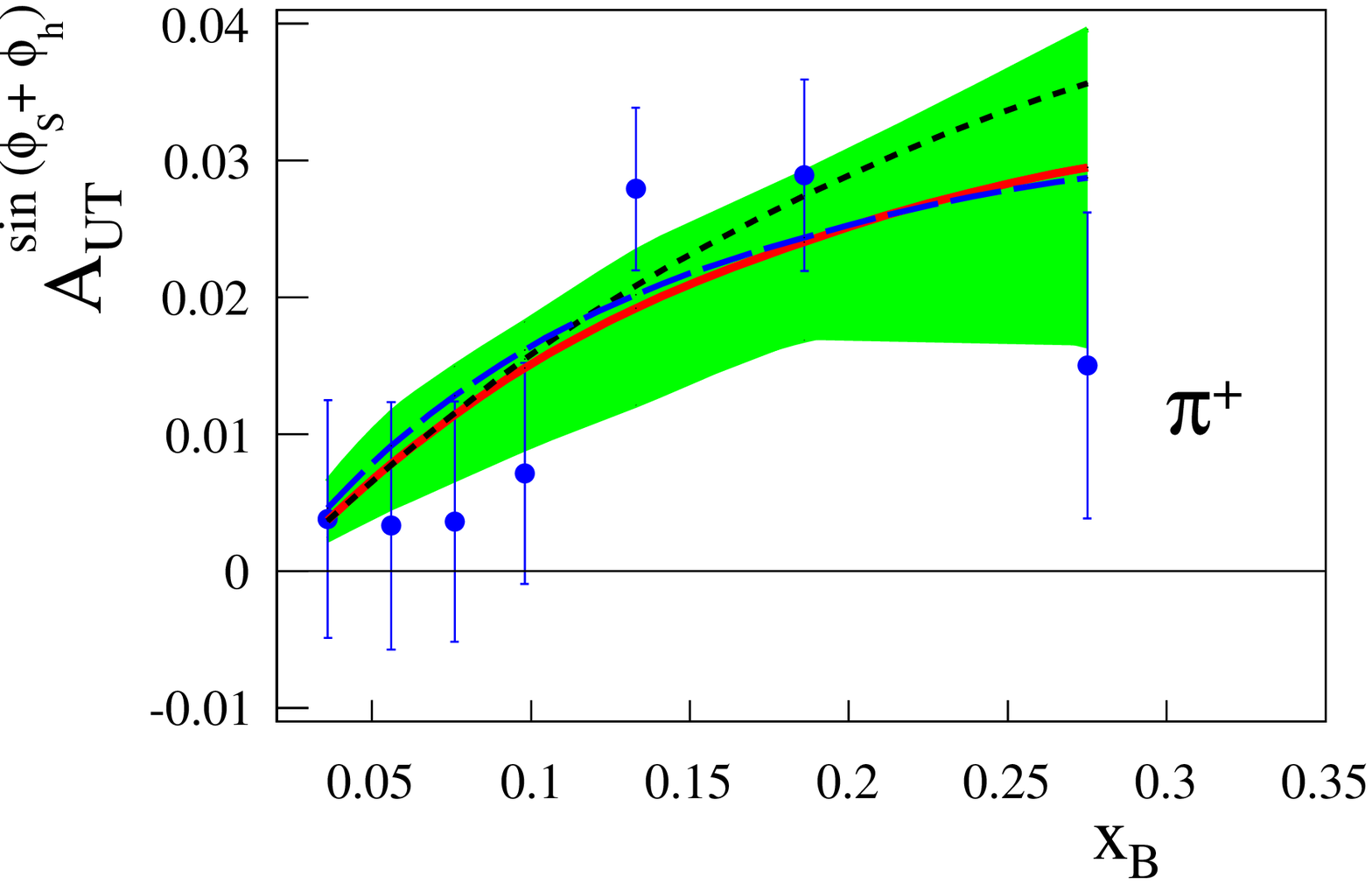}
\includegraphics[width=5.5cm]{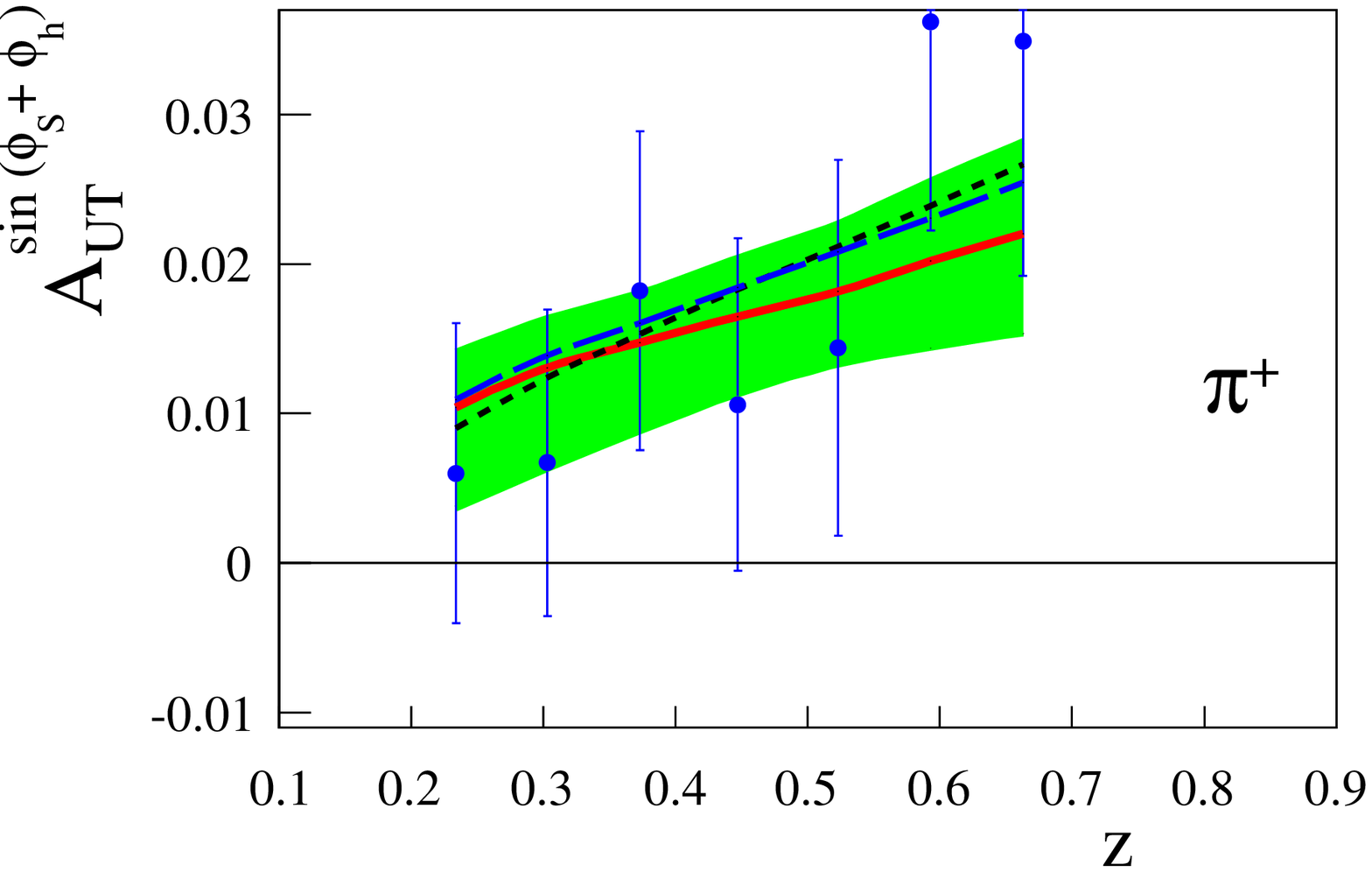}
\includegraphics[width=5.5cm]{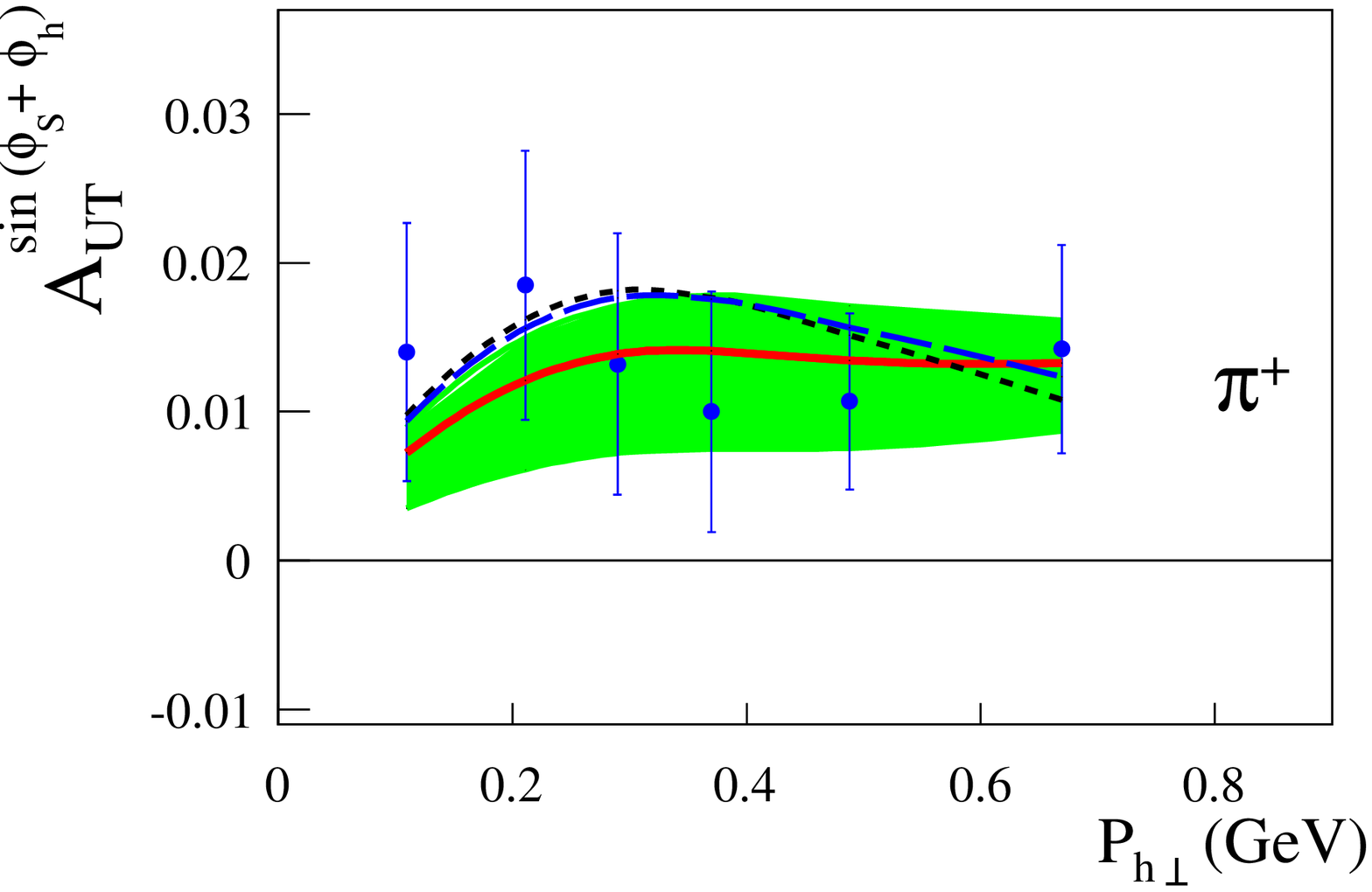}
\includegraphics[width=5.5cm]{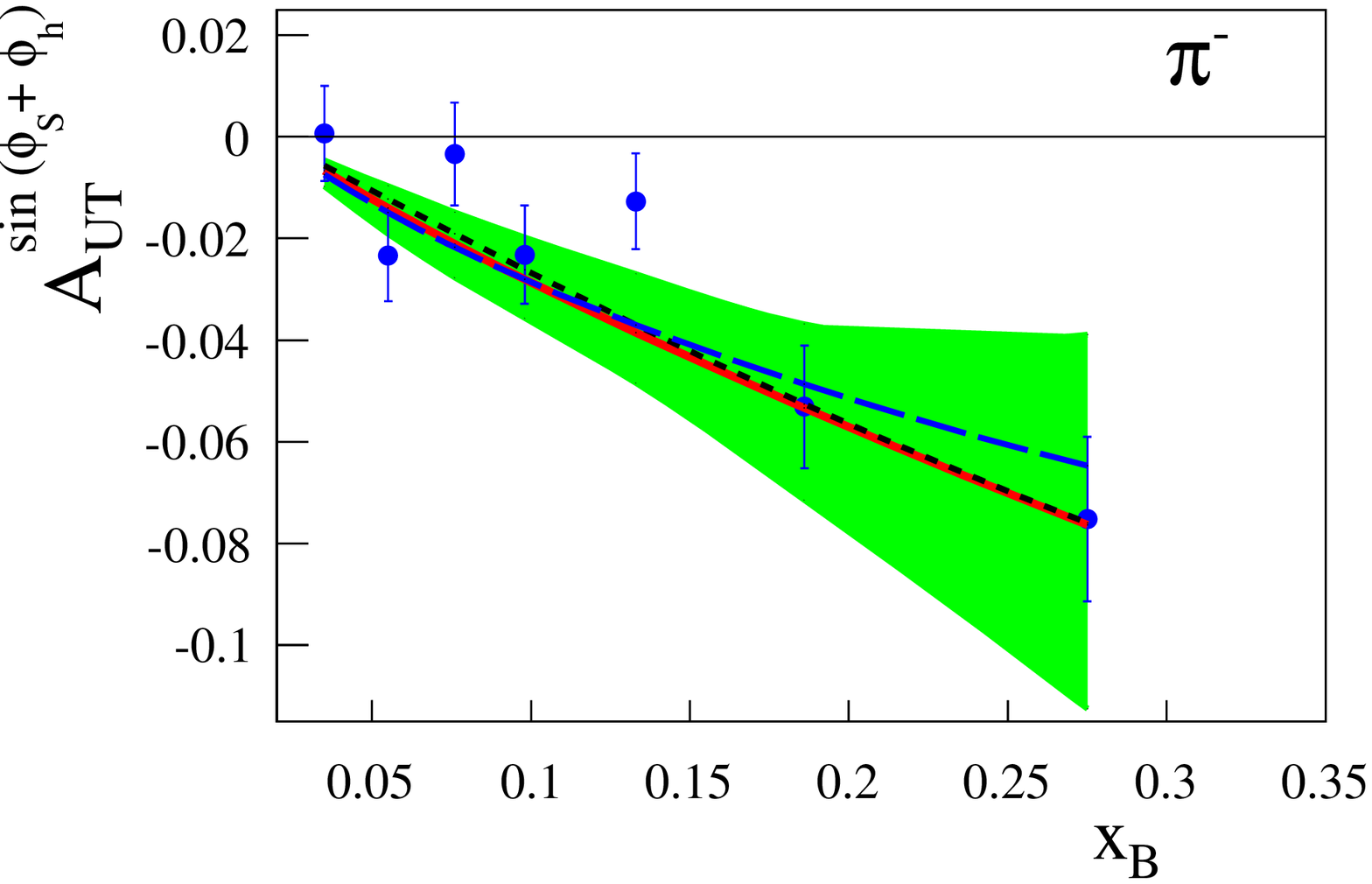}
\includegraphics[width=5.5cm]{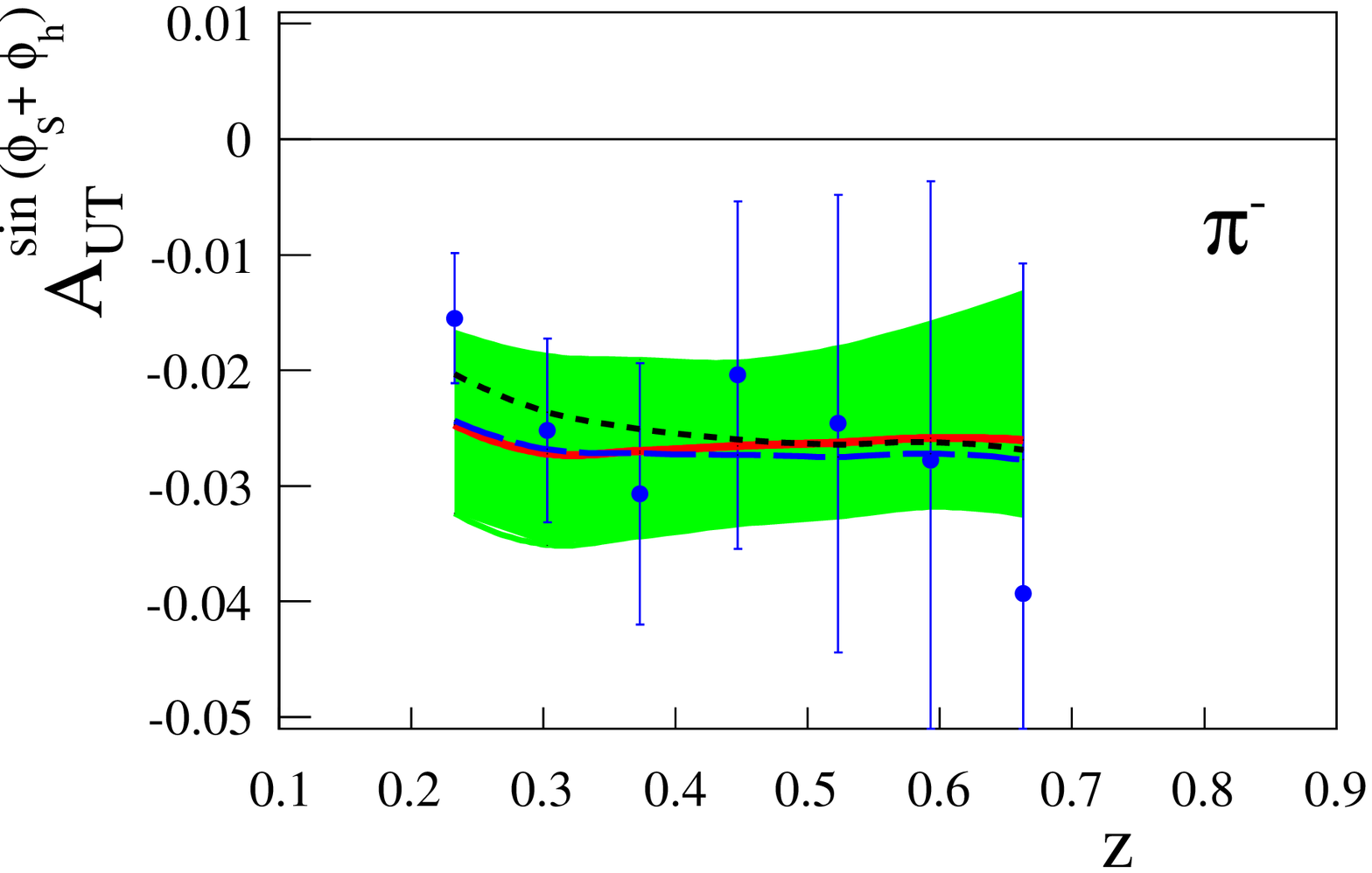}
\includegraphics[width=5.5cm]{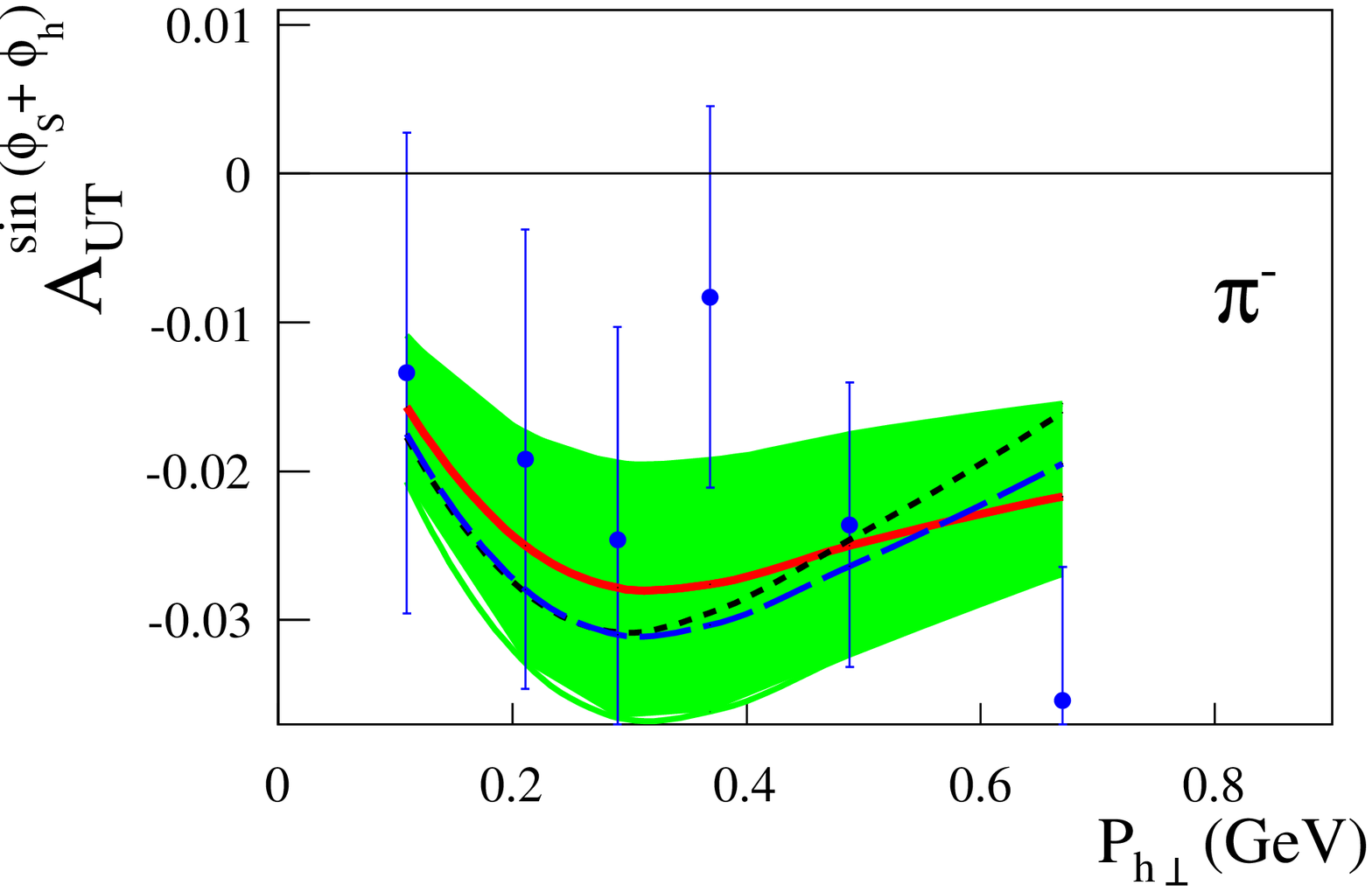}
\caption{Collins asymmetries measured by HERMES Collaboration
 \cite{Airapetian:2009ae} as a function of
$x_B$, $z$, $P_{h\perp}$ in production of $\pi^+$ (left panels) and $\pi^-$ (right panels).
The solid line corresponds to the full NLL$'$ calculation, the dashed line to the LL calculation, and the
dotted to the calculation without TMD evolution. Calculations are performed with parameters from Table~\ref{parameters}. The shaded region corresponds to our estimate of 90\% C.L. error band.}
\label{fig:hermes_comp}
\end{figure}

\begin{figure}[tbp]
\centering
\includegraphics[width=5.5cm]{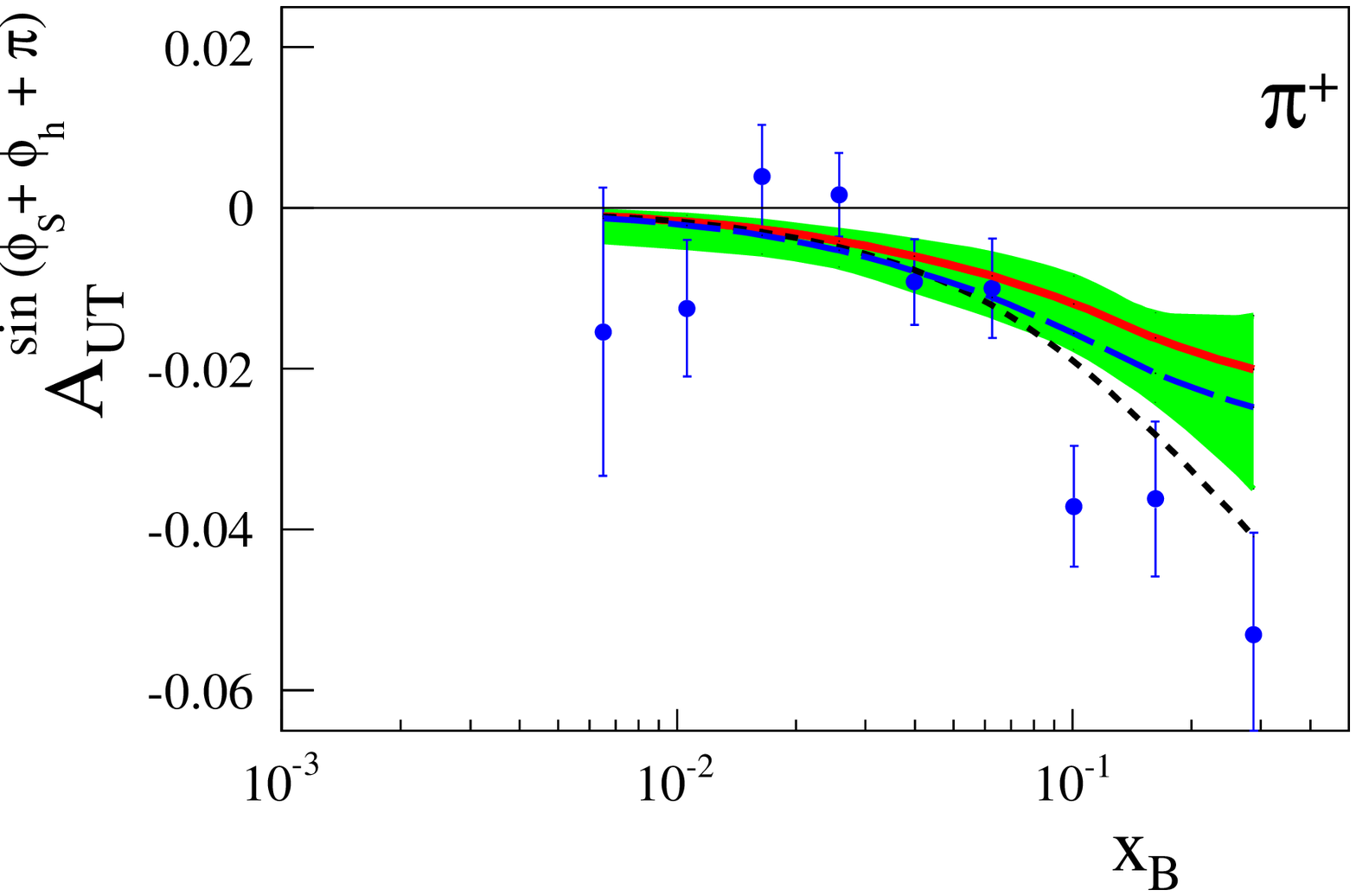}
\includegraphics[width=5.5cm]{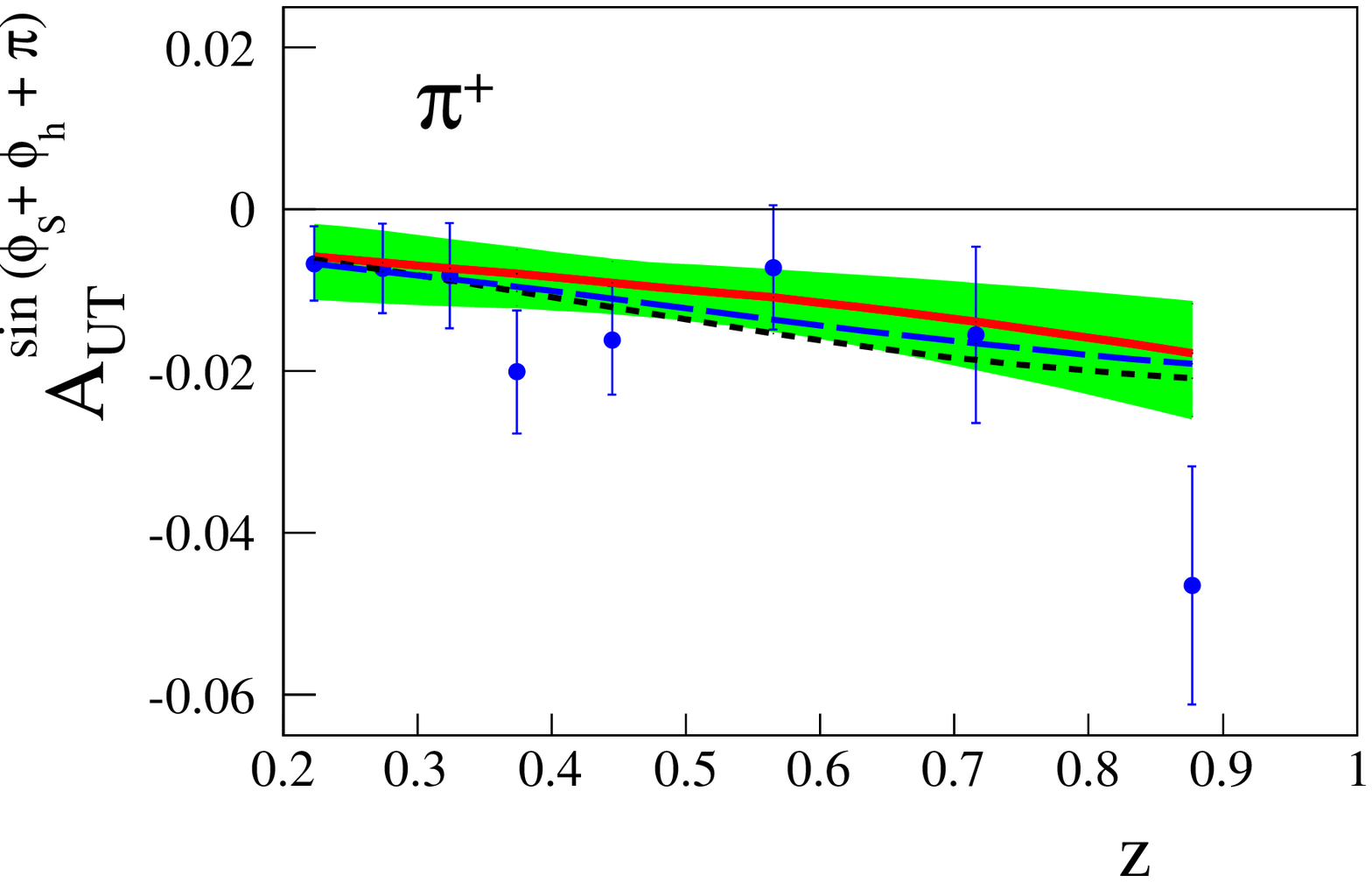}
\includegraphics[width=5.5cm]{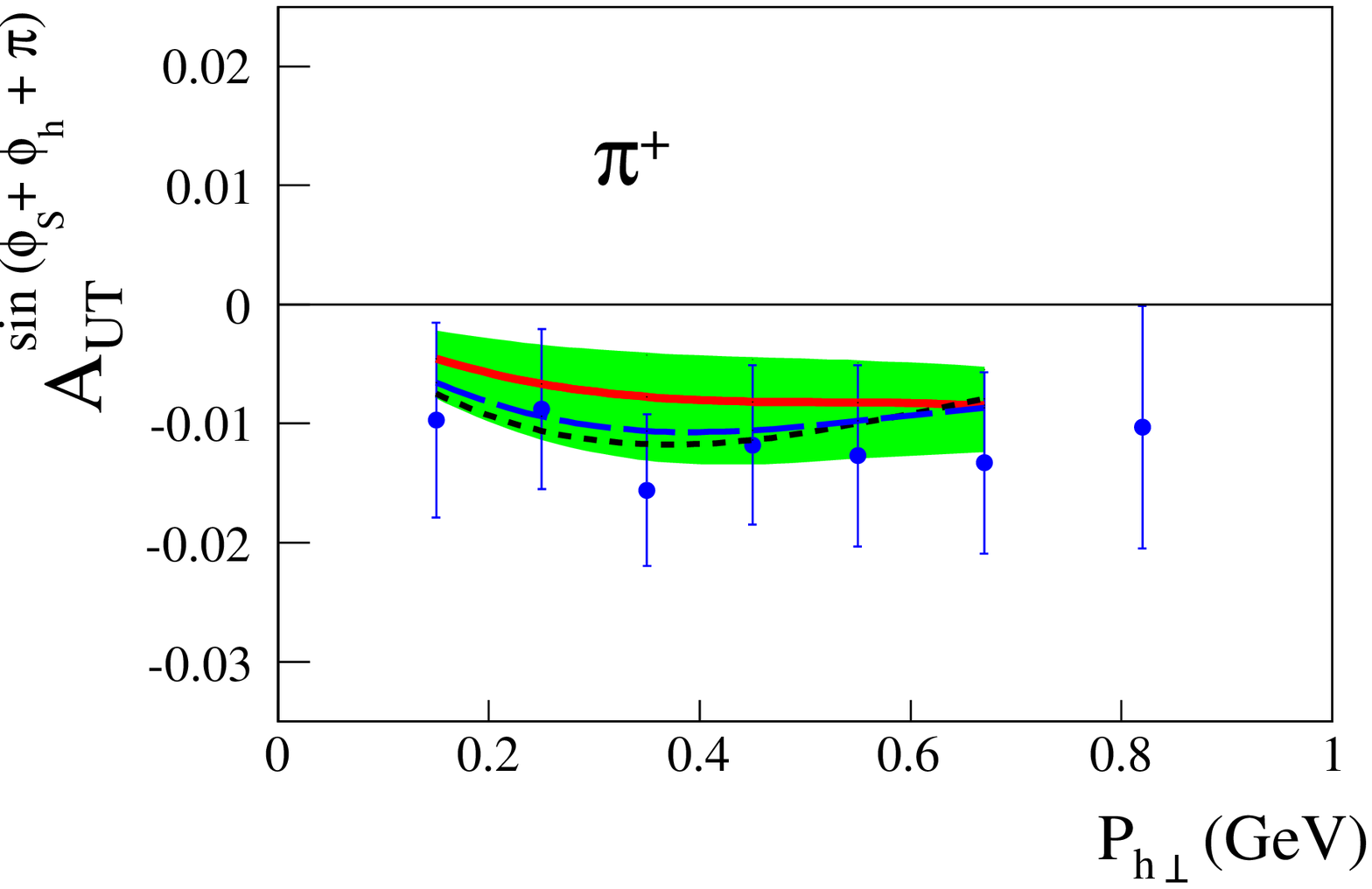}
\includegraphics[width=5.5cm]{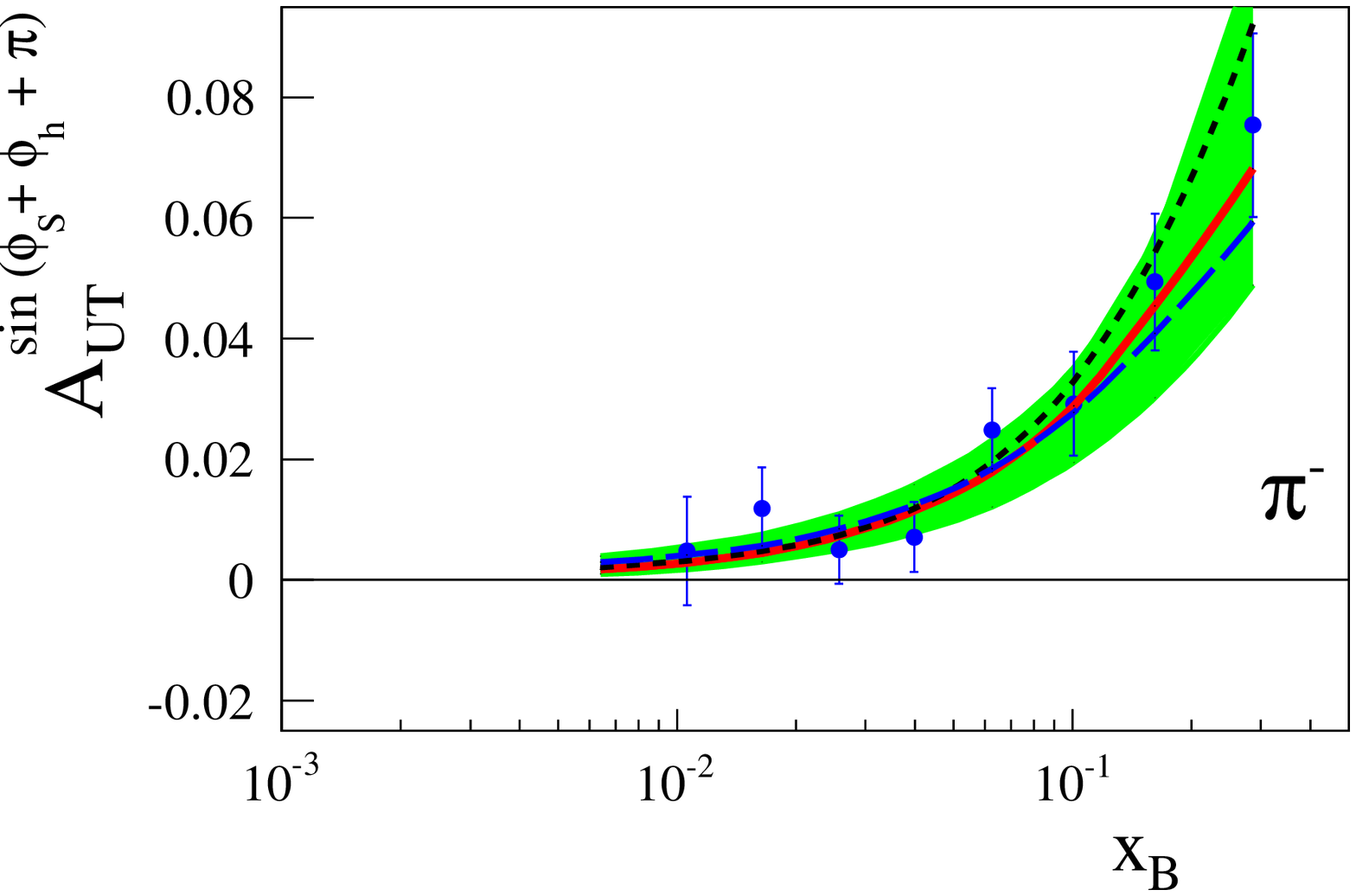}
\includegraphics[width=5.5cm]{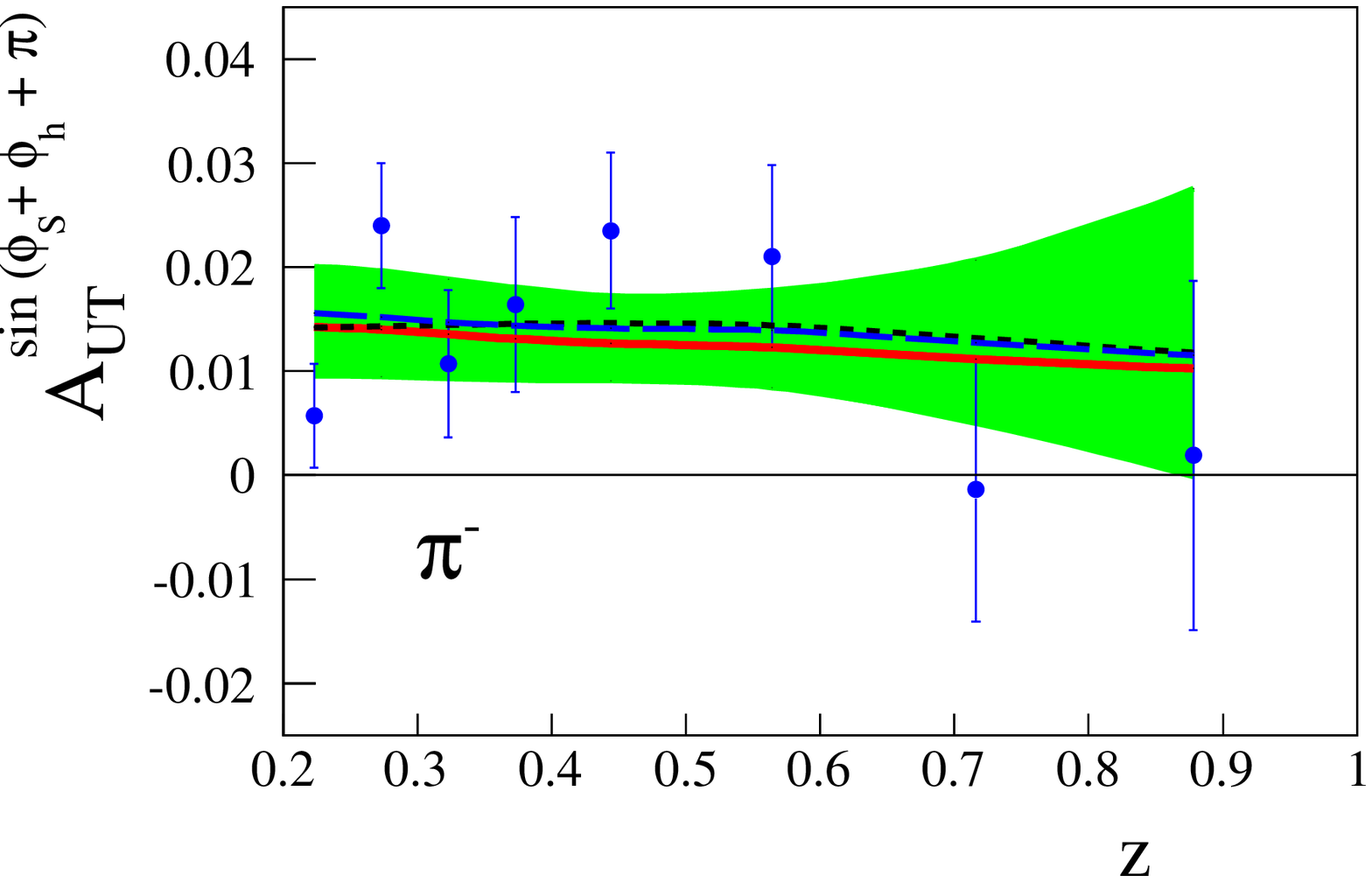}
\includegraphics[width=5.5cm]{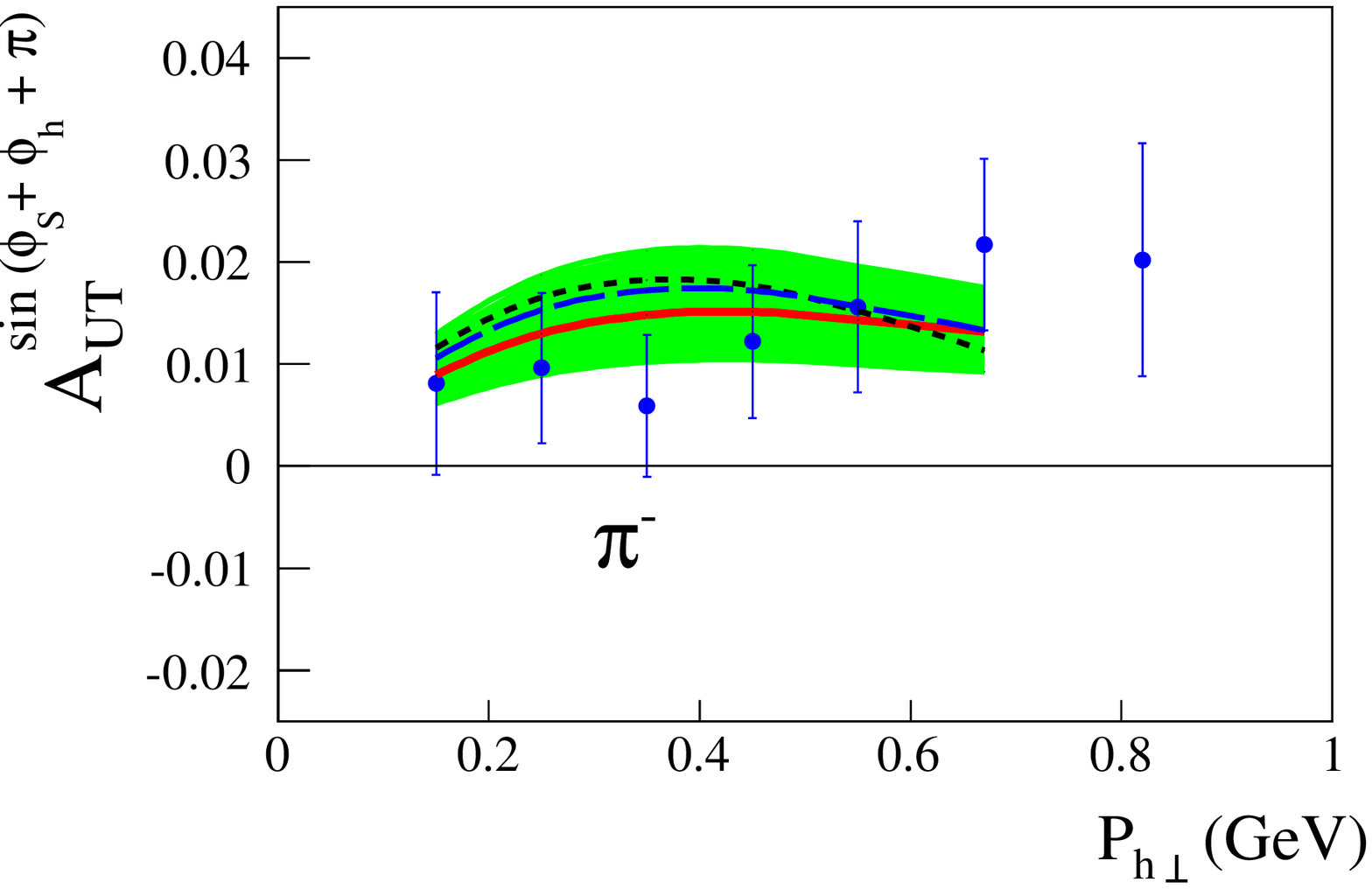}
\caption{Description of Collins asymmetries measured by the COMPASS Collaboration on NH$_3$ proton
\cite{Adolph:2014zba} as a function of
$x_B$, $z$, $P_{h\perp}$ in production of $\pi^+$ (left panels) and $\pi^-$ (right panels).
The solid line corresponds to the full NLL$'$ calculation, the dashed line to the LL calculation, and the
dotted to the calculation without TMD evolution. Calculations are performed with parameters from Table~\ref{parameters}. The shaded region corresponds to our estimate of 90\% C.L. error band.}
\label{fig:compass_comp}
\end{figure}

We show NLL$'$, LL, and no TMD evolution results for asymmetry as function of $x_B$, $z$, $P_{h\perp}$ for HERMES in Fig.~\ref{fig:hermes_comp}. The computation at Leading Logarithmic accuracy (LL) is done by using only $A^{(1)}$ in the perturbative Sudakov form factor and $C^{(0)}$ coefficient function. No TMD evolution implies that the perturbative Sudakov form factor and parameter $g_2$ are set to zero, accordingly we use DSS LO for fragmentation functions and CTEQ6LO for distribution functions and set the scale to $Q_0^2=2.4$ GeV$^2$.  The dotted line in Fig.~\ref{fig:hermes_comp} shows result without TMD evolution. 
One can see that at low energy, the results are quite similar for all three calculations. This happens due to the fact that in {\em ratios} most of the numerical effects of evolution {\em cancel} out. The precision of existing SIDIS experimental data is such that it does not allow to distinguish among different theoretical accuracies used to calculate TMDs. It happens due to the fact that both energy and $Q^2$ are quite low for SIDIS. The difference grows as we consider COMPASS data in Fig~\ref{fig:compass_comp}, and the data becomes sensitive to the choice of accuracy. One can also see from  Fig.~\ref{fig:hermes_comp}  that the difference in different precisions (no evolution, LL, NLL') is comparable with the error band of the NLL' extraction. One can conclude that {\em results} of phenomenological extraction using different precisions will be very similar if low energy experiments are used. In fact low energy experimental data are dominated by non-perturbative physics and even {\em tree level} approximation (no TMD evolution) grasps well the features of the underlying physics. We will in fact see that our results compare very well to the results of Torino-Cagliari-JLab group \cite{Anselmino:2013vqa}.

\begin{figure}[tbp]
\centering
\includegraphics[width=7cm]{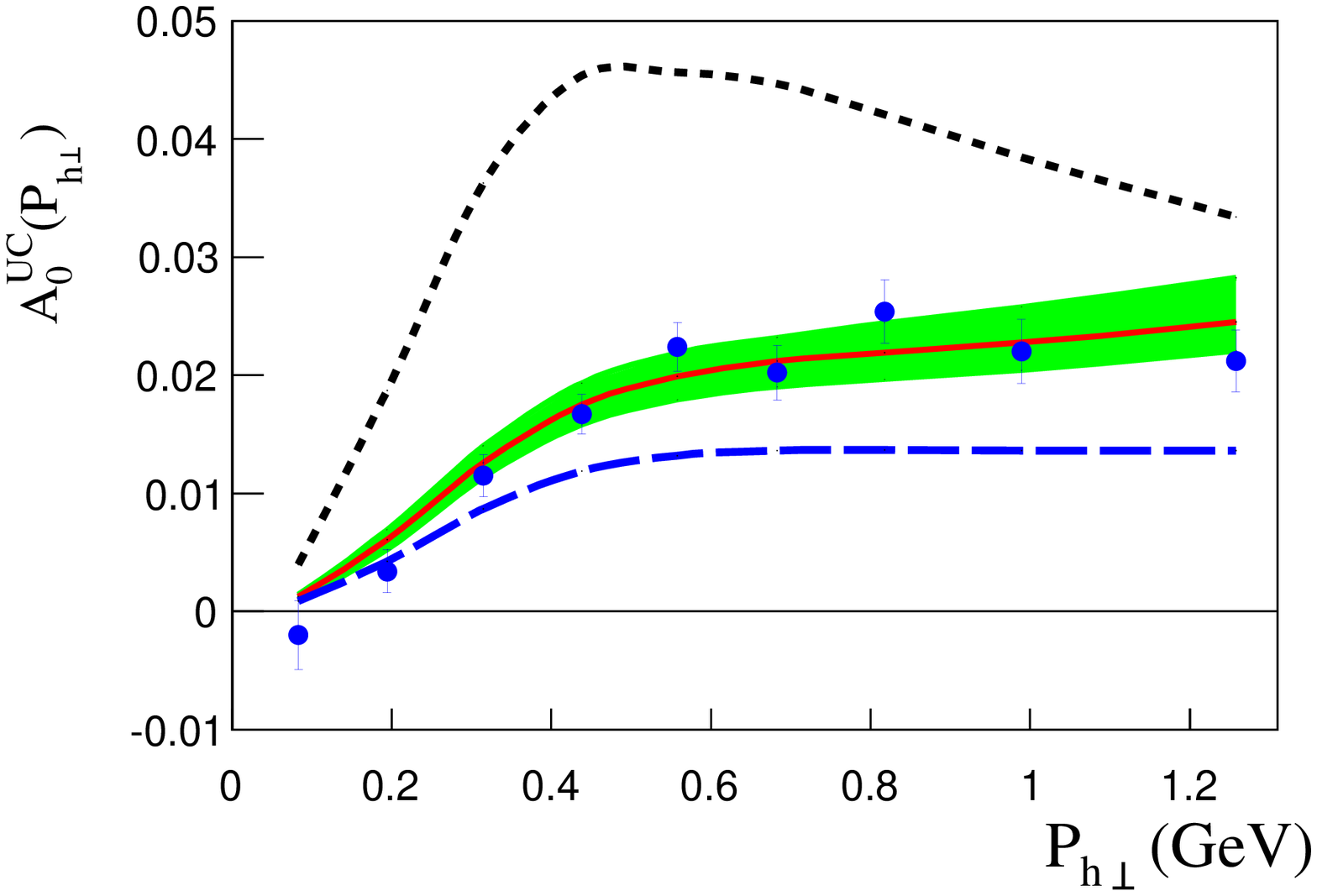}(a)
\includegraphics[width=7cm]{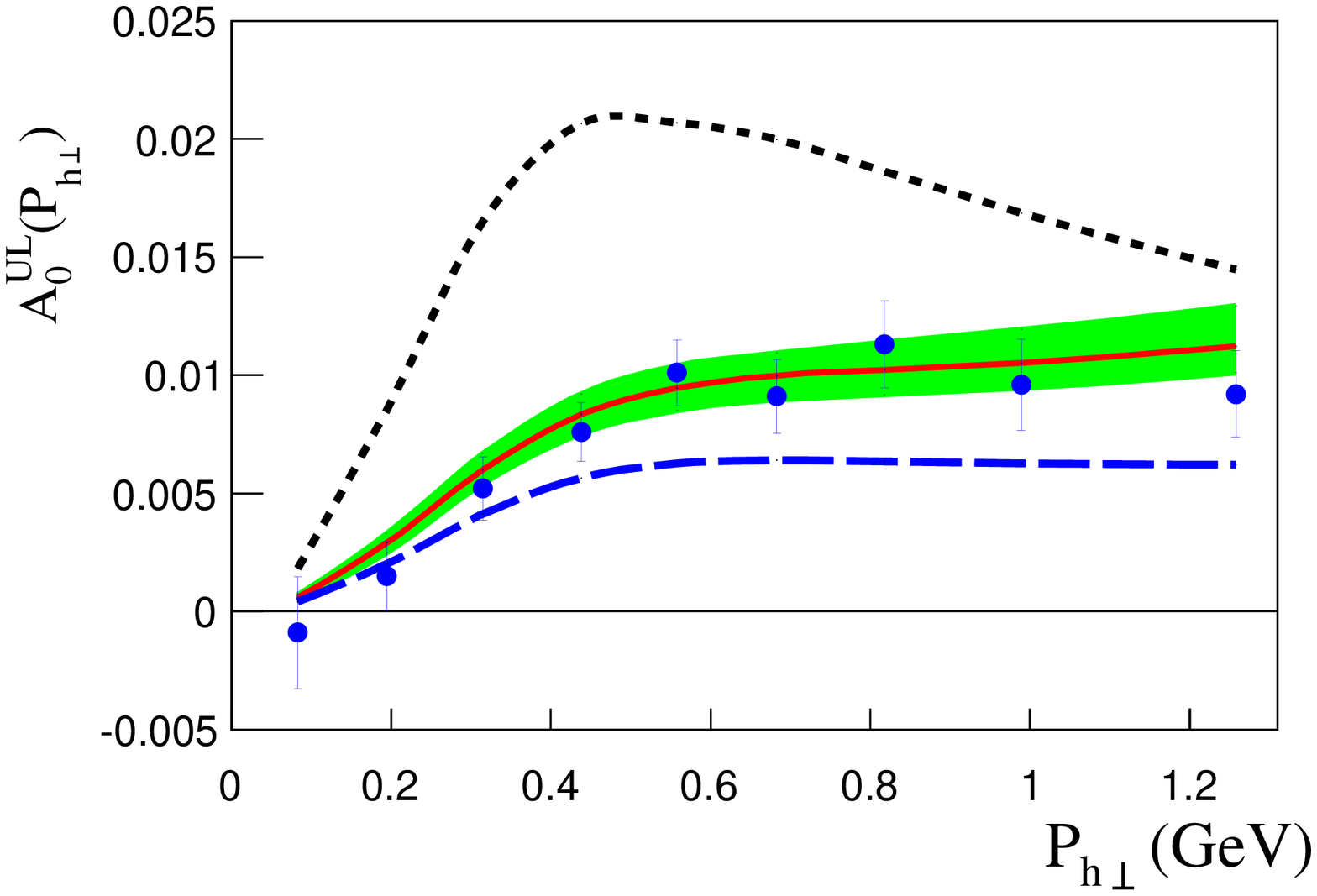}(b)
\caption{Collins asymmetries measured by {\em BABAR}~\cite{TheBABAR:2013yha} collaboration as a function of
$P_{h\perp}$ in production of unlike sign ``U" over like sign ``L" (a) and charged ``C" (b) pion pairs at $Q^2 = 110$ GeV$^2$.
The solid line corresponds to the full NLL$'$ calculation, the dashed line to the LL calculation, and the
dotted to the calculation without TMD evolution. Calculations are performed with parameters from Table~\ref{parameters}. }
\label{fig:babar}
\end{figure}

\begin{figure}[tbp]
\centering
\includegraphics[width=5.5cm]{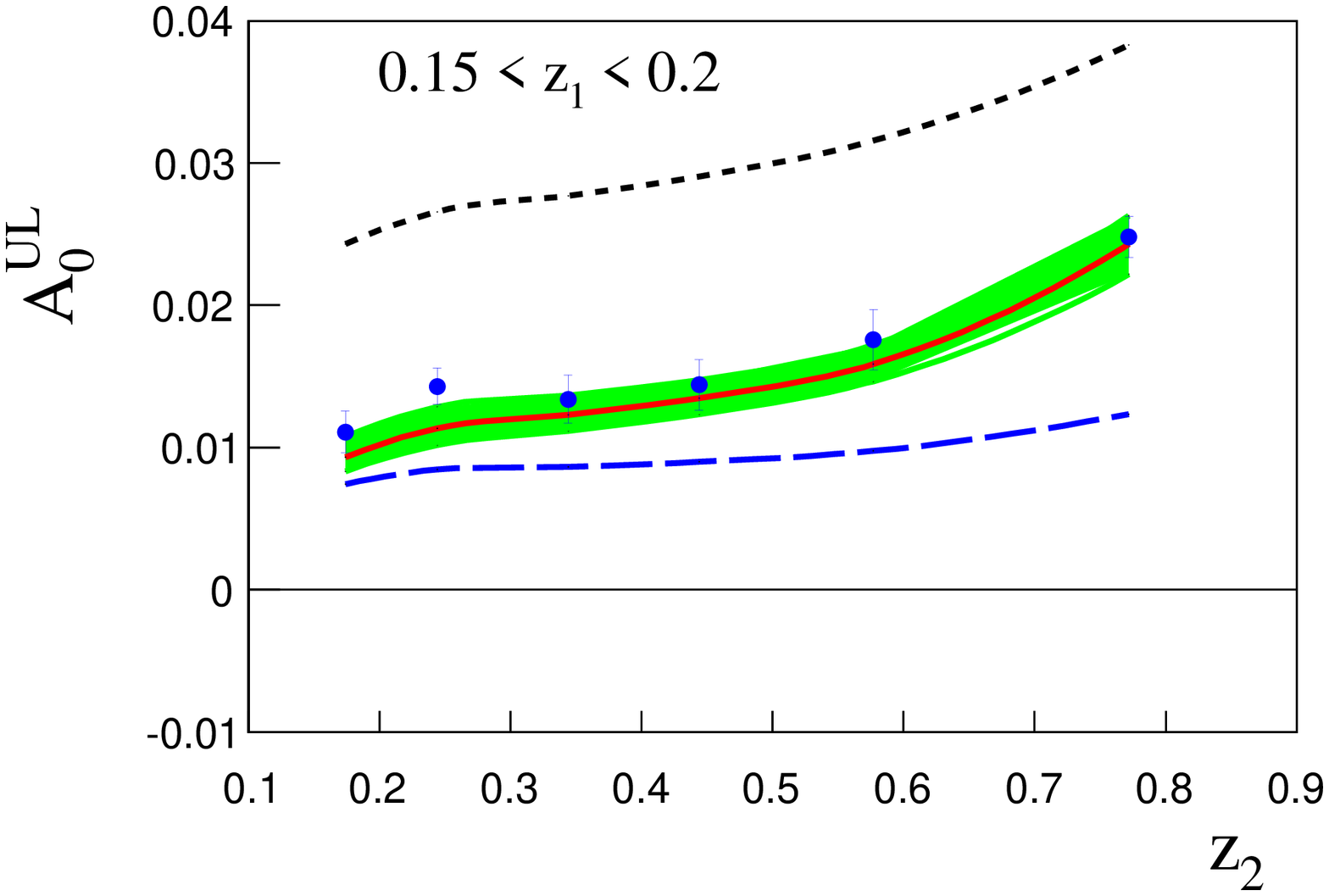}
\includegraphics[width=5.5cm]{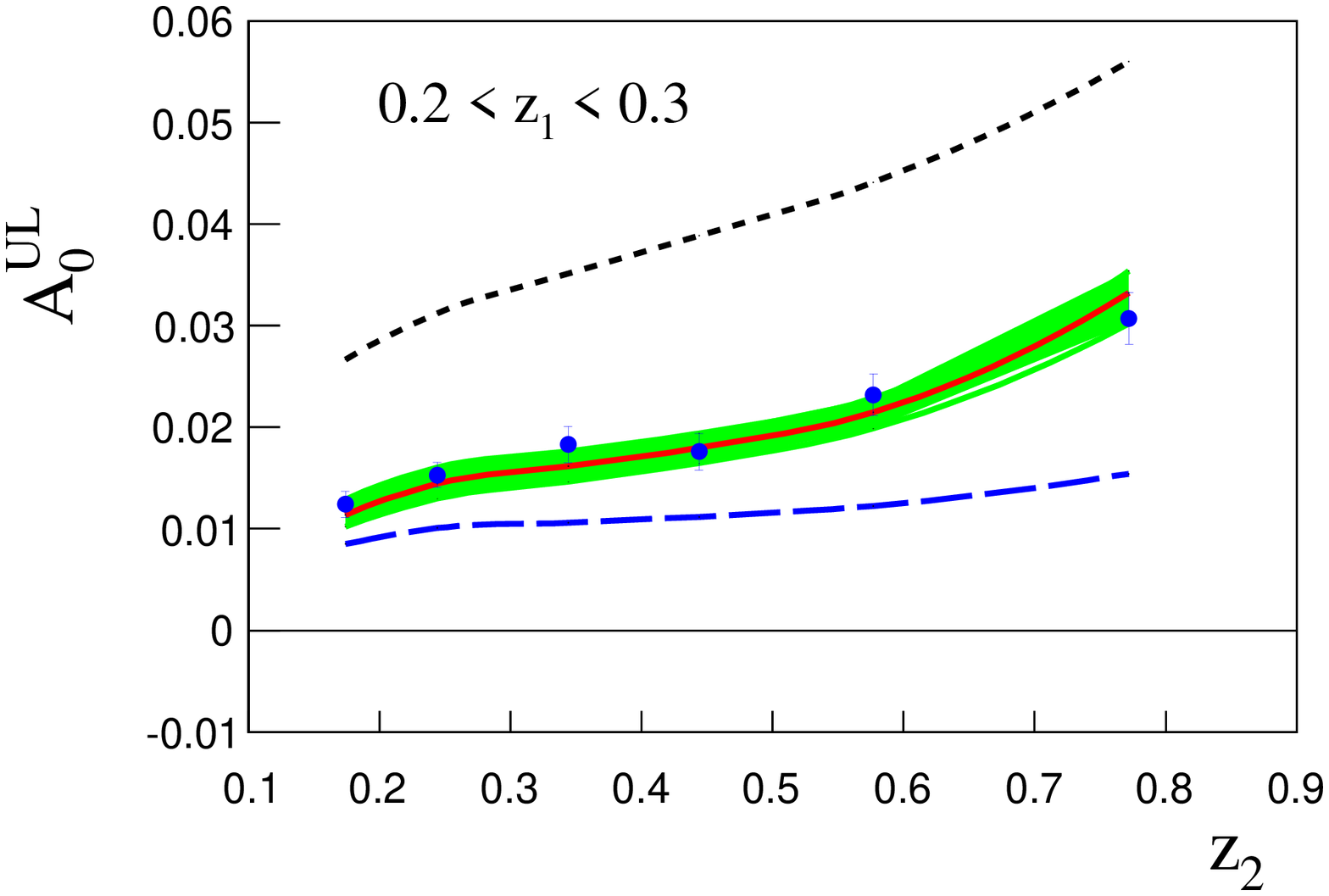}
\includegraphics[width=5.5cm]{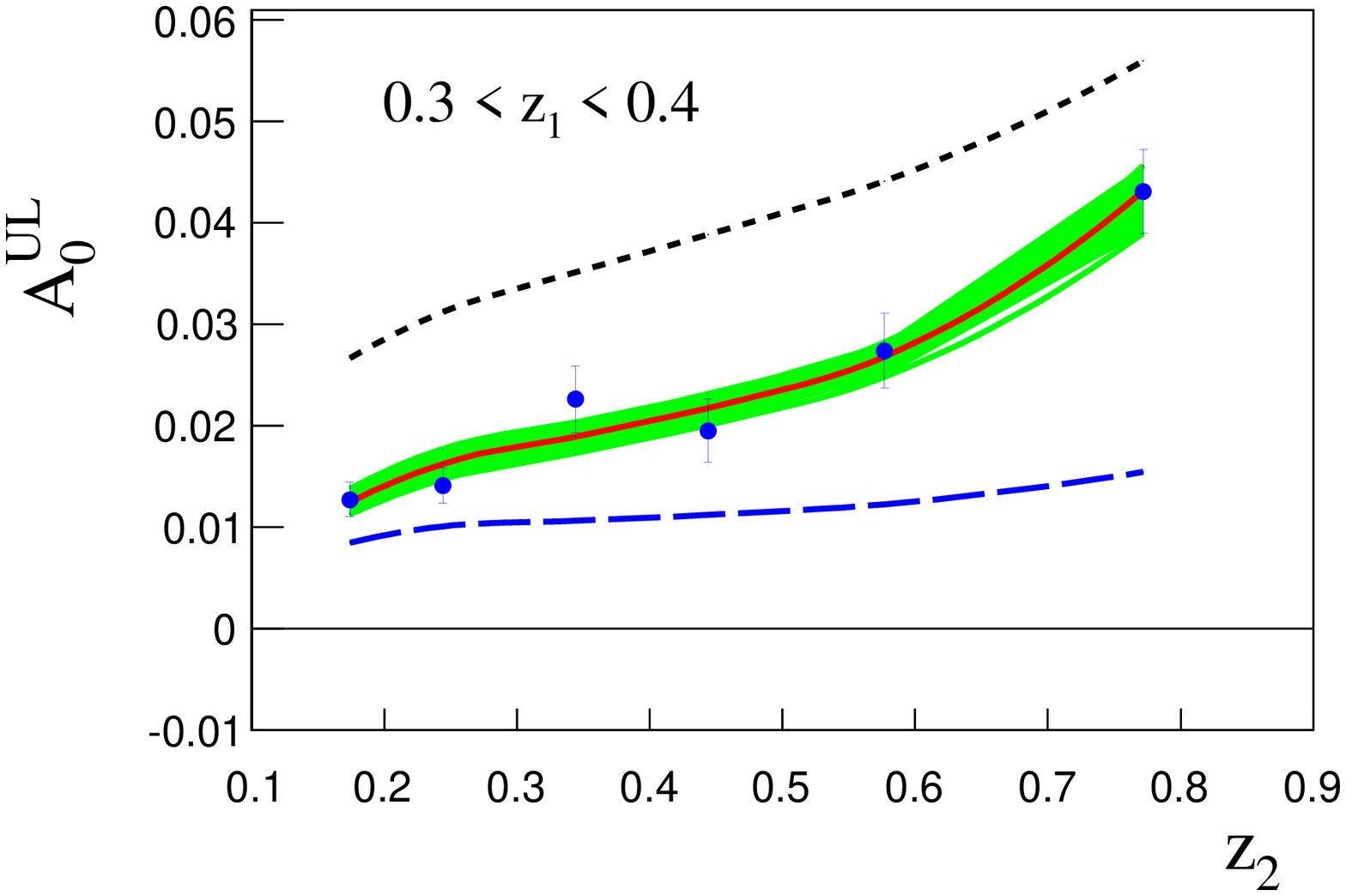}
\includegraphics[width=5.5cm]{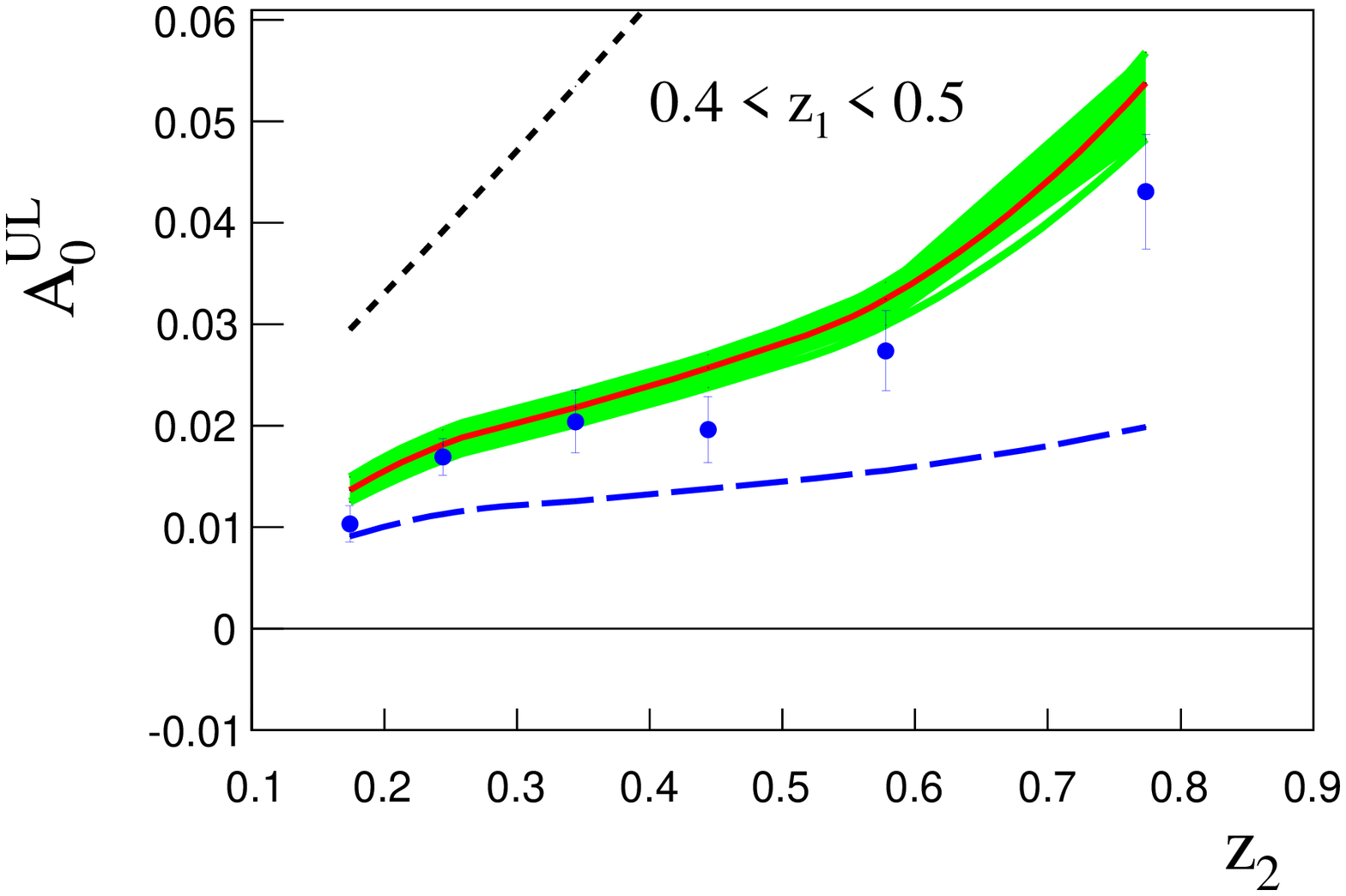}
\includegraphics[width=5.5cm]{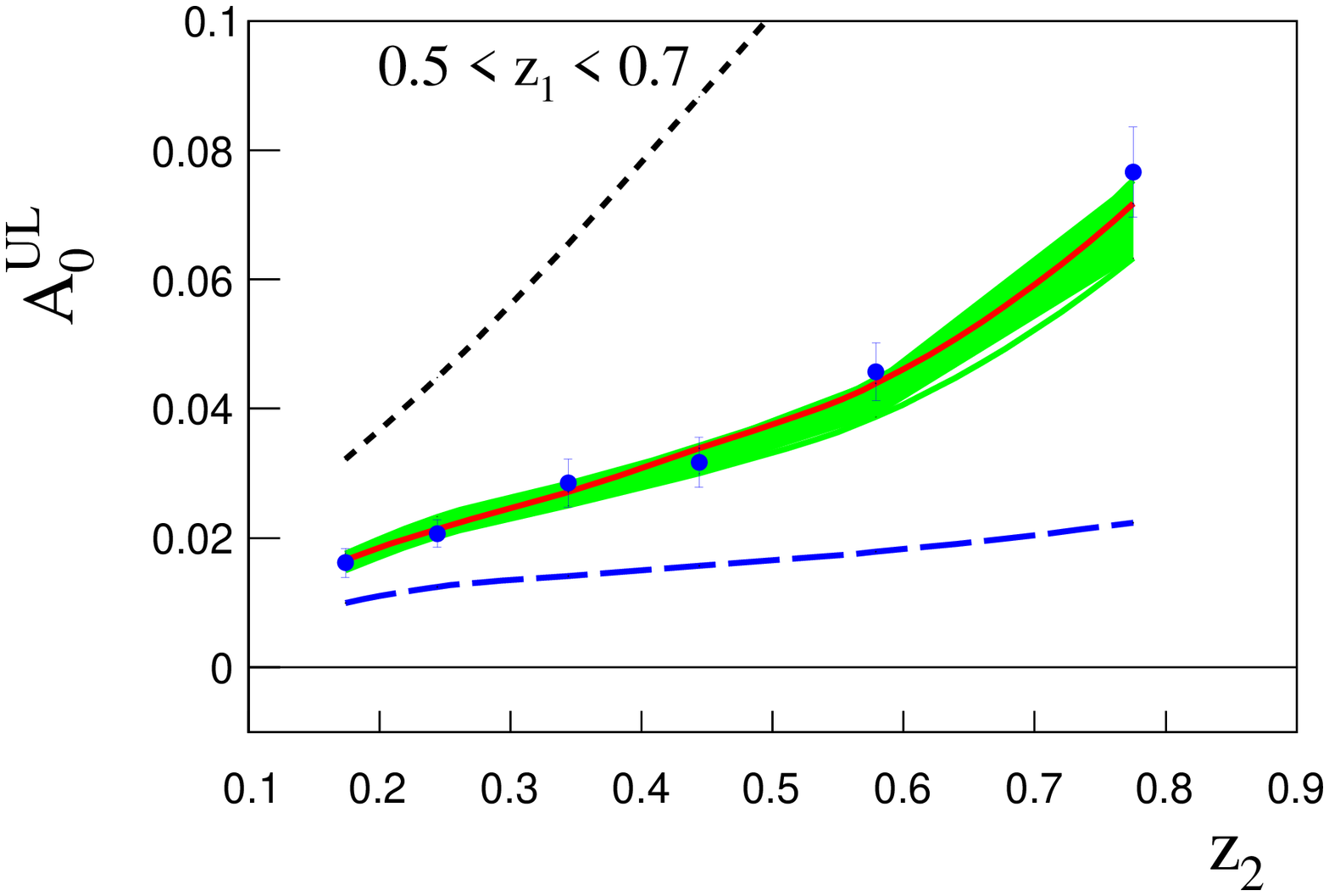}
\includegraphics[width=5.5cm]{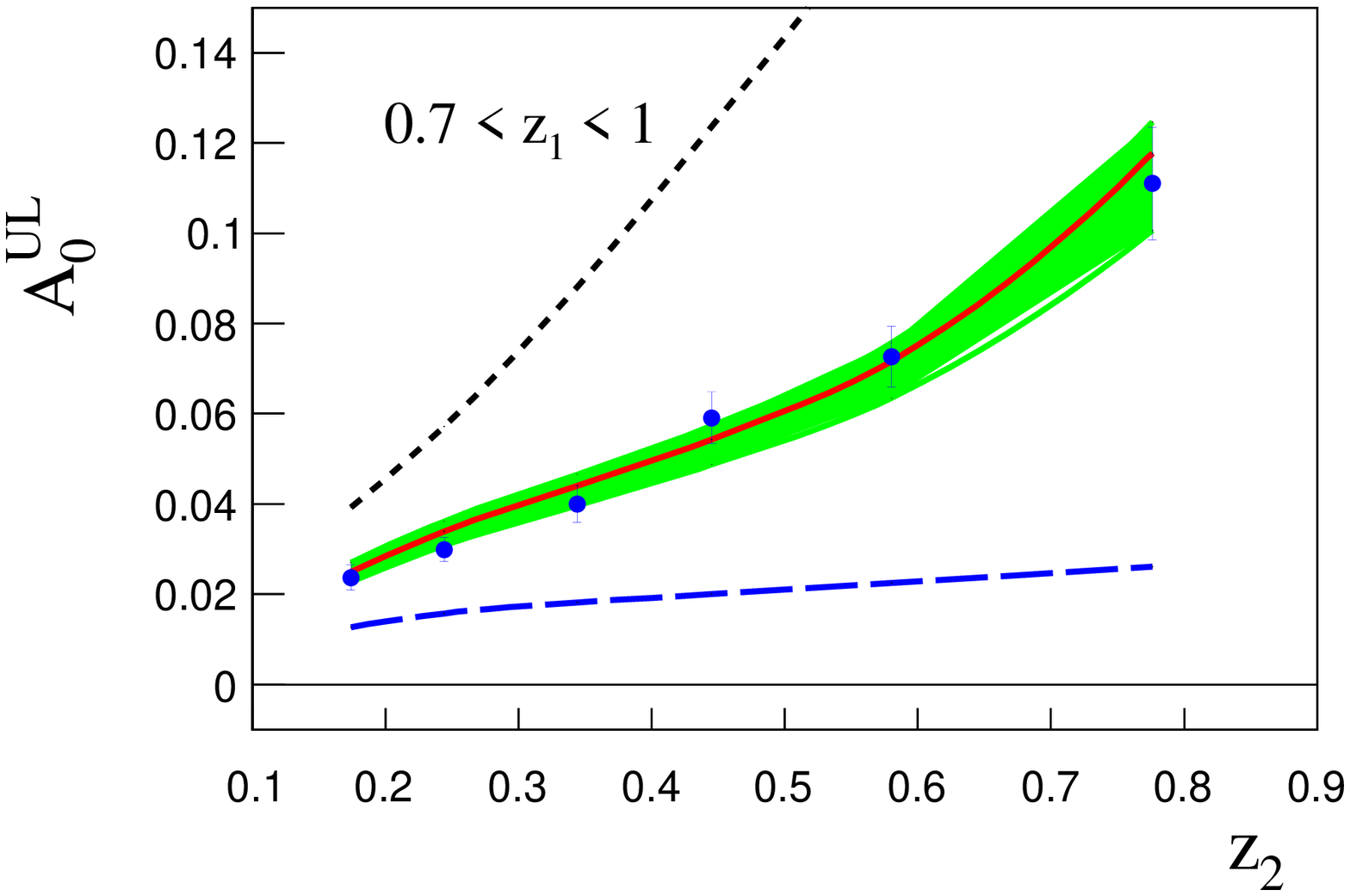}
\caption{Collins asymmetries measured by {\em BABAR}~\cite{TheBABAR:2013yha} Collaboration as a function of
$z_{h2}$  in production of unlike sign ``U" over like sign ``L" pion pairs at $Q^2 = 110$ GeV$^2$.
The solid line corresponds to the full NLL$'$ calculation, the dashed line to the LL calculation, and the
dotted to the calculation without TMD evolution. Calculations are performed with parameters from Table~\ref{parameters}. The shaded region corresponds to our estimate of 90\% C.L. error band.}
\label{fig:babarz}
\end{figure}

In Fig.~\ref{fig:babar} we show theoretical computations for $e^+e^-$ without TMD evolution (dotted line), LL accuracy (dashed line),
and the complete NLL$'$ accuracy (solid line). The difference between these computations diminishes when we include higher orders, it means that the theoretical uncertainty
improves. We conjecture that the difference between NLL$'$ and NNLL will be smaller than difference between NLL$'$ and LL and thus be comparable to experimental errors. One can also observe that asymmetry at $Q^2=110$ GeV$^2$ is suppressed by factor 2 -- 3  with respect to tree-level calculations due to the Sudakov form factor. One can also conclude that NLL accuracy is essential for $e^+e^-$ data. Notice that we present calculations with fixed parameters determined by the  NLL$'$ fit. The difference between different curves shows sensitivity to the theoretical accuracy and to inclusion of higher order. The observation that calculation without TMD evolution or LL cannot describe the data with these parameters does not mean that a {\em fit} of the data without TMD evolution or LL is impossible. In fact such fits are most probably possible and could yield results of similar quality of description of the data. There is no doubt however that higher order computations such as NLL have an advantage of having better control of theoretical uncertainty. The fact that we utilize NLO collinear distributions is very encouraging, these distributions describe inclusive data sets much better than LO distributions. We also observe that $e^+e^-$ experiments are very sensitive to the inclusion of higher order corrections. This can be clearly seen from Fig.~\ref{fig:babarz} where we compute Collins asymmetries measured by {\em BABAR}~\cite{TheBABAR:2013yha} Collaboration as a function of
$z_{h2}$ in different bins of $z_{h1}$. One can see that importance of higher orders increases with increasing value of  $z_h$.

%
%
\subsection{Predictions for unpolarized multiplicities in SIDIS and $e^+e^-$}
\begin{figure}[tbp]
\centering
\includegraphics[width=8cm]{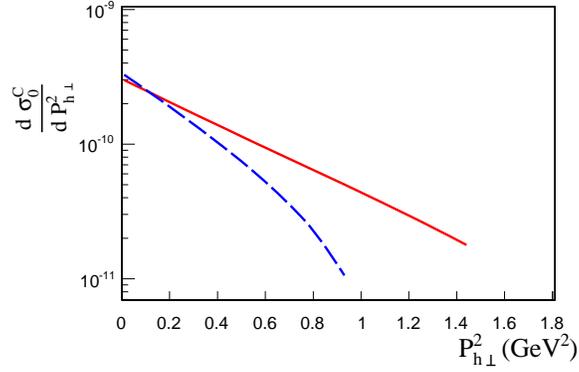} 
\caption{Prediction for unpolarized cross-section   in $e^+e^-$ at $Q^2=110$ GeV$^2$ to be measured by BELLE and  
{\em BABAR} Collaborations (solid line) and BESIII Collaboration at $Q^2= 13$ GeV$^2$ (divided by a factor 110, dashed line)
   as a function of $P_{h\perp}^2$.}
\label{fig:babar_unpol}
\end{figure}

We predict unpolarized cross section of charged pion production to be measured by BELLE,  
{\em BABAR} and BESIII Collaborations and given by the formula
\bea
\frac{d \sigma_0^C}{d P_{h\perp}^2} \equiv \frac{1}{\langle 1+\cos^2\theta \rangle}\frac{d^3\sigma^{e^+e^-\to h_1 h_2 + X}}{dz_{h1}dz_{h2} dP_{h\perp}^2}
= \frac{N_c \pi^2 \alpha_{\rm em}^2}{2 Q^2} Z_{uu}^{h_1h_2}
 \ ,
\label{e+e-_unpol}
\eea
where $h_1,h_2$ can be any charged pion, $z_1 = z_2 = 0.3$. The prediction is given in Fig.~\ref{fig:babar_unpol}. As one can see we predict that measured cross-section will be wider for BELLE and  
{\em BABAR} Collaborations $Q^2= 110$ GeV$^2$   with respect to BESIII Collaboration $Q^2= 13$ GeV$^2$. At the same time BESIII Collaboration $Q^2= 13$ GeV$^2$ cross section will be larger than that measured by  BELLE and  
{\em BABAR} Collaborations at $Q^2= 110$ GeV$^2$. In Fig.~\ref{fig:babar_unpol} we divide the predicted  cross-section for BESIII Collaboration by a factor $110$ in order to compare widths with expected cords-section at BELLE and  
{\em BABAR} Collaborations.
Effective widening of the cross-section with growth of $Q^2$ is a sign of TMD evolution and the future data from BELLE and  
{\em BABAR} Collaborations will be crucial for our understanding of the evolution.

\begin{figure}[tbp]
\centering
\includegraphics[width=8cm]{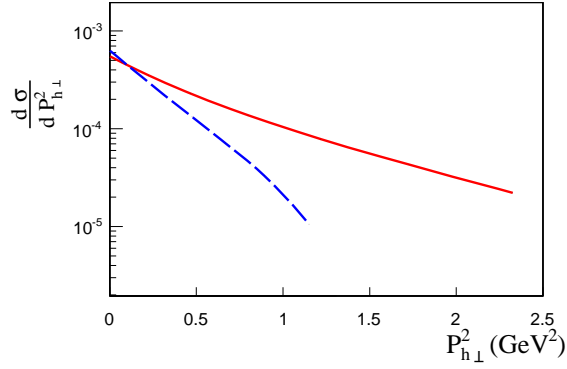} 
\caption{Prediction for unpolarized cross-section for an Electron-Ion Collider of energy $\sqrt{s} = 70$ GeV at $Q^2=100$ GeV$^2$ (solid line) and $Q^2=10$ GeV$^2$  (divided by a factor 200, dashed line)  
   as a function of $P_{h\perp}^2$.}
\label{fig:eic_unpol}
\end{figure}

Similar behavior is shown in  Fig.~\ref{fig:eic_unpol} of the unpolarized cross-section predicted for EIC at $\sqrt{s} = 70$ GeV and at $Q^2=10$ GeV$^2$ and $Q^2=100$ GeV$^2$, choosing $\langle z_h\rangle = 0.36$ and $\langle y \rangle = 0.53$. We plot 
\bea
\frac{d^4\sigma}{dx_B dy dz_h d^2 P_{h\perp}}  
 =    \pi   \sigma_0(x_B, y, Q^2)
 F_{UU}   \, .
  \label{eq:unpol_cross}
\eea
The ultimate test of the TMD evolution will be in measurements of unpolarized cross-sections. We highly encourage BELLE,  
{\em BABAR}    and BESIII Collaborations to perform the analysis of the data on unpolarized cross-sections. Such measurements will allow us to test predictions of TMD evolution and will allow for a better understanding of unpolarized TMD fragmentation functions that can be measured directly only at $e^+e^-$ facilities.

 The universality of TMD evolution will be further tested in future measurements at Electron Ion Collider. EIC can easily span several decades in $Q^2$ and allow for a much better understanding of the nucleon 3D structure. The data of EIC combined with that of Jefferson Lab 12 will cover a very wide region of $x$ and provide a multi-binning data needed for future phenomenological analysis. We plan to study the impact of EIC and Jefferson Lab 12 data in future publications. 

%
%
\subsection{Predictions for future experiments in SIDIS and $e^+e^-$}
\begin{figure}[tbp]
\centering
\includegraphics[width=8cm]{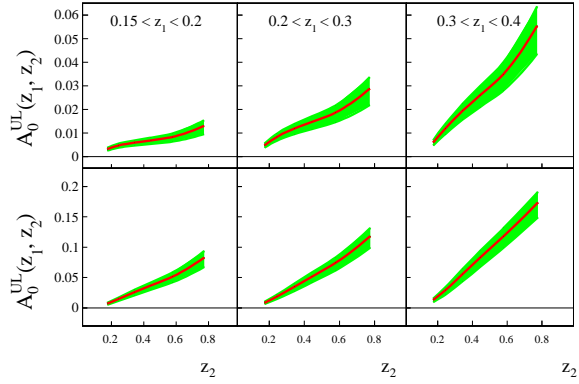} 
\caption{Predictions for UL Collins asymmetries in $e^+e^-$ at $Q^2=13$ GeV$^2$ to be measured by BESIII 
\cite{Guan:2015oaa}  as a function of $z_{h2}$ in different bins of $z_{h1}$.}
\label{fig:besiii}
\end{figure}

BESIII is collecting data \cite{Guan:2015oaa} in $e^+e^-$ at $Q^2 \simeq 13$ GeV$^2$. The preliminary reults are compatible with {\em bigger} asymmetries predicted by two of us in Ref.~\cite{Sun:2013hua}. Here we present updated predictions assuming the same binning as {\em BABAR} and the following values of $\langle \sin^2\theta\rangle/\langle 1+\cos^2\theta \rangle = 0.65$ at each bin, we also integrate the result in the following region of $P_{h\perp} < 1$ GeV. Actual values of asymmetry will depend on details of binning and kinematics. The predictions are presented in Fig.~\ref{fig:besiii}. We give predictions for $A_0^{UL}$ asymmetries, we predict enhancement of the asymmetry by factor $2-3$, compare to Fig.~\ref{fig:epem_babar} (a). Note that our predictions will have to be scaled with actual experimental values of $\langle \sin^2\theta\rangle/\langle 1+\cos^2\theta \rangle_{exp}$ from BESIII.

Measurements of Collins asymmetries are going to be performed at Jefferson Lab 12 GeV upgrade~\cite{Dudek:2012vr} and the planned
Electron Ion Collider~\cite{Boer:2011fh,Accardi:2012qut,Aschenauer:2014twa}.
The high precision of Jefferson Lab 12 measurements will eventually allow for better determination of transversity distributions
in high-$x$ region and low-$x$ region along with higher span in $Q^2$ will be covered by EIC.

\begin{figure}[tbp]
\centering
\includegraphics[width=8cm]{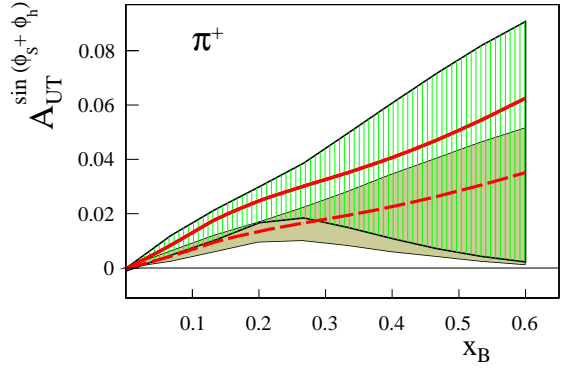} 
\caption{Predictions for   Collins asymmetry    as a function of $x_B$ for $\pi^+$ production on protons in SIDIS at $Q^2=10$ GeV$^2$ (solid line and vertical-line hashed region) and $Q^2=100$ GeV$^2$  (dashed line and shaded region) to be measured by EIC at energy of $\sqrt{s}=70$ GeV.}
\label{fig:eic}
\end{figure}

Electron Ion Collider is going to allow studies of evolution in $Q^2$ and energy $\sqrt{s}$ od single spin asymmetries. It is going to provide a big leverage arm in 
$Q^2$ and will have variable center of mass energy  $\sqrt{s}$. We present here predictions of Collins asymmetry as function of $x_B$ for two different values of
$Q^2=10$ GeV$^2$ and $Q^2=100$ GeV$^2$ in Fig.~\ref{fig:eic}. Note that $x_b$ and  $Q^2$ are correlated via $Q^2 = s x_B y$, we also fix average values of $z_h$ and $P_{h\perp}$, $\langle z_h\rangle = 0.36$ and $\langle P_{h\perp}\rangle = 0.4$ GeV. One can see from Fig.~\ref{fig:eic} that we predict a moderate decrease of the asymmetry with $Q^2$. Measurements in low-$x$ region are going to provide information on sea quark transversity. Our current extraction neglects sea quarks, so the asymmetry becomes very small in low-$x$ region.

\begin{figure}[tbp]
\centering
\includegraphics[width=8cm]{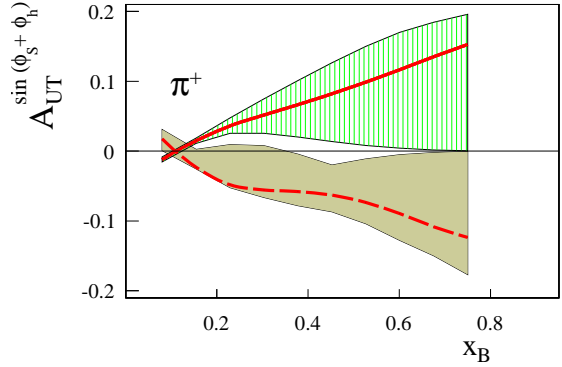} 
\caption{Predictions for   Collins asymmetry    as a function of $x_B$ for $\pi^+$ production at Jefferson Lab 12 GeV on proton target (solid line and vertical-line hashed region) and effective neutron target (dashed line and shaded region).}
\label{fig:jlab12}
\end{figure}

The Jefferson Lab 12 GeV program is going to extend our knowledge of the underlying distributions in the large-$x$ region. 
Both proton and neutron ($^3$He) targets will provide information of distributions of $u$ and $d$-quarks.
We present predictions for JLab 12 at 11 GeV incident electron beam on proton and $^3$He (effective neutron) targets
in Fig.~\ref{fig:jlab12}. One can see that we predict sizable asymmetries of order of 10\%, future data is going to highly improve the knowledge of transversity in large-$x$ region, currently the error band is very big, see Fig.~\ref{fig:jlab12}. In order to give predictions in Fig.~\ref{fig:jlab12} we fixed the average kinematical variables, $\langle y\rangle = 0.57$, $\langle z_h\rangle = 0.5$, and $\langle P_{h\perp}\rangle = 0.38$ GeV.

%
%
\subsection{Comparison to other extractions}

\begin{figure}[tbp]
\centering
\includegraphics[width=8cm]{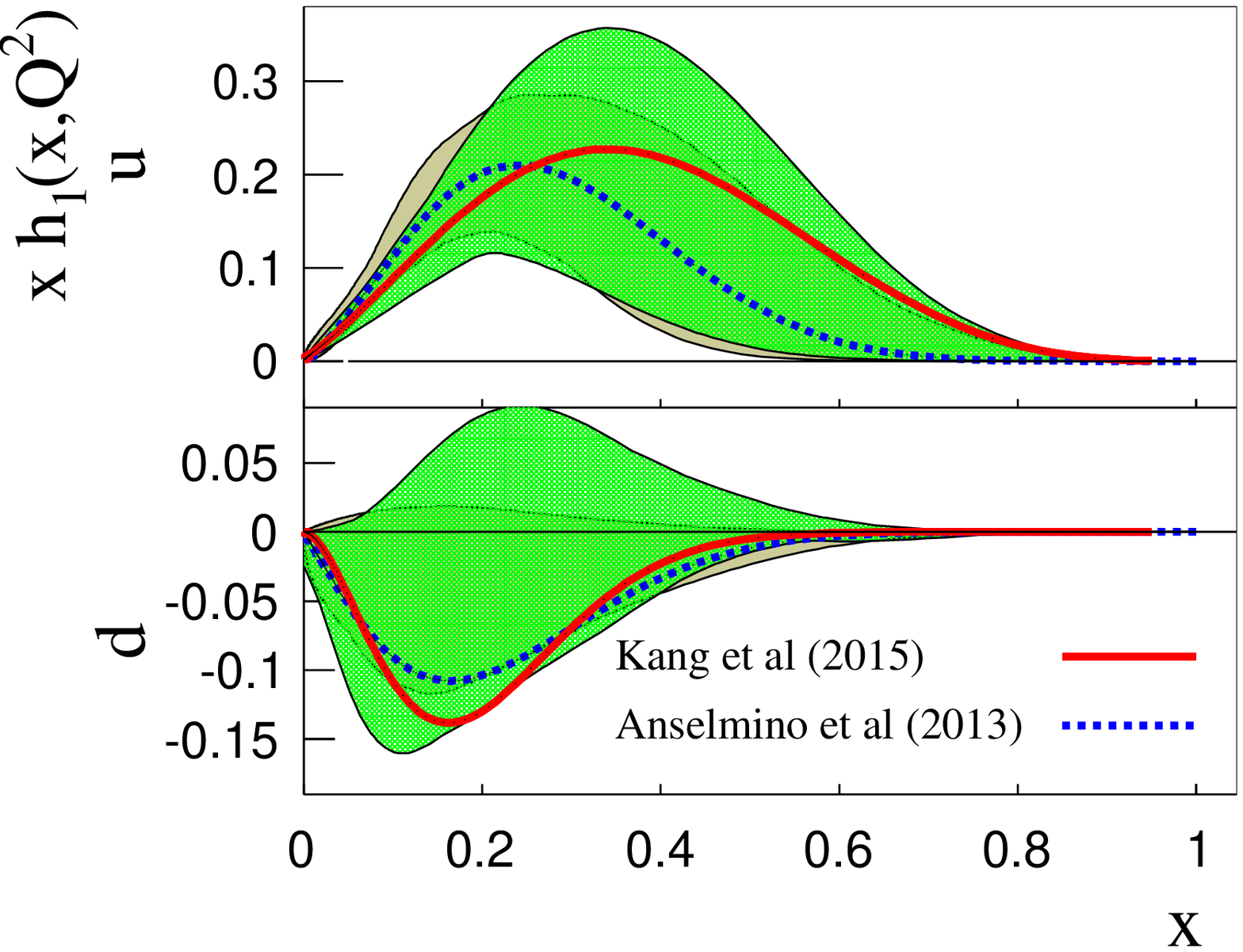}(a)\hspace{0cm}
\includegraphics[width=8cm]{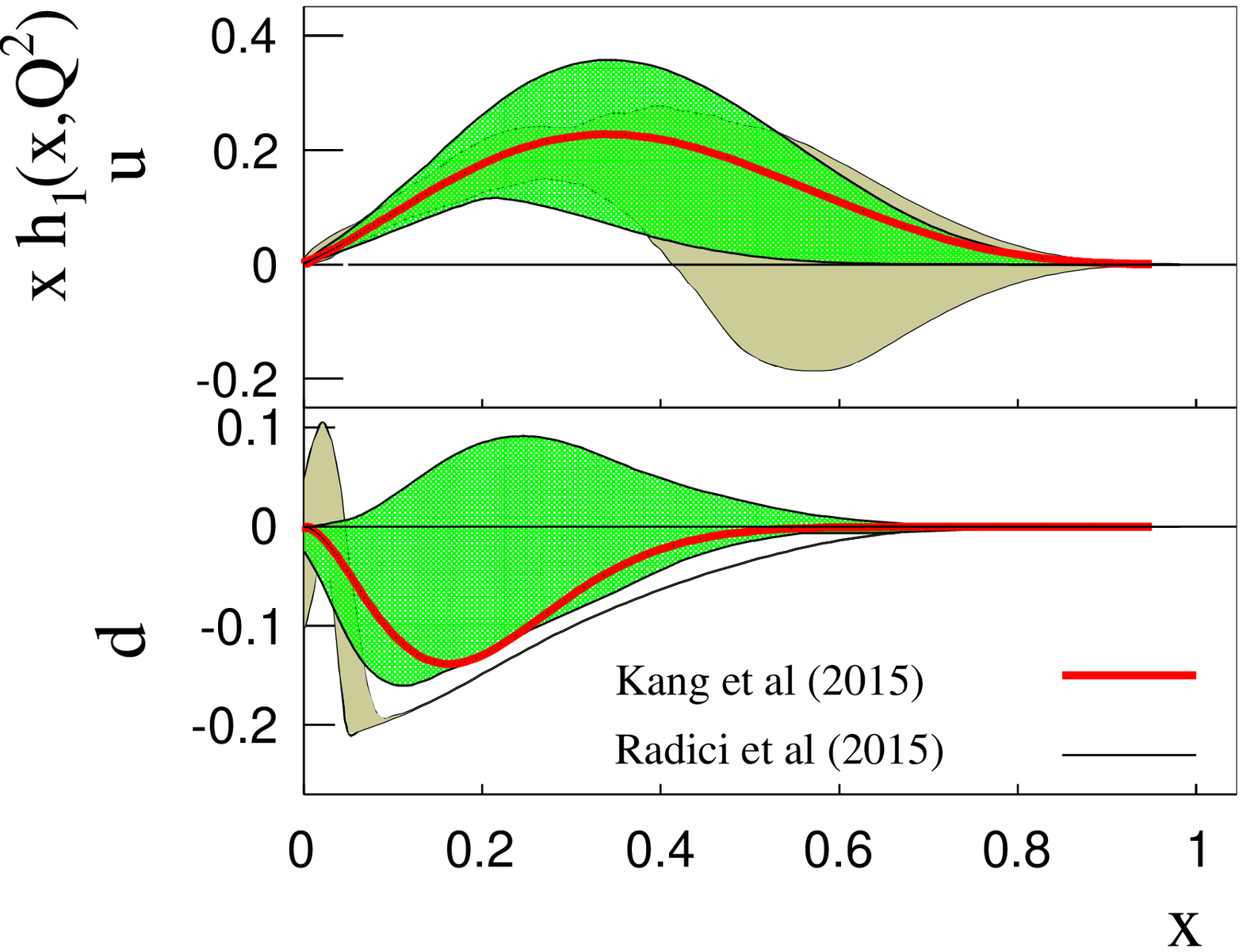}(b)
\caption{ (a) Comparison of extracted transversity (solid lines  and shaded region)  $Q^2 = 2.4$ GeV$^2$ with  Torino-Cagliari-JLab 2013 extraction \cite{Anselmino:2013vqa} (dashed lines and shaded region). \\
(b) Comparison of extracted transversity (solid lines  and shaded region) at $Q^2 = 2.4$ GeV$^2$ with  Pavia 2015 extraction \cite{Radici:2015mwa} (shaded region).}
\label{fig:comparison_torinopavia}
\end{figure}

Tree level extraction of transversity and Collins fragmentation functions was performed by Torino-Cagliari-JLab group in papers ~\cite{Anselmino:2007fs,Anselmino:2008jk,Anselmino:2013vqa}. In Fig.~\ref{fig:comparison_torinopavia} (a) we present comparison of extracted transversity at NLL and result of Ref.~\cite{Anselmino:2013vqa}. We also compare to extraction of transversity via dihadron fragmentation method \cite{Radici:2015mwa} Fig.~\ref{fig:comparison_torinopavia} (b). One can see that all three extractions give consistent results in the explored region of $x_B$. Within error bands of each extraction results are compatible with each other. One can see that the experimental data indeed show some tension, saturation of Soffer bound, for $d$-quark in high-$x$ region as predicted in Ref.~\cite{Ralston:2008sm}. This saturation happens in the region not explored by the current experimental data, so future data from Jefferson Lab 12 will be very important to test the Soffer bound and to constrain the transversity and tensor charge.

The functions themselves are slightly different as can be seen by comparing solid and dashes lines in Fig.~\ref{fig:comparison_torinopavia} (a). In fact
Ref.~\cite{Anselmino:2013vqa} uses tree level TMD expression (no TMD evolution) for extraction and we use NLL TMD formalism. Results should be different even though in asymmetries, as we saw, at low energies results with NLL TMD are comparable with tree level. At higher energies and $Q^2$ situation changes and extracted functions must be different. At the same time one should remember  TMD evolution does not act as a universal $Q^2$ suppression factor. A complicated
Fourier transform should be performed that mixes $Q^2$ and $b$ dependence and thus the resulting functions are different in shape, but comparable in magnitude. It is also very encouraging that tree level TMD extractions yielded results very similar  to our NLL extraction. This makes the previous phenomenological results valid even though the appropriate TMD evolution  was not taken into account. It also means that we need to have experimental data on unpolarized cross sections differential in $P_{h\perp}$. As we have seen the effects of evolution should be evident  in the data and those measurements will help to establish validity of the modern formulation of TMD evolution.

\begin{figure}[tbp]
\centering
\includegraphics[width=8cm]{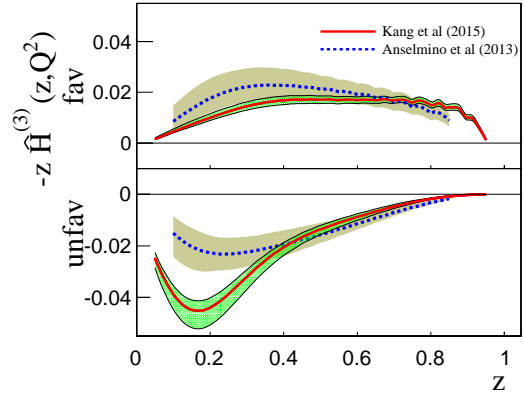} 
\caption{ Comparison of extracted Collins fragmentation functions (solid lines) at $Q^2 = 2.4$ GeV$^2$ with  Torino-Cagliari-JLab 2013 extraction \cite{Anselmino:2013vqa} (dashed lines and shaded region). }
\label{fig:comparison_collinstorino}
\end{figure}

 We compare extracted Collins fragmentation functions   $-z H^{(3)}(z)$ in Fig.~\ref{fig:comparison_collinstorino}   at $Q^2 = 2.4$ GeV$^2$  with extraction of Torino-Cagliari-JLab 2013 \cite{Anselmino:2013vqa}. The resulting Collins FF have the same signs but shapes and sizes are slightly different. Indeed one could expect it as far as $Q^2$ of $e^+e^-$ is different and evolution effect must be more evident. At the same time those functions for both tree level and NLL TMD give the same (or similar) theoretical asymmetries that are well compared to the experimental data of SIDIS and $e^+e^-$. The favored Collins fragmentation function is much better determined by the existing data, as one can see from Fig.~\ref{fig:comparison_collinstorino} that the functions at $Q^2 = 2.4$ GeV$^2$ are compatible within error bands. The unfavored fragmentation functions are different, however those functions are not very well determined by existing experimental data.

\begin{figure}[tbp]
\centering
\includegraphics[width=6cm]{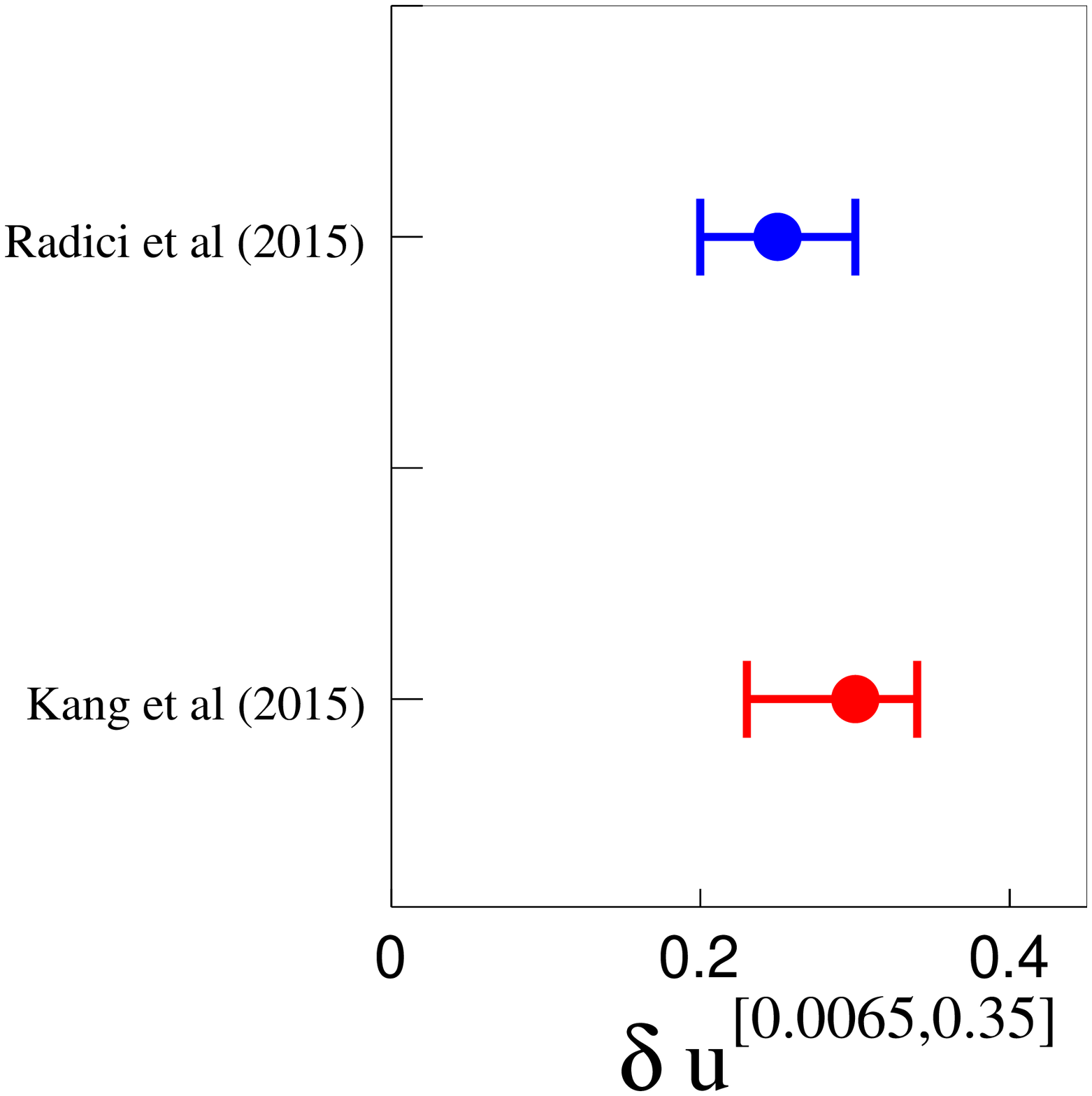} \hspace{1cm}
\includegraphics[width=6cm]{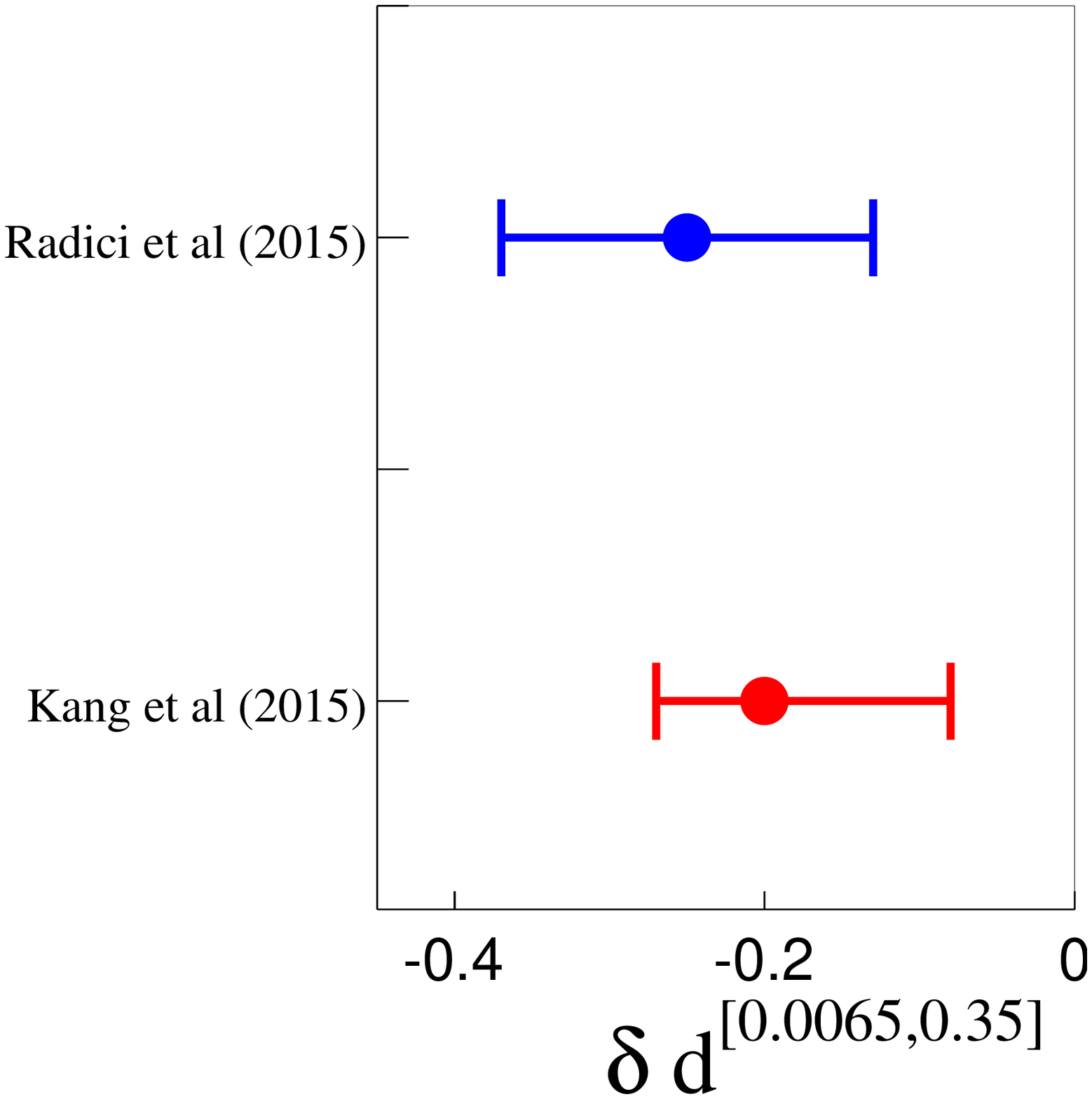} 
\caption{ Comparison of tensor charge $\delta q^{[0.0065,0.35]}$  for $u$-quark and $d$-quark from this paper at 68\% C.L. (Kang et al 2015) and result from Ref.~\cite{Radici:2015mwa}  (Radici et al 2015) at 68\% C.L. Both results are at $Q^2=10$ GeV$^2$.}
\label{fig:comparison_tensorpavia}
\end{figure}

We also compare the tensor change from our and other extractions 
in Fig.~\ref{fig:comparison_tensorpavia}.  The contribution to tensor 
charge of Ref.~\cite{Radici:2015mwa} is found by extraction using 
the so-called dihadron fragmentation function that couples to collinear transversity distribution.
 The corresponding functions have DGLAP type evolution known at LO and were used in 
Ref.~\cite{Radici:2015mwa}.
The results plotted in Fig.~~\ref{fig:comparison_tensorpavia} corresponds 
to our estimates of the contribution to $u$-quark and $d$-quark in the 
region of $x$ $[0.065,0.35]$ at $Q^2 = 10$ GeV$^2$ at 68\% C.L. (label 1) 
and the contribution to $u$-quark and $d$-quark in the same region of $x$ 
and the same $Q^2$ using the so-called flexible scenario, 
$\alpha_s(M_Z^2) = 0.125$, of Ref.~\cite{Radici:2015mwa}. 
One can see that our extraction has an {\em excellent} precision for 
both $u$-quark and $d$-quark. 
 The fact that the central values and 
errors of extracted tensor charges are in a good agreement in both methods, 
ours and Ref.~\cite{Radici:2015mwa}, is very positive and allows for future 
investigations of transversity including all available data in a global fit.

\begin{figure}[tbp]
\centering
\includegraphics[width=6cm]{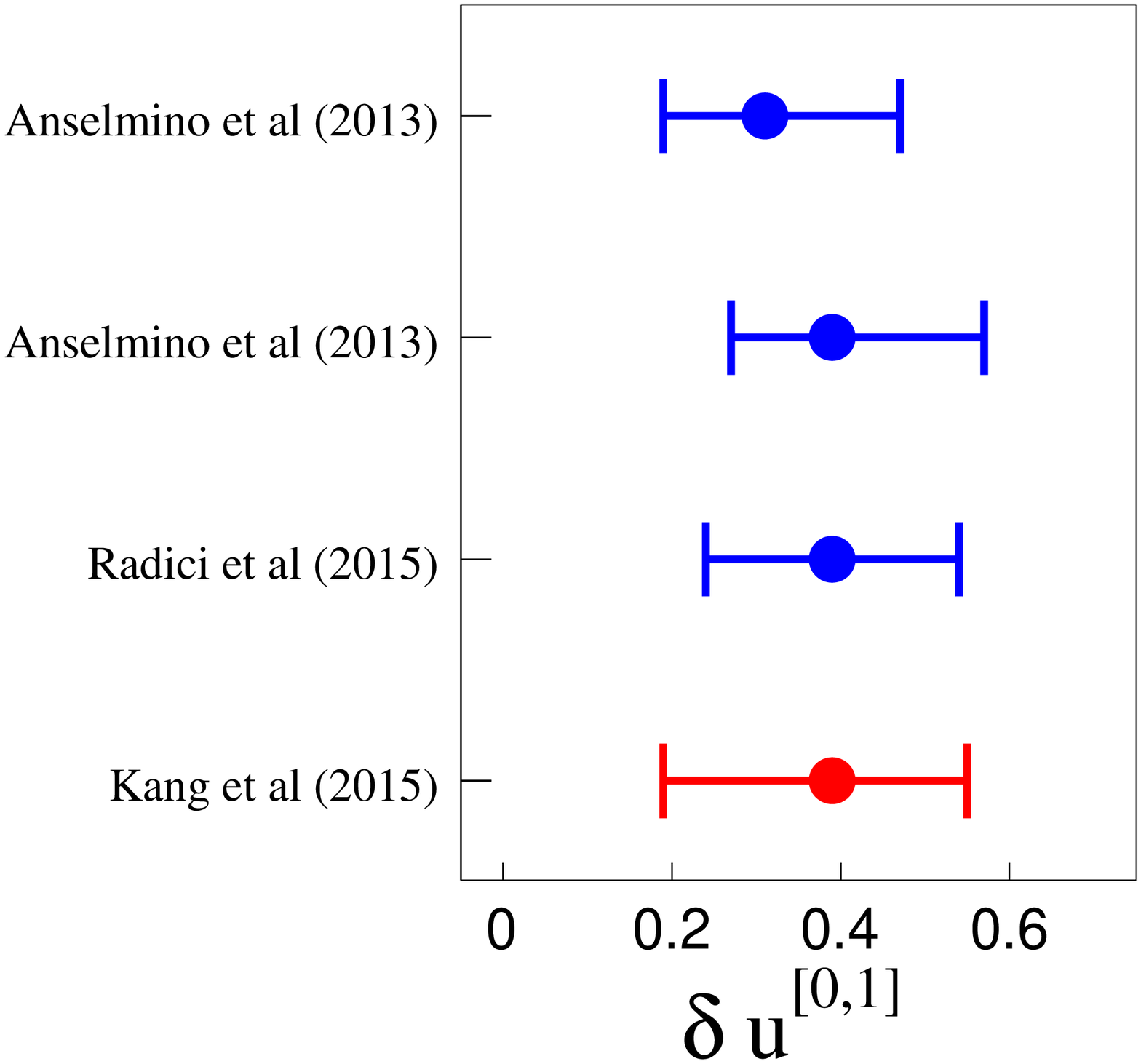} \hspace{1cm}
\includegraphics[width=6cm]{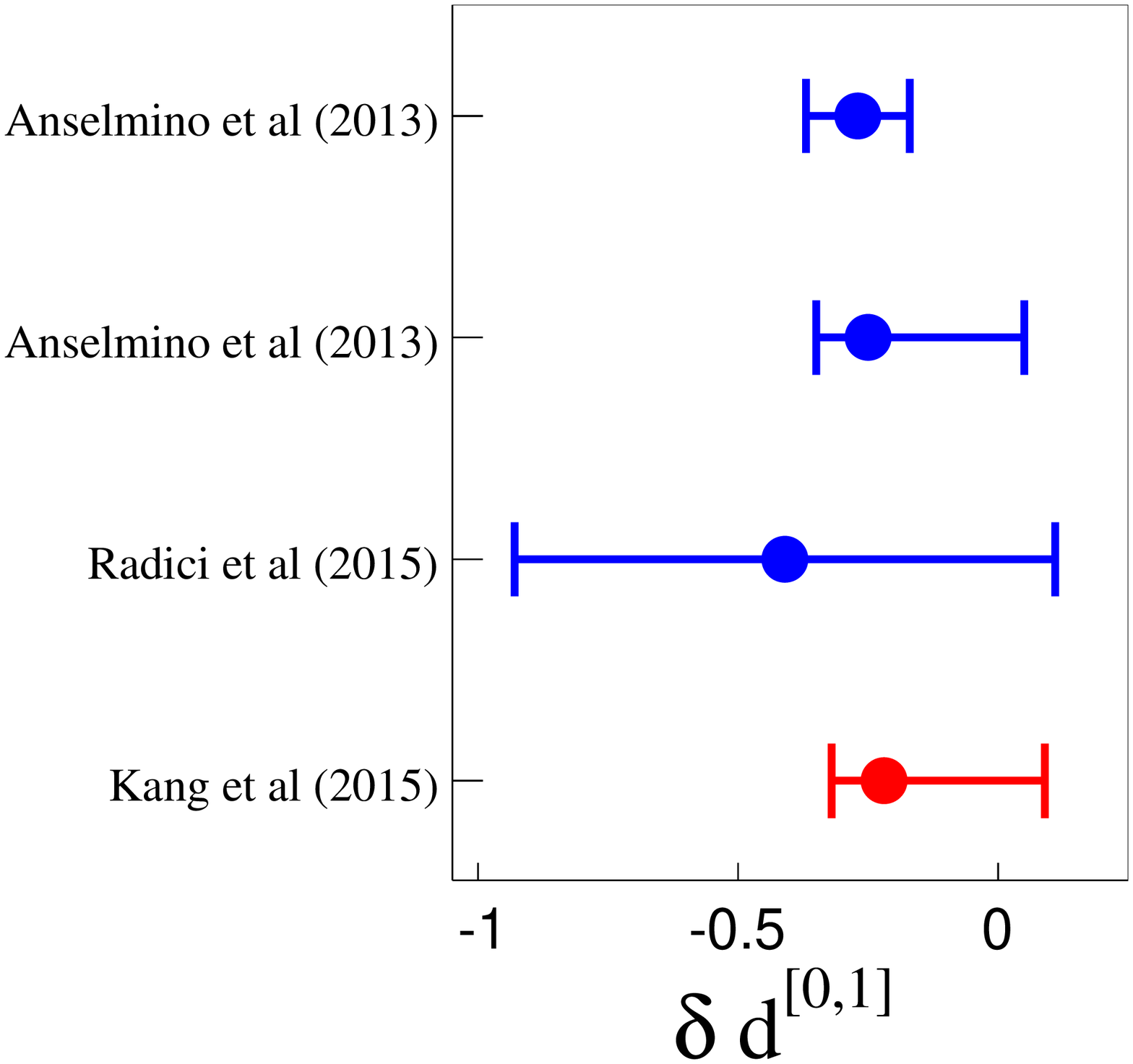} 
\caption{ Comparison of tensor charge  $\delta q^{[0,1]}$  for $u$-quark and $d$-quark in the whole region of $x$ from this paper at 90\% C.L. (Kang et al 2015) at $Q^2=10$ GeV$^2$ and result from Ref.~\cite{Radici:2015mwa}  (Radici et al 2015) at at 68\% C.L. and $Q^2=1$ GeV$^2$,
and Ref.~\cite{Anselmino:2013vqa} at 95\% C.L. standard and polynomial fit (Anselmino et al 2013) at $Q^2=0.8$ GeV$^2$.}
\label{fig:comparison_tensorpaviatorino}
\end{figure}

Our results compare well with extractions from Ref.~\cite{Anselmino:2013vqa}. Even though correct TMD evolution was not used in Ref.~\cite{Anselmino:2013vqa} the effects of DGLAP evolution of collinear distributions were taken into account and the resulting fit is of good quality, 
$\chi^2/{d.o.f.} =0.8$ for the so-called standard parametrization of Collins fragmentation functions. In fact the probability that the model of Ref.~\cite{Anselmino:2013vqa} correctly describes the data is $P(0.8*249,249) = 99$\%. The tensor charge was estimated at 95\% C.L. using two different parametrizations for Collins fragmentation functions, the so-called standard parametrization that utilized similar to our parametrization and the polynomial parametrization. In Fig.~\ref{fig:comparison_tensorpaviatorino} we compare our results with calculations from Ref.~\cite{Anselmino:2013vqa} at 95\% C.L. at $Q^2=0.8$ GeV$^2$ and calculations at 68 \% at $Q^2=1$ GeV$^2$ of Ref.~\cite{Radici:2015mwa}. Even though we compare tensor charge at different values of $Q^2$ its evolution is quite slow, so the good agreement of all three methods is a good sign. We conclude that tensor charge perhaps is very stable with respect to evolution effects that are included in phenomenological extractions. It also means that phenomenological results of Ref.~\cite{Anselmino:2013vqa} and other extractions without TMD evolution are valid phenomenologically. One should remember, of course, that TMD evolution is more complicated if compared to DGLAP evolution (even though formal solutions are simpler in TMD case). The usage of non perturbative kernels make it very important to actually demonstrate that the proper evolution is indeed exhibited by the experimental data. Once correct evolution and non perturbative Sudakov factor are established the results of Ref.~\cite{Anselmino:2013vqa}  should be improved by utilizing the appropriate TMD evolution that we have formulated in this paper.

\begin{figure}[tbp]
\centering
\includegraphics[width=6cm]{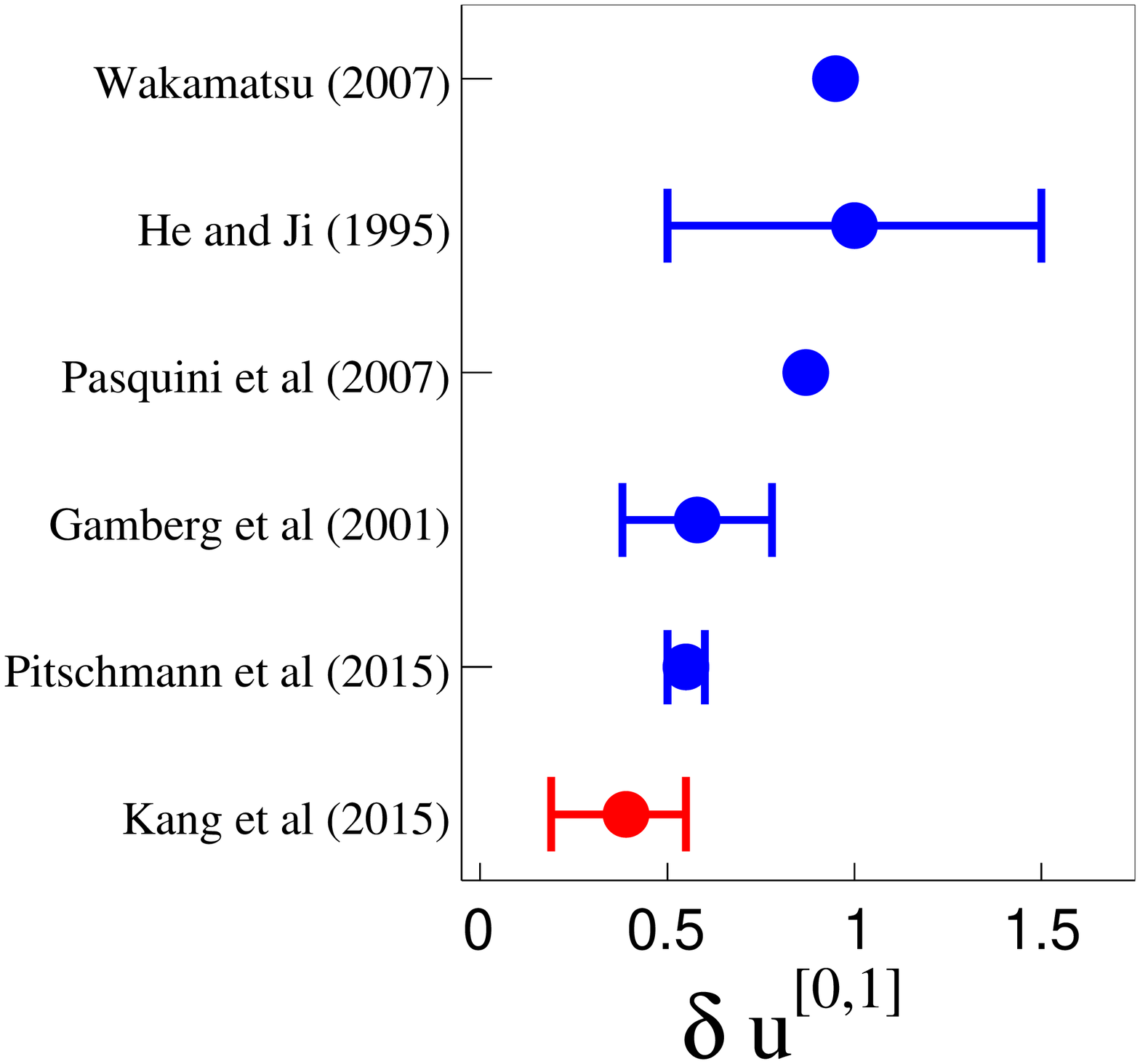} \hspace{1cm}
\includegraphics[width=6cm]{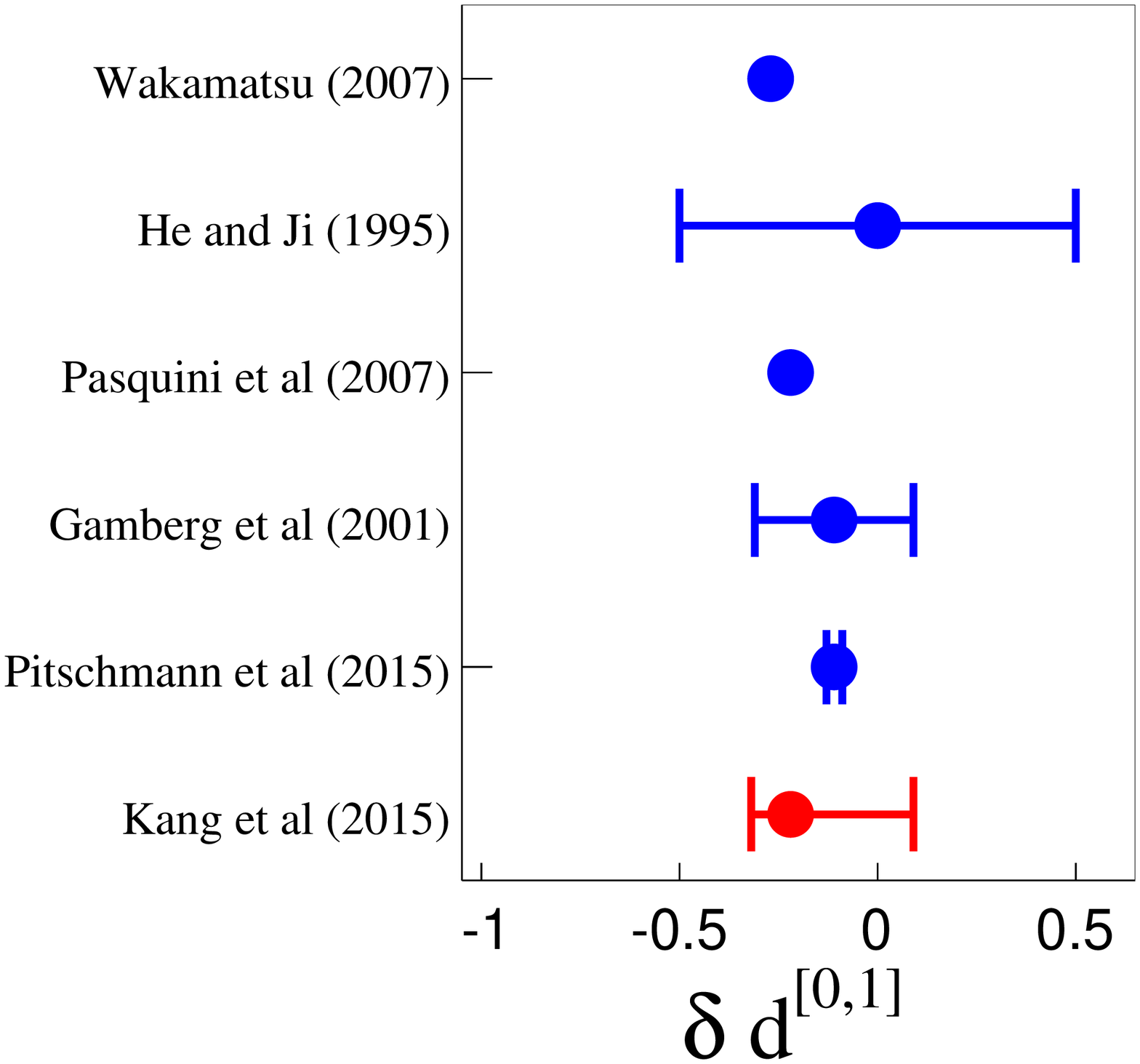} 
\caption{ Comparison of tensor charge  $\delta q^{[0,1]}$  for $u$-quark and $d$-quark in the whole region of $x$ from this paper at 90\% C.L. (Kang et al 2015) at $Q^2=10$ GeV$^2$ and results from Refs.~\cite{Pitschmann:2014jxa,Gamberg:2001qc,Pasquini:2006iv,He:1994gz,Wakamatsu:2007nc}.}
\label{fig:comparison_models}
\end{figure}

In Fig.~\ref{fig:comparison_models} we compare tensor charge  $\delta q^{[0,1]}$  for $u$ and $d$-quarks  from this paper at 90\% C.L.  at $Q^2=10$ GeV$^2$ and results from various model estimates of Refs.~\cite{Pitschmann:2014jxa,Gamberg:2001qc,Pasquini:2006iv,He:1994gz,Wakamatsu:2007nc}. One can see that our results are close to results of Ref.~\cite{Gamberg:2001qc} that actually used  the approximate mass degeneracy of the light axial vector
mesons ($a_1$(1260), $b_1$(1235) and $h_1$(1170)) and pole dominance
to calculate the tensor charge. DSE calculations of tensor charge of Ref.~\cite{Pitschmann:2014jxa} are also close to our results.

Finally we present our estimates for the isovector nucleon tensor charge $g_T= \delta u - \delta d$:
\begin{eqnarray}
 g_{T} &=&  +0.61^{+0.26}_{-0.51} \ , 
\end{eqnarray}
at 90\% C.L. and 
\begin{eqnarray}
 g_{T} &=&  +0.61^{+0.15}_{-0.25} \ , 
\end{eqnarray}
at 68\% C.L.at  $Q^2=10$ GeV$^2$. This result can be compared to lattice QCD calculations.

\begin{figure}[tbp]
\centering
\includegraphics[width=6cm]{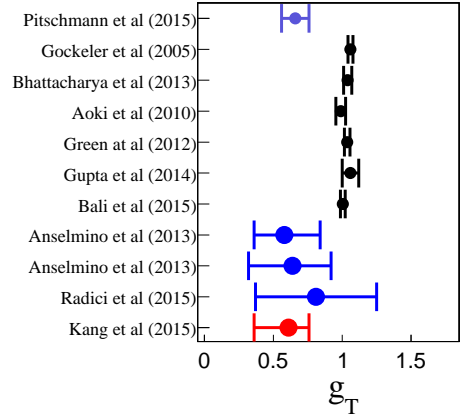}   
\caption{ Comparison of the isovector nucleon tensor charge $g_T$   from this paper at 68\% C.L. (Kang et al 2015) at $Q^2=10$ GeV$^2$ and result from Ref.~\cite{Radici:2015mwa}  (Radici et al 2015) at 68\% CL and  $Q^2=4$ GeV$^2$,
and Ref.~\cite{Anselmino:2013vqa} at 95\% CL standard and polynomial fit (Anselmino et al 2013) at $Q^2=0.8$ GeV$^2$. Other points are lattice computation  at $Q^2=4$ GeV$^2$ of Bali et al Ref.~\cite{Bali:2014nma},  Gupta et al Ref.~\cite{Gupta:2015tpa}, Green et al Ref.~\cite{Green:2012ej}, Aoki et al Ref.~\cite{Aoki:2010xg}, Bhattacharya et al  ref.~\cite{Bhattacharya:2013ehc}, Gockeler et al Ref.~\cite{Gockeler:2005cj}. Pitschmann et al is DSE calculation at $Q^2=4$ GeV$^2$  Ref.~\cite{Pitschmann:2014jxa}.}
\label{fig:comparison_gt}
\end{figure}

In Fig.~\ref{fig:comparison_gt} we compare our result with extraction of Radici et al Ref.~\cite{Radici:2015mwa} at  $Q^2=4$ GeV$^2$, Anselmino et al  Ref.~\cite{Anselmino:2013vqa} standard and polynomial   at $Q^2=0.8$ GeV$^2$, and a series of lattice computations.
Bali et al Ref.~\cite{Bali:2014nma} estimate $g_T$ at $m_\pi \simeq 150$ MeV using RQCD  with $2$ flavor NPI Wilson-clover fermions,
Gupta et al Ref.~\cite{Gupta:2015tpa}  use $2+1+1$ flavor HISQ lattices generated by the MILC collaboration with lowest $m_\pi = 130$ MeV,
Green et al Ref.~\cite{Green:2012ej} use  $2+1$ flavor BMW clover-improved Wilson action with pion masses between 149 and 356 MeV, Aoki et al use gauge configurations generated by the RBC and UKQCD Collaborations with $(2+1)$-flavor QCD with domain wall fermions, PNDME Collaboration Bhattacharya at al  \cite{Bhattacharya:2013ehc} use wo ensembles of highly improved staggered quarks lattices generated by the MILC collaboration with $2 + 1 + 1$ dynamical flavors at a lattice spacing of $0.12$ fm and with light- quark masses corresponding to pions with masses 310 and 220 MeV. references to other calculations of $g_T$ on lattice can be found for instance in Ref.~\cite{Bhattacharya:2013ehc}. Ref.~\cite{Gockeler:2005cj}  uses $n_f=2$ lattice QCD, based on clover-improved Wilson fermions. One can see from Fig.~\ref{fig:comparison_gt} that all phenomenological extractions indicate {\em small} values for the isovector nucleon tensor charge compared to lattice QCD. 
DSE computations of $g_T$ at $Q^2=4$ GeV$^2$ were performed in Ref.~\cite{Pitschmann:2014jxa} and the result is different from most of lattice computations and closer to phenomenological extraction from the data.

The value of $g_T$ extracted from the data may influence searches of BSM physics that depend on $g_T$ \cite{DelNobile:2013sia,Bhattacharya:2011qm}. One can see that our determination of $g_T$ is the most precise existing extraction from experimental data.

The isoscalar nucleon tensor charge $g_T^0= \delta u + \delta d$ can
be readily computed using our results.
 We present result for $g_T^0$ for completeness
\begin{eqnarray}
g_T^0 = +0.17^{+0.47}_{-0.30} \, ,
\end{eqnarray}
at 90\% C.L. at  $Q^2=10$ GeV$^2$.

Refs.~\cite{Pobylitsa:2000tt,Pobylitsa:2002fr,Pobylitsa:1996rs} explores large-$N_c$ behavior of parton distributions in QCD and predicts that
\begin{eqnarray}
|h_1^u(x) - h_1^d(x)| \gg |h_1^u(x) + h_1^d(x)| \, ,
\end{eqnarray}
we indeed observe that transversity for $u$ and $d$-quarks are of similar magnitude and opposite signs and $g_T > g_T^0$ and thus our results are compatible with large-$N_c$ predictions.

%
%
\section{Summary}
\label{sec:summary}
In this paper, we have performed a global analysis of the Collins azimuthal asymmetries
in $e^+e^-$ annihilation and SIDIS processes, for the first time, with full QCD dynamics taken into 
account, including the appropriate TMD evolution effects at the NLL$'$ order and perturbative
QCD corrections at the NLO. The valence quark contributions to the nucleon
tensor charge were estimated based on our analysis. Let us summarize the major results
of this comprehensive study.

First, the full QCD evolution effects are crucial to describe the Collins asymmetries in the
back-to-back di-hadron productions in $e^+e^-$ annihilations, where current data come
from the $B$-factories at the center of mass energy around $10.6$ GeV. At this energy range,
the TMD evolution has significant effect on the asymmetry distributions as functions
of the transverse momentum, and the longitudinal momentum fractions carried by the hadrons
in the fragmentation processes. 
These features have been clearly demonstrated in Figs.~\ref{fig:babar}-\ref{fig:babarz}.
In particular, the transverse momentum dependence illustrates the effects
coming from the Sudakov resummation form factors where the perturbative part
plays an important role due to large value of the resolution scale  $Q\simeq 10.6$ (GeV). The associated scale evolution effects
in the $\hat H^{(3)}(z)$ is another important aspect in the calculations. The evolution 
kernel is different from that of the unpolarized fragmentation function, and it changes the functional form dependence
of $z_{h1}$ and $z_{h2}$. In addition, there is cancellation between favored and unfavored
Collins fragmentation functions, not only the shape but also the size are modified with the 
full evolution effects taken into account. 

Second, because of relative narrow $Q^2$ range in the current SIDIS data, the
evolution effects are not so evident as compared to that in $e^+e^-$ annihilation
processes. This was shown in Figs.~\ref{fig:hermes_comp} and \ref{fig:compass_comp}.
However, we would like to emphasize that, in order to precisely constrain
the quark transversity distributions, we need to perform the complete QCD evolution
in the theoretical calculations of the asymmetries to compare to the experimental data. 
This will become more important with high precision data from future experiments
at the Jefferson Lab 12 GeV upgrade~\cite{Dudek:2012vr} and the planned
Electron Ion Collider~\cite{Boer:2011fh,Accardi:2012qut,Aschenauer:2014twa}.

Third, the quark transversity distributions from our analysis are comparable to 
previous determinations, including the leading order analysis of the same Collins
asymmetries in SIDIS and $e^+e^-$ annihilation processes, and the di-hadron
fragmentation channel in DIS and $e^+e^-$ processes, see Fig.~\ref{fig:comparison_torinopavia}. 
In particular, the consistency between the Collins asymmetry analysis and the di-hadron fragmentation
analysis is a strong encouragement toward a future global fit to include all 
experimental data to constrain the quark transversity distributions. 

We observe, however, the Collins fragmentation functions from our analysis
are quite different from those determined from the leading order analysis in Ref.~\cite{Anselmino:2013vqa}, although
they are in the same order of magnitude. To further test the evolution effects, we 
emphasize the importance of future experiment measurements, in particular, in
the energy range different from $B$-factories, such as those from the BEPC II at the experiment BESIII. 
We have made predictions for these experiments in Figs.~\ref{fig:babar_unpol} and \ref{fig:besiii}. We hope the data will become 
available soon, and can be included into the global fit in the near future. We encourage  BELLE,  
{\em BABAR}    and BESIII Collaborations to perform the analysis of the data on unpolarized cross-sections as such data are curtail for our understanding of TMD fragmentation functions.

Finally, we summarize the nucleon tensor charge contribution from our analysis,
\begin{eqnarray}
\delta u^{[0.0065,0.35]} &=&  +0.30_{-0.12}^{+0.08} \ ,\\
\delta d^{[0.0065,0.35]} &=&  -0.20_{-0.11}^{+0.28}   \ ,
\end{eqnarray}
at 90\% C.L. at  $Q^2=10$ GeV$^2$, in the kinematic
range covered by the current experiments.

\begin{eqnarray}
\delta u^{[0.0065,0.35]} &=&  +0.30^{+0.04}_{-0.07} \ ,\\
\delta d^{[0.0065,0.35]} &=&  -0.20_{-0.07}^{+0.12}   \ , 
\end{eqnarray}
at 68\% C.L. at  $Q^2=10$ GeV$^2$.

If we extend to the complete $x$ range by assuming the 
model dependence in our fit, we would obtain 
\begin{eqnarray}
\delta u^{[0,1]} &=&  +0.39_{-0.20}^{+0.16} \ ,\\
\delta d^{[0,1]} &=& -0.22_{-0.10}^{+0.31} \ ,
\end{eqnarray}
at 90\% C.L. at  $Q^2=10$ GeV$^2$ and
\begin{eqnarray}
\delta u^{[0,1]} &=&  +0.39_{-0.11}^{+0.07} \ ,\\
\delta d^{[0,1]} &=& -0.22_{-0.08}^{+0.14} \ ,
\end{eqnarray}
at 68\% C.L. at  $Q^2=10$ GeV$^2$. We emphasize that the above
constraints depend on the functional form in our analysis, and the
numbers quoted here should be taken cautiously. It is, nevertheless, 
interesting to compare to previous determinations.
In Fig.~\ref{fig:comparison_torinopavia} we show the comparisons of the nucleon
tensor charges between our results and other determinations,
together with some model calculations and the lattice 
computations.

Much of improvements can be made in the future. First, more experimental
data are in horizon from the 12 GeV upgrade of Jefferson Lab experiments, which 
actually will cover large-$x$ region and is of crucial importance to constrain the quark transversity
distribution in that region. Since the nucleon tensor charge contribution
is an integral of the quark transversity distribution, future Jefferson Lab data will be 
very important to reduce the uncertainties quoted above, and the uncertainties
we can not address at the moment, such as the kinematic extension to obtain
$\delta q^{[0,1]}$. 

The TMD evolution and the procedure to perform the global 
analysis will be an important part in the future analysis for other observables, for
example, the Sivers asymmetries in SIDIS. We plan to carry out this analysis in a future publication.

A number of improvements can be pursued in the theoretical part of the formalism. In this paper, we
have taken the approximate evolution kernel for the twist-three quark-gluon-quark 
correlation contribution to the fragmentation function $\hat H^{(3)}(z)$. 
For a complete analysis, we should include other terms in this evolution 
equation. Although it may not be possible to have a closed evolution 
equations for both $\hat H^{(3)}(z)$ and the related twist-three fragmentation
functions $H_D(z_1,z_2)$, one should be able to estimate the contributions
from these additional terms. Second, with more experimental data available, we 
shall include the flavor dependence in the non-perturbative form factors
in the Collins fragmentation function in the CSS resummation formalism. 
In this paper, we have assumed that they are flavor independent. The flavor dependence of distribution and fragmentation functions will be explored 
in the future analysis with more data available, in particular, the data on
the transverse momentum dependence of the asymmetries in $e^+e^-$ annihilation
processes.

As a final remark, we would like to emphasize that our results and the methodology 
in the analysis will play an important role in phenomenological applications of perturbative
QCD to the vast experimental data on SIDIS, Drell-Yan and $e^+e^-$ and  in extraction of the relevant TMD parton
distributions of the nucleon.

\section{acknowledgements }

We thank  D.~Boer, J.~C.~Collins, L.~Gamberg, J.~Qiu, W.~Vogelsang, D.~Richards, and C.~-P.~Yuan for discussions and suggestions.
This material is based upon work supported by the U.S. Department of Energy,
Office of Science, Office of Nuclear Physics, under contracts No.~DE-AC02-05CH11231 (P.S., F.Y.), No.~DE-AC52-06NA25396 (Z.K.), and No.~DE-AC05-06OR23177 (A.P.).

\appendix
\section{One-loop Calculation of the Collins Asymmetry in SIDIS}

To study the perturbative corrections and extract the hard factor in the
Eq.~\eqref{eq:fut_b}, we need to carry out a calculation for $\widetilde F_{\rm collins}$
at one-loop order. The leading order expression and the virtual diagram
contributions follow that in the previous calculations for, e.g., the Sivers
single spin asymmetry in SIDIS~\cite{Kang:2011mr,Sun:2013dya}. For the real gluon radiation, we use the
results in Ref.~\cite{Yuan:2009dw}  
\begin{eqnarray}
  F_{\rm collins}^\beta \Bigg|_{P_{h\perp}\ll Q}
      &=&
      \frac{z_hP_{h\perp}^\beta}{(\vec{P}_{h\perp}^2)^2}
      \frac{\alpha_s}{2\pi^2}C_F\int_{x_B}^1 \frac{dx}{x}\int_{z_h}^1\frac{dz}{z}\sum_q e_q^2 h_1^q(x)
      (z) \left\{ \hat{H}_{h/q}^{(3)}(z) \delta(\hat \xi-1)\left[\frac{2\xi^2}{(1-\xi)_+}\right]\right.\nonumber\\
&&+  \delta(\xi-1) \left[ -2\hat\xi\left(z^3\frac{\partial}{\partial z}\frac{\hat{H}_{h/q}^{(3)}(z)}{z^2}\right)
      +\hat{H}_{h/q}^{(3)}(z)\frac{2\hat\xi^2}{(1-\hat\xi)_+}\right]\nonumber\\
&&\left.+2\delta(\hat\xi-1)\delta(\xi-1)\hat{H}_{h/q}^{(3)}(z)\ln\frac{z_h^2Q^2}{\vec{P}_{h\perp}^2}
\right\} \ ,
\end{eqnarray}
where $\xi = x_B/x$, $\hat\xi = z_h/z$ and we only keep the most important diagonal contributions from $\hat{H}_{h/q}^{(3)}(z)$, and
the contributions from $\hat H_D(z_{1},z_{2})$ can be found from Ref.~\cite{Yuan:2009dw}.
By applying the Fourier transform (some of the useful integrals are
listed in the Appendix of Ref.~\cite{Kang:2011mr} and Eq.~(42) of Ref.~\cite{Dai:2014ala}), we obtain the following result
for $\widetilde{F}_{\rm collins}^\beta(Q,b)$,
\begin{eqnarray}
\widetilde{F}_{\rm collins}^\beta\Bigg|_{real}&=& \frac{\alpha_s}{2\pi}C_F\left(\frac{ib^\alpha}{2}\right) \int_{x_B}^1 \frac{dx}{x}\int_{z_h}^1\frac{dz}{z}\sum_q e_q^2 h_1^q(x)\hat{H}_{h/q}^{(3)}(z)
\left\{\left(-\frac{1}{\epsilon}+\ln\frac{c_0^2\hat\xi^2}{b^2\mu^2}\right)\right.\nonumber\\
&&\times \left[\delta(1-\hat\xi)\left(\frac{2\xi}{(1-\xi)_+}\right)+\delta(1-\xi)\left(\frac{2\hat\xi}{(1-\hat\xi)_+} +2\delta(\hat\xi-1)\right)\right]\nonumber\\
&&+2\delta(1-\xi)\delta(1-\hat\xi)\left[\frac{1}{\epsilon^2}-\frac{1}{\epsilon}\ln\frac{Q^2}{\mu^2}
+\frac{1}{2}\left(\ln\frac{Q^2}{\mu^2}\right)^2-\frac{1}{2}\left(\ln\frac{Q^2b^2}{c_0^2}\right)^2-\frac{\pi^2}{12}\right]\nonumber\\
&&\left.-2\delta(1-\xi)\delta(1-\hat\xi)\left(-\frac{1}{\epsilon}+\ln\frac{c_0^2}{b^2\mu^2}\right)\right\} \ ,
\end{eqnarray}
where we have partial integrated out the derivative terms in the previous
equation to simplify the above expression.
Clearly, the real diagrams contributions contain soft divergence ($1/\epsilon^2$),
which will be cancelled by the virtual diagrams contributions.
The virtual diagram contributes to a factor,
\begin{equation}
\frac{\alpha_s}{2\pi}C_F\left[-\frac{2}{\epsilon^2}-\frac{3}{\epsilon}+\frac{2}{\epsilon}\ln\frac{Q^2}{\mu^2}
+\frac{1}{6}\pi^2+3\ln\frac{Q^2}{\mu^2}-\left(\ln\frac{Q^2}{\mu^2}\right)^2-8\right]  \ .
\end{equation}
After canceling out these divergences, we have the total
contribution at one-loop order,
\begin{eqnarray}
\widetilde{F}_{\rm collins}^\beta&=&\frac{\alpha_s}{2\pi} \int_{x_B}^1 \frac{dx}{x}\int_{z_h}^1\frac{dz}{z}\sum_q e_q^2 h_1^q (x)\hat{H}_{h/q}^{(3)}(z)\left(\frac{ib^\alpha}{2}\right)
\left\{\left(-\frac{1}{\epsilon}+\ln\frac{c_0^2\hat\xi^2}{b^2\mu^2}\right)\right.\nonumber\\
&&\times \left(\hat{P}_{q\to q}^{\rm c}(\hat\xi)\delta(1-\xi)+{P}_{q\to q}^{h_1}(\xi)\delta(1-\hat \xi)\right)\nonumber\\
&&\left.+\delta(1-\xi)\delta(1-\hat\xi)C_F\left[3\ln\frac{Q^2b^2}{c_0^2}-\left(\ln\frac{Q^2b^2}{c_0^2}\right)^2-8\right]\right\} \ ,
\end{eqnarray}
where ${P}$ represents the associated splitting kernels. They can be derived from the above
results,
\begin{eqnarray}
{P}_{q\to q}^{h_1}(\xi)&=&C_F \left[ \frac{2\xi}{(1-\xi)_+}+\frac{3}{2}\delta(1-\xi) \right]\ , \\
\hat{P}_{q\to q}^{\rm c}(\hat \xi)&=&C_F \left[ \frac{2\hat\xi}{(1-\hat\xi)_+}+\frac{3}{2}\delta(1-\hat\xi)+\cdots \right]\ , 
\label{eq:collins_splitting}
\end{eqnarray}
where we only list the part we have shown in the above from the contribution from $\hat H^{(3)}(z)$ term.
In general, the evolution of twist-three correlation functions involves multiple
parton correlation contributions, for which there is no homogenous form. 

To demonstrate the TMD factorization and calculate the hard factor in the TMD
factorization, we have to calculate the transverse momentum dependence in the 
quark transversity distribution and the Collins fragmentation function at one-loop 
order. For the transversity
distribution, we have 
\begin{eqnarray}
h_1(x_B,k_\perp)\Bigg|_{real}=\frac{\alpha_s}{2\pi^2}C_F\frac{1}{k_\perp^2}\int_{x_B}^1 \frac{dx}{x}h_1(x)\left[\frac{2\xi}{(1-\xi)_+}+
\delta(1-\xi)\left(\ln\frac{x_B^2\zeta^2}{k_\perp^2}-1\right)\right] \ ,
\end{eqnarray}
for the un-subtracted distribution in the JMY scheme. Adding the virtual contribution,
\begin{equation}
h_1(x_B,k_\perp)\Bigg|_{virtual}=h_1(x_B)\frac{\alpha_s}{2\pi}C_F\left[-\frac{1}{\epsilon^2}-\frac{5}{2\epsilon}
+\frac{1}{\epsilon}\ln\frac{x_B^2\zeta^2}{\mu^2}+\frac{x_B^2\zeta^2}{\mu^2}-\frac{1}{2}\ln^2\left(\frac{x_B^2\zeta^2}{\mu^2}\right)
-\frac{5}{12}\pi^2-2\right] \ ,\label{eq:a8}
\end{equation}
we obtain the total contribution for the un-subtracted quark transversity TMD at the one-loop order,
\begin{eqnarray}
h_1^{unsub}(x_B,b ;\zeta,\mu)\Bigg|_{\alpha_s}&=&\frac{\alpha_s}{2\pi}\int_{x_B}^1\frac{dx}{x}h_1(x)\left\{
\left(-\frac{1}{\epsilon}+\ln\frac{c_0^2}{b^2\bar\mu^2}\right)P^{h_1}_{q\to q}(\xi)-C_F\delta(1-\xi)\ln\frac{c_0^2}{b^2\mu^2}\right.\nonumber\\
&&\left.+C_F\delta(1-\xi)\left[\frac{3}{2}\ln\frac{b^2\mu^2}{c_0^2}+\ln\frac{x_B^2\zeta^2}{\mu^2}-\frac{1}{2}\ln^2\left(\frac{x_B^2\zeta^2b^2}{c_0^2}\right)
-2-\frac{\pi^2}{2}\right]\right\} \ ,
\end{eqnarray}
in the JMY scheme. Therefore, the subtracted TMD quark transversity distribution can be written
as
\begin{eqnarray}
h_1^{sub (JMY)}(x_B,b ;\zeta,\mu)\Bigg|_{\alpha_s}&=&\frac{\alpha_s}{2\pi}\int_{x_B}^1\frac{dx}{x}h_1(x,\bar\mu)\left\{
\ln\frac{c_0^2}{b^2\bar\mu^2}P^{h_1}_{q\to q}(\xi)\right.\nonumber\\
&&\left.+C_F\delta(1-\xi)\left[\left(\frac{3}{2}+\ln\rho\right)\ln\frac{b^2\mu^2}{c_0^2}+\ln\frac{x_B^2\zeta^2}{\mu^2}-\frac{1}{2}\ln^2\left(\frac{x_B^2\zeta^2b^2}{c_0^2}\right)
-2-\frac{\pi^2}{2}\right]\right\} \ ,
\end{eqnarray}
where we have also applied the renormalization for the integrated transversity distribution.
By setting $x_B^2\zeta^2=\rho\mu_b^2$ and $\mu=\mu_b$ as the initial scales for the
TMD evolutions, we obtain the $C$-coefficient and the
hard function ${\cal H}_{1q}$ of Eq.~(\ref{tmdh}) as
\begin{equation}
\widetilde{\cal H}_{1q}^{(JMY)}=1+\frac{\alpha_s}{2\pi}C_F\left[\ln\rho-\frac{1}{2}\ln^2\rho-\frac{\pi^2}{2}-2\right], ~~\delta C_{q\to q}(\xi,\mu_b)=\delta(1-\xi)\left(1+{\cal O}(\alpha_s^2)\right) \ .
\end{equation}
Similarly, we can carry out the calculations in the JCC scheme, for which we have the
TMD quark transversity at one-loop order,
\begin{eqnarray}
h_1^{sub (JCC)}(x_B,b;\zeta,\mu)\Bigg|_{\alpha_s}&=&\frac{\alpha_s}{2\pi}\int_{x_B}^1\frac{dx}{x}h_1(x,\bar\mu)\left\{
\ln\frac{c_0^2}{b^2\bar\mu^2}P^{h_1}_{q\to q}(\xi)\right.\nonumber\\
&&\left.+C_F\delta(1-\xi)\left[\frac{3}{2}\ln\frac{b^2\mu^2}{c_0^2}+\frac{1}{2}\ln^2\left(\frac{\zeta_c^2}{\mu^2}\right)
-\frac{1}{2}\ln^2\left(\frac{\zeta_c^2b^2}{c_0^2}\right)\right]\right\} \ .
\end{eqnarray}
Applying the above result into Eq.~(\ref{tmdh}) and setting $\zeta_c=\mu=\mu_b$ as the initial
scales for the TMD evolutions, we have 
\begin{equation}
\widetilde{\cal H}_{1q}^{(JCC)}=1+{\cal O}(\alpha_s^2), ~~\delta C_{q\to q}(\xi,\mu_b)=\delta(1-\xi)\left(1+{\cal O}(\alpha_s^2)\right) \ .
\end{equation}
The above calculations can be extended to the Collins fragmentation function. The
transverse momentum dependence can be calculated from perturbative
QCD, and written in terms of the twist-three fragmentation function,
\begin{eqnarray}
H_1^\perp(z_h,p_\perp)\Bigg|_{real}&=&\frac{\alpha_s}{2\pi^2}C_F\frac{1}{(p_\perp^2)^2}
\int \frac{dz}{z} \left[ -2\hat\xi\left(z^3\frac{\partial}{\partial z}\frac{\hat{H}_{h/q}^{(3)}(z)}{z^2}\right)\right.\nonumber\\
&&\left.
      +\hat{H}_{h/q}^{(3)}(z)\left(\frac{2\hat\xi^2}{(1-\hat\xi)_+}+\delta(1-\hat\xi)\left(\ln\frac{\zeta^2}{p_\perp^2}-2\right)\right)\right]\ .
\end{eqnarray}
Fourier transforming into $b$-space and adding the virtual diagram contribution (similar to
that in Eq.(\ref{eq:a8})), we obtain the un-subtracted Collins fragmentation function at one-loop order,
\begin{eqnarray}
\widetilde{H}_1^{\perp \alpha \, unsub}(x_B,b;\zeta,\mu)\Bigg|_{\alpha_s}&=&\frac{ib^\alpha}{2}
\frac{\alpha_s}{2\pi}\int\frac{dz}{z}\hat{H}_{h/q}^{(3)}(z)\left\{
\left(-\frac{1}{\epsilon}+\ln\frac{c_0^2\hat\xi^2}{b^2\bar\mu^2}\right)\hat{P}^{c}_{q\to q}(\hat\xi)-C_F\delta(1-\hat\xi)\ln\frac{c_0^2}{b^2\mu^2}\right.\nonumber\\
&&\left.+C_F\delta(1-\hat\xi)\left[\frac{3}{2}\ln\frac{b^2\mu^2}{c_0^2}+\ln\frac{\hat\zeta^2}{\mu^2}-\frac{1}{2}\ln^2\left(\frac{\hat\zeta^2b^2}{c_0^2}\right)-2-\frac{\pi^2}{2}\right]\right\} \ ,
\end{eqnarray}
in the JMY scheme. For the subtracted Collins fragmentation function, we have,
\begin{eqnarray}
\widetilde{H}_1^{\perp \alpha \, sub\,  (JMY)}(x_B,b;\zeta,\mu)\Bigg|_{\alpha_s}&=&\frac{ib^\alpha}{2}
\frac{\alpha_s}{2\pi}\int\frac{dz}{z}\hat{H}_{h/q}^{(3)}(z)\left\{\ln\frac{c_0^2\hat\xi^2}{b^2\bar\mu^2}
\hat{P}^{c}_{q\to q}(\hat\xi)\right.\nonumber\\
&&\left.+C_F\delta(1-\hat\xi)\left[\left(\frac{3}{2}+\ln\rho\right)\ln\frac{b^2\mu^2}{c_0^2}+\ln\frac{\hat\zeta^2}{\mu^2}-\frac{1}{2}\ln^2\left(\frac{\hat\zeta^2b^2}{c_0^2}\right)-2-\frac{\pi^2}{2}\right]\right\} \ .
\end{eqnarray}
Similarly, we obtain the subtracted Collins fragmentation function in the JCC scheme,
\begin{eqnarray}
\widetilde{H}_1^{\perp \alpha \,sub \, (JCC)}(x_B,b;\zeta,\mu)\Bigg|_{\alpha_s}&=&\frac{ib^\alpha}{2}
\frac{\alpha_s}{2\pi}\int\frac{dz}{z}\hat{H}_{h/q}^{(3)}(z)\left\{\ln\frac{c_0^2\hat\xi^2}{b^2\bar\mu^2}
\hat{P}^{c}_{q\to q}(\hat\xi)\right.\nonumber\\
&&\left.+C_F\delta(1-\hat\xi)\left[\frac{3}{2}\ln\frac{b^2\mu^2}{c_0^2}+\frac{1}{2}\ln^2\left(\frac{\hat\zeta_c^2}{\mu^2}\right)-\frac{1}{2}\ln^2\left(\frac{\hat\zeta^2b^2}{c_0^2}\right)\right]\right\} \ .
\end{eqnarray}
From the above results, we derive the associated $C$-functions,
\begin{eqnarray}
\widetilde{\cal H}_c^{(JMY)}&=&1+\frac{\alpha_s}{2\pi}C_F\left[\ln\rho-\frac{1}{2}\ln^2\rho-\frac{\pi^2}{2}-2\right], 
~~\delta \hat C_{q\to q}(\hat\xi,\mu_b)=\delta(1-\hat\xi)+\frac{\alpha_s}{2\pi}C_F\hat{P}^c_{q\to q}(\hat \xi)\ln\hat \xi^2\ ,\nonumber\\
\widetilde{\cal H}_c^{(JCC)}&=&1+{\cal O}(\alpha_s^2), ~~\delta \hat C_{q\to q}(\hat\xi,\mu_b)=\delta(1-\hat\xi)
+\frac{\alpha_s}{2\pi}C_F\hat{P}^c_{q\to q}(\hat \xi)\ln\hat \xi^2 \ .
\end{eqnarray}
Finally, we can obtain the hard factors in both schemes.
For example, in the Ji-Ma-Yuan scheme,
\begin{equation}
H_{\rm collins}^{({\rm DIS})JMY}(Q;\mu)=H_{UU}^{({\rm DIS})JMY}(Q;\mu)=  1+  
\frac{\alpha_s}{2\pi}C_F\left[\ln\frac{Q^2}{\mu^2}+\ln\rho^2\ln\frac{Q^2}{\mu^2}-\ln\rho^2+\ln^2\rho+\pi^2-4\right] \ .\label{hqjmy}
\end{equation}
Note that the hard part is the same for $F_{UU}$ and $F_{UT}$ that is why we used the same notation $H$ in Eqs.~(\ref{eq:fuu_b},\ref{eq:fut_b})
Similarly, for the Collins-11 TMD scheme, we have
\begin{equation}
H_{\rm collins}^{({\rm DIS})JCC}(Q;\mu)=H_{UU}^{({\rm DIS})JCC}(Q;\mu)=  1+  
\frac{\alpha_s}{2\pi}C_F\left[3\ln\frac{Q^2}{\mu^2}-\ln^2\left(\frac{Q^2}{\mu^2}\right)-8\right] \ .\label{hqjcc}
\end{equation}
These hard factors can be calculated from the factorization of $\widetilde{F}_{\rm collins}^\alpha$, or from
simply the virtual graphs for both the cross sections and the parton distribution and fragmentation functions. 
We will get the consistent results.

In the end, the $C$-functions in Eq.~(53) can be calculated from the above results,
\begin{eqnarray}
\delta C^{(\rm SIDIS)}(\xi)&=&\delta C(\xi)\times \widetilde{\cal H}_{1q}\times\sqrt{H_{\rm collins}^{(\rm SIDIS)}(\mu=Q)}\ ,\\
\delta \hat C^{(\rm SIDIS)}(\hat\xi)&=&\delta \hat{C}(\hat\xi)\times \widetilde{\cal H}_{c}\times\sqrt{H_{\rm collins}^{(\rm SIDIS)}(\mu=Q)}\ ,
\end{eqnarray}
where the scheme dependence is cancelled out between ${\cal H}_{1q}$ and $H_{\rm collins}^{(\rm SIDIS)}$.
In particular, the $\rho$ dependence disappear in the JMY scheme when applying the above 
formulas to calculate the $C$-functions in the standard CSS resummation.
Similarly, we can calculate the $C$-functions for the $e^+e^-$ annihilation processes,
\begin{eqnarray}
\hat {C}^{(\rm e^+e^-)}(\hat \xi)&=&\hat C(\hat \xi)\times \widetilde{\cal D}_{q}\times\sqrt{H_{uu}^{({\rm e^+e^-})}(\mu=Q)}\ ,\\
\delta \hat C^{(\rm e^+e^-)}(\hat\xi)&=&\delta \hat{C}(\hat\xi)\times \widetilde{\cal H}_{c}\times\sqrt{H_{\rm collins}^{({\rm e^+e^-})}(\mu=Q)}\ .
\end{eqnarray}
Again, the scheme dependence is cancelled out between the last two factors in the above equations.
Comparing the SIDIS and $e^+e^-$ processes, we also find out that the difference comes
from the hard factors.

 \bibliography{\BibPath/collins-final}
\end{document}